\documentclass[sigconf, nonacm]{acmart}
\AtBeginDocument{%
  }

\usepackage{algorithm}
\usepackage[noend]{algpseudocode}
\usepackage{circledsteps}
\usepackage{amsmath}
\usepackage{mathtools}
\usepackage{amsthm}
\usepackage{dsfont}
\usepackage{multirow}
\usepackage{euscript}
\usepackage{booktabs}
\usepackage{physics}
\usepackage{pifont}
\usepackage{capt-of}
\usepackage{cleveref}
\usepackage{listings}
\usepackage{spverbatim}
\usepackage{enumitem}
\usepackage{mdframed}
\usepackage{makecell}
\usepackage{float}
\usepackage{xparse}
\usepackage{xifthen} % For easier string comparison
\usepackage{makecell} % add this in your preamble
\usepackage{etoolbox}
\newtoggle{techreport}
\toggletrue{techreport}

\newcommand{\str}[1]{\texttt{#1}}

\Crefname{figure}{Fig.}{Figs.}
\usepackage[cal=boondox,scr=txupr]{mathalpha}
\usepackage{titlesec}

\setcopyright{acmlicensed}
\copyrightyear{2018}
\acmYear{2018}
\acmDOI{XXXXXXX.XXXXXXX}
\acmConference[Conference acronym 'XX]{Make sure to enter the correct
  conference title from your rights confirmation email}{June 03--05,
  2018}{Woodstock, NY}
\acmISBN{978-1-4503-XXXX-X/2018/06}

\newcommand\vldbdoi{XX.XX/XXX.XX}
\newcommand\vldbpages{XXX-XXX}
\newcommand\vldbvolume{14}
\newcommand\vldbissue{1}
\newcommand\vldbyear{2020}
\newcommand\vldbauthors{\authors}
\newcommand\vldbtitle{\shorttitle} 
\newcommand\vldbavailabilityurl{}
\newcommand\vldbpagestyle{plain}

\begin{document}

\settopmatter{printfolios=true}  % enables page numbers in top matter
\pagestyle{plain}     
\title{\vspace{-0.2cm}Featurized-Decomposition Join: \\Low-Cost Semantic Joins with Guarantees}
%\title{\textcolor{cadmiumgreen}{J}\textcolor{bargain}{AI}\textcolor{cadmiumgreen}{N}: Cheap \textcolor{bargain}{AI}-Powered \textcolor{cadmiumgreen}{J}OI\textcolor{cadmiumgreen}{N} with Guarantees}
%\title{Cut Costs, Not Accuracy with BARG\textcolor{bargain}{AI}N:\\ AI-Powered Data Processing with Guarantees }
%\title{Cut Costs, Not Accuracy:\\ Guaranteed Accurate AI-Powered Data Processing with BARG\textcolor{bargain}{AI}N}
%\title{BARG\textcolor{bargain}{AI}N:\\ Guaranteed Accurate AI-Powered Data Processing for less}

\newcommand{\name}{FDJ}

\newcommand{\techreport}[1]{#1}
\newcommand{\nontechreport}[1]{}%\textcolor{blue}{#1}}

\newcommand{\rone}[1]{#1}
\newcommand{\rtwo}[1]{#1}
\newcommand{\rthree}[1]{#1}
\newcommand{\rall}[1]{#1}

\newcommand{\sep}[1]{\textcolor{blue}{sep:#1}}
\newcommand{\shreya}[1]{\textcolor{teal}{shreya:#1}}
\renewcommand{\mathfrak}[1]{\mathscr{#1}}
\renewcommand{\undef}[1]{\textcolor{blue!70!black}{\texttt{#1}}}

\definecolor{bargain}{rgb}{0.5859375, 0.01171875, 0.01171875}
\definecolor{cadmiumgreen}{rgb}{0.0, 0.42, 0.24}
\definecolor{burgundy}{rgb}{0.5, 0.0, 0.13}
\definecolor{pink_figma}{rgb}{0.8828125, 0.4765625, 0.84765625}%226,122, 217}
\settopmatter{printacmref=false}
\pagestyle{plain}
%%
%% The "author" command and its associated commands are used to define the authors and their affiliations.

\techreport{
\author{Sepanta Zeighami}
\affiliation{%
  \institution{UC Berkeley}
  %\streetaddress{P.O. Box 1212}
  %\state{Ireland}
}
\email{zeighami@berkeley.edu}

\author{Shreya Shankar}
\affiliation{%
  \institution{UC Berkeley}
}
\email{shreyashankar@berkeley.edu}

\author{Aditya Parameswaran}
\affiliation{%
  \institution{UC Berkeley}
}
\email{adityagp@berkeley.edu}
}

\setlength{\abovedisplayskip}{0pt}
\setlength{\belowdisplayskip}{0pt}
\setlength{\abovedisplayshortskip}{0pt}
\setlength{\belowdisplayshortskip}{0pt}

\setlength{\textfloatsep}{0pt}
\setlength{\floatsep}{0pt}
\setlength{\intextsep}{0pt}
\setlength{\dbltextfloatsep}{0pt}
\setlength{\dblfloatsep}{0pt}
\setlength{\abovecaptionskip}{0pt}
\setlength{\belowcaptionskip}{0pt}
\titlespacing*{\section}{0pt}{0.8ex}{0.5ex}
\titlespacing*{\subsection}{0pt}{0.9ex}{0.7ex}
\titlespacing*{\subsubsection}{0pt}{0.8ex}{0.7ex}

%\addtolength{\topmargin}{-0.1in}
%\addtolength{\textheight}{0.1in}

%\addtolength{\tabcolsep}{-0.4em}

\NewDocumentEnvironment{revcomment}{m o}% m = mandatory, o = optional
{
  \par\vspace{0.5em}%\smallskip
  \IfValueT{#2}{%
    \IfEqCase{#2}{%
      {true}{\noindent\rule{\linewidth}{0.4pt}}
    }[\PackageWarning{revcomment}{Unknown flag `#2', no rule added.}]
  }
  %\par\vspace{-0.5em}
  \noindent\textbf{#1. }\itshape
}
{
  \par%\medskip
}
%\clearpage
\if 0
\setcounter{section}{0}
\pagenumbering{roman}
\renewcommand{\thesection}{\Roman{section}}
\renewcommand{\thefigure}{\Roman{figure}}
\renewcommand{\thetable}{\Roman{table}}
\input{revision_letter}
%\input{rebuttal}
\setcounter{figure}{0}
\setcounter{table}{0}
\renewcommand{\thefigure}{\arabic{figure}}
\renewcommand{\thetable}{\arabic{table}}
\renewcommand{\thesection}{\arabic{section}}
\clearpage
\pagenumbering{arabic}
\setcounter{section}{0} 
\setcounter{page}{1} 
\fi

\begin{abstract}
%\vspace{-0.1cm}
Large Language Models (LLMs) are being increasingly used within data systems to process large datasets with text fields. 
A broad class of such tasks involves a \textit{semantic join}---joining two tables based on a natural language predicate per pair of tuples, evaluated using an LLM. Semantic joins generalize tasks such as entity matching and record categorization, as well as more complex text understanding tasks.
%A broad class of such data processing tasks involve a \textit{Semantic Join}: creating a Join of two tables by applying a user-provided natural language filter to the cross product of the two tables using an LLM.
%\textit{Semantic Join}s represent a broad class of tasks including entity matching, categorizing records and create a Join of tables based on complex relationships specified in natural language. 
A naive implementation is expensive as it requires invoking an LLM for every pair of rows in the cross product. Existing approaches mitigate this cost by first applying embedding-based semantic similarity to filter candidate pairs, deferring to an LLM only when similarity scores are deemed inconclusive.
However, these methods yield limited gains in practice, since semantic similarity may not reliably predict the join outcome.
We propose Featurized-Decomposition Join (\name{} for short), a novel approach for performing semantic joins that significantly reduces cost while preserving quality. 
\name{} automatically extracts features and combines them into a logical expression in conjunctive normal form that we call a \textit{featurized decomposition} to effectively prune out non-matching pairs. 
A featurized decomposition extracts key information from text records and performs inexpensive comparisons on the extracted features. 
We show how to use LLMs to automatically extract reliable features and compose them into logical expressions while providing statistical guarantees on the output result---an inherently challenging problem due to dependencies among features. %Our method can achieve an \textit{asymptotic} cost improvement, reducing cost from $O(n^2)$ of existing solutions to $O(n)$ for a join of two tables with $n$ rows. 
Experiments on real-world datasets show \textbf{up to 10 times reduction in cost} compared with the state-of-the-art while providing the same quality guarantees.%, with this advantage growing with data size. 
\end{abstract}

\maketitle

\if 0
\pagestyle{\vldbpagestyle}
\begingroup\small\noindent\raggedright\textbf{PVLDB Reference Format:}\\
\vldbauthors. \vldbtitle. PVLDB, \vldbvolume(\vldbissue): \vldbpages, \vldbyear.\\
\href{https://doi.org/\vldbdoi}{doi:\vldbdoi}
\endgroup
\begingroup
\renewcommand\thefootnote{}\footnote{\noindent
This work is licensed under the Creative Commons BY-NC-ND 4.0 International License. Visit \url{https://creativecommons.org/licenses/by-nc-nd/4.0/} to view a copy of this license. For any use beyond those covered by this license, obtain permission by emailing \href{mailto:info@vldb.org}{info@vldb.org}. Copyright is held by the owner/author(s). Publication rights licensed to the VLDB Endowment. \\
\raggedright Proceedings of the VLDB Endowment, Vol. \vldbvolume, No. \vldbissue\ %
ISSN 2150-8097. \\
\href{https://doi.org/\vldbdoi}{doi:\vldbdoi} \\
}\addtocounter{footnote}{-1}\endgroup

\ifdefempty{\vldbavailabilityurl}{}{
\vspace{.3cm}
\begingroup\small\noindent\raggedright\textbf{PVLDB Artifact Availability:}\\
The source code, data, and/or other artifacts have been made available at \url{\vldbavailabilityurl}.
\endgroup
}
\fi

\vspace{-0.2cm}
\section{Introduction}\label{sec:intro}
%Large Language Models (LLMs) are being increasingly used as a building block in data systems to process large text datasets. A broad class of such tasks involve a \textit{Semantic join}---joining two tables by applying a user-specified natural language filter to their cross product using an LLM. %, where the column to be joined contains different textual representation of the same entity (e.g., due to the use of acronyms and synonyms), 
A \textit{semantic join} between two tables applies a user-specified natural language filter to their cross product using a Large Language Model (LLM). This operation is now supported by many industrial data systems~\cite{snowflake25, alloydb, databricks-llm, duckdb-llm} as well as open-source ones~\cite{pg-ai, flock, shankar2024docetl, patel2024lotus}. Semantic joins enable a wide range of downstream applications. These include traditional entity matching tasks~\cite{narayan2022can, shankar2024docetl, liu2024declarative} such as identifying when two marketplace listings describe the same product~\cite{kopcke2010evaluation, mudgal2018deep}; large-scale multi-label classification, such as matching doctors' notes for patients to a long list of medical reactions~\cite{patel2024lotus, d2023biodex}; and cross-referencing complex documents based on entities they refer to~\cite{mahari2024lepard, pavlick2016gun, police}. 
\if 0
A \textit{semantic join} between two tables applies a user-specified natural language filter to their cross product using a Large Language Model (LLM). This operation is currently supported by many industrial database systems~\cite{snowflake25, alloydb, databricks-llm, duckdb-llm} as well as open-source ones~\cite{pg-ai,flock, shankar2024docetl, patel2024lotus}. Semantic joins enable many downstream tasks: traditional entity matching tasks~\cite{narayan2022can, shankar2024docetl, liu2024declarative},  e.g., matching different product listings in a marketplace that refer to the same product based on their description \cite{kopcke2010evaluation, mudgal2018deep}; large scale multi-label classification tasks, e.g., matching patient records to medical reactions from a long list of possible reactions \cite{patel2024lotus, d2023biodex}; joining text records based on complex relationships, e.g., mapping legal court cases to legal precedents they cite \cite{mahari2024lepard}; and cross referencing complex documents based on entities they refer to \cite{pavlick2016gun, police}. 
\fi
A real-world instance of the last category that we use as our running example is the \textit{police record} matching task from the Police Records Access Project, a collaborative effort that we are involved in with journalists and public defenders \cite{police}. Here, different police and legal documents (e.g., eyewitness reports, testimonials, court filings, investigation reports) that refer to the same incident need to be matched---a semantic self-join that requires the LLM to determine whether two documents refer to the same incident. Such documents are long and complex, and identifying matching documents requires understanding the incidents they describe, e.g., knowing their date, location, type of police activity, and people involved. This entity matching is useful for journalists analyzing police accountability, for example, to analyze the outcome of court cases for incidents involving police activity with use of force---without semantic joins, the entity matching had traditionally been done by journalists by hand \cite{police}.

Although highly valuable, performing semantic joins is computationally expensive. A naive approach requires invoking an LLM on every record pair in the cross product to determine whether each pair satisfies the join condition. However, making $\Omega(n^2)$ LLM calls for $n$ data records is prohibitive even for moderately sized datasets. LOTUS~\cite{patel2024lotus} instead leverages a model cascade approach, by first computing the cosine similarity between the vector embedding of the two records as a \textit{proxy score} for matching. Record pairs with low similarity are discarded as non-matches, those with high similarity are accepted as matches, and  pairs with inconclusive scores are evaluated by an LLM. This decision is made by setting \textit{cascade thresholds} on proxy scores to determine when to defer to an LLM. To avoid accuracy loss compared to using an LLM on all record pairs, cascade thresholds are set to statistically guarantee the output quality is close to that of using an LLM on every record pair. More recently, BARGAIN~\cite{zeighami2025cut} refines model cascades by improving when to defer to an LLM (i.e., how to set cascade thresholds), thereby reducing the frequency of expensive model calls. %BARGAIN also provide stronger statistical quality guarantees: their join outputs achieve high recall and precision compared to using the LLM on all pairs. However, Lotus’s guarantees hold only asymptotically and in the limit. Thus, as observed in both our experiments and prior work~\cite{zeighami2025cut}, it often fails to meet its theoretical targets in practice.

However, the effectiveness of such model cascade solutions depends on the quality of the embedding vectors. When proxy scores derived from the embeddings closely approximate LLM outputs, they can reduce cost; however, if embeddings poorly encode the join criteria, proxy scores become unreliable and most join decisions will need to be made by an LLM. In practice, text records are often lengthy and contain both relevant and irrelevant information for the join. For instance, in police records, each document may include multiple names, locations, dates, and incident descriptions, as well as irrelevant information such as boilerplate headers. Embedding this information into a single vector fails to capture the relevant details with sufficient fidelity (as is theoretically suggested by \cite{weller2025theoretical}). Thus, semantic similarity becomes a poor indicator of the join, and methods that rely on semantic similarity between embeddings end up using the LLM on most record pairs, providing limited savings. In our police records example, BARGAIN defers to the LLM on around 80\% of all record pairs, thus costing close to that of naively comparing all pairs. We see similar results on other real-world datasets.

\begin{figure}
\vspace{-0.4cm}
    \centering
    \includegraphics[width=\linewidth]{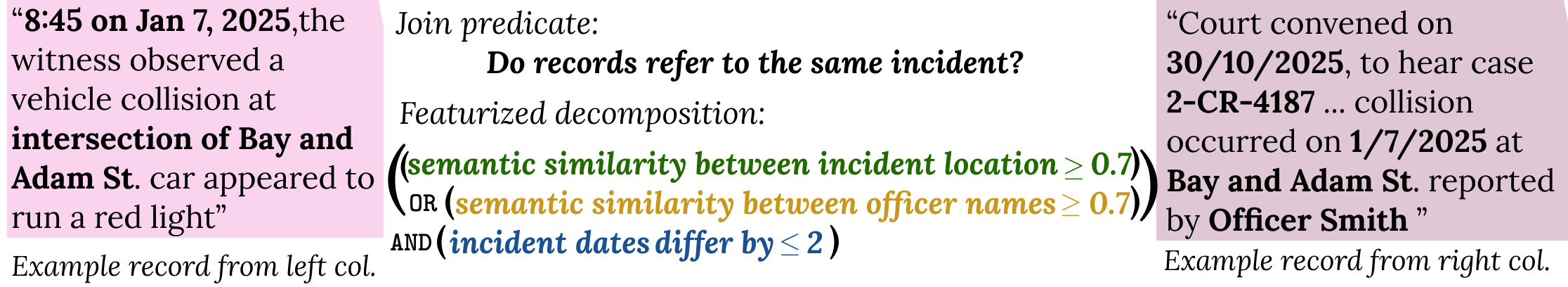}
    \caption{Example of a Featurized Decomposition}
    \label{fig:intro_ex}
\end{figure}

In this paper, we present \textit{Featurized Decomposition Join} (\name{}), a novel method for optimizing semantic joins. Our key insight for reducing join cost is to logically rewrite the join condition into inexpensive predicates that extract features from text records and use the features to perform the join. Extracting features can be done cheaply---it requires only a linear pass over the records with an LLM---relative to performing a quadratic number of pairwise comparisons with an LLM to obtain the join result. 
Thus, a logical expression that accurately represents the join condition using feature-based predicates can significantly reduce the join cost. We call such a logical expression a \textit{featurized decomposition}:  
a logical expression in conjunctive normal form (CNF) consisting of predicates that extract and compare features using inexpensive distance functions (e.g., lexical, arithmetic, or embedding-based). Fig.~\ref{fig:intro_ex} illustrates a featurized decomposition for our running example which combines predicates on three features---date, location, and police officer---to decide if two records refer to the same incident. Featurized decompositions can be complex, since the join condition may depend on multiple features and the decomposition must be robust to variations, incorrect extractions, or missing feature values.  

%---we theoretically show when this is possible. % \sep{might need to remove this}. %NEED TO SAY DISCJUNTIVE

%. perform the join, \name{} first extract such features from the documents, and then designs predicates that can be evaluated on the extracted features. For example, it tests whether the extracted dates are similar; if they are not the join condition is expected to not be satisfied. %\name{} automatically discovers and constructs reliable featurized decompositions that provide statistical guarantees on the precision and recall of the output join results while minimizing costs. 

%We automatically construct featurized decompositions that best represent the join condition. 

An accurate featurized decompositions can perform a semantic join with low cost, but automatically constructing one is challenging for two main reasons. 
%The main challenge is to automatically construct a decomposition that achieves this low cost while guaranteeing the required output precision and recall.
%We formalize the problem of automatically constructing a decomposition that achieves low cost while guaranteeing the required output precision and recall and demonstrate that it is NP-hard. %As such, automatically constructing reliable featurized decompositions is difficult.  
%In practice, there are two challenges in solving this problem. 
First, finding relevant features and extracting them correctly is difficult. A naive approach is to ask an LLM to produce all relevant features directly. % to ask an LLM to generate a featurized decomposition directly. 
However, LLMs often miss important features or produce redundant ones, and may not extract the feature values correctly or consistently across records. Meanwhile, applying LLMs to extract features incurs costs and must be done judiciously. 
%---LLMs are error prone and problems are exacerbated by lack of consistency in text datasets across records (e.g., different records may contain different information). 
Second, even after finding a set of relevant features and extracting them correctly, designing a featurized decomposition using such features while providing theoretical guarantees on output quality is non-trivial. In fact, we show that the problem of constructing the lowest cost decomposition with guarantees is NP-hard. %Given a set of features, constructing a decomposition is a CNF construction problem, previously studied for designing blocking schemes for entity resolution \cite{michelson2006learning, kejriwal2013unsupervised, bilenko2006adaptive, kejriwal2015dnf}. However, existing solutions do not provide quality guarantees and may produce a logical expression with low recall or precision. 
Quality guarantees are particularly important to ensure reliability given that features are automatically identified and extracted by LLMs which inevitably introduces errors that must be accounted for.  %. 
%featurized decompositions apply to any natural language join instruction and text records, which requires automatically constructing feature extraction functions and predicates that use those feature from text records and the join task. . 
Meanwhile, guaranteeing quality is challenging as it requires an in-depth analysis of the quality of the featurized decomposition when constructing it. Existing model cascades approaches provide statistical guarantees for the simpler setting of finding a single threshold on the proxy score \cite{zeighami2025cut, kang2020approximate, patel2024lotus}. Providing statistical guarantees when constructing a CNF requires a non-trivial high-dimensional generalization of such statistical results. % that is non-trivial. 

To automatically construct a reliable featurized decomposition, \name{} begins by extracting a comprehensive set of candidate features through an iterative process. At each iteration, it evaluates a running set of features using a sample of results labeled by the LLM, identifies records where the features fail to accurately estimate the join outcome and generates new features to address these gaps. Next, \name{} uses these features to form the decomposition. %---automatically determining which features to use, defining predicates over them, and combining these predicates into a logical expression. 
This construction phase is also guided by labeled samples: \name{} evaluates alternative decompositions, estimates their join costs, and selects the one with the lowest cost that is estimated to meet the desired quality targets. A crucial part of this step is ensuring that the chosen decomposition satisfies quality guarantees. We theoretically guarantee output quality by appropriately choosing the threshold parameters in our predicates---recall that each predicate in a featurized decomposition evaluates whether the distance between extracted features is below a certain threshold (e.g., whether extracted dates are within \textit{two} days, see Fig.~\ref{fig:intro_ex}). We provide a tight theoretical analysis of our threshold selection procedure, generalizing existing bounds in the model cascade literature that set one-dimensional thresholds \cite{zeighami2025cut, kang2020approximate, patel2024lotus} to high dimensions.  

Using the above methodology, \name{} automatically constructs a featurized decomposition and uses it to perform the join while providing statistical guarantees on the precision and recall of the output. Our experimental evaluation across 6 different real-world datasets show \name{} provides significant cost savings. %, Table~\ref{tab:res_overview} presents average and maximum cost savings we observed in our experiments across various real-world datasets. 
\textbf{Over BARGAIN applied to semantic joins, \name{} has about half the cost on average, but reduces the cost up to 10 times on some datasets}. We also compared \name{} with the \textit{optimal cascade} that sets optimal cascade thresholds on proxy scores. This oracle approach provides a lower bound on the cost of any cascade-based solution that relies on semantic similarity alone but is infeasible in practice as it requires setting thresholds by looking at LLM outputs for all pairs. \textbf{\name{} is up to 8 times cheaper than the optimal cascade. }

\textbf{Contributions}. To summarize, our contributions are as follows:
\begin{itemize}[leftmargin=1em]
\vspace{-0.2cm}
\item We propose \textit{featurized decomposition} as a new mechanism for optimizing semantic joins.
\item  We present Featurized-Decomposition Join (\name{}), a solution for cost-optimized semantic joins using featurized decomposition. 
\item  As part of \name{}, we present novel statistical results on how to set multiple thresholds in cascade-like architectures in high-dimensions, generalizing the results of \cite{zeighami2025cut, kang2020approximate, patel2024lotus}.
\item  We present experimental results across various real-world datasets, showing up to 10 times reduction in cost over state-of-the-art.
\end{itemize}

\vspace{-0.2cm}
\section{Preliminaries}\label{sec:setup}
%\vspace{-0.1cm}
\textbf{Semantic Joins}. We are given two sets of strings (text columns, documents) and a natural language predicate. 
Our goal is to find the subset of the cross product of the two sets on which the natural language predicate evaluates to true, where the evaluation is done by an LLM.
We let $\mathcal{L}$ be a user-specified LLM, $\mathtt{L}$ and $\mathtt{R}$ be two sets of strings and let $\mathtt{p}$ be the natural language predicate, called the \textit{join condition}. $\mathtt{p}$ is a parameterized string (or a langex~\cite{patel2024lotus}) with parameters $\str{r}\in\mathtt{R}$ and $\str{l}\in\mathtt{L}$ as input, e.g., ``\texttt{Do \{l\} and \{r\} describe the same incident?}''. %, where $\str{l}$ and $\str{r}$ can be any value from $\str{L}$ and $\str{R}$ respectively). 
We define $\mathcal{L}_\mathtt{p}:\mathtt{L}\times \mathtt{R}\rightarrow \{0, 1\}$ to be a function that determines whether the join condition is satisfied for a pair from $\mathtt{L}\times \mathtt{R}$. Specifically, given  $(\str{l},\str{r})\in\str{L}\times\str{R}$, $\mathcal{L}_\mathtt{p}$ first substitutes $\str{l}$ and $\str{r}$ in $\str{p}$ and then passes the resulting string as a prompt to the LLM $\mathcal{L}$ to obtain a boolean output. We say two records, $(\str{l}, \str{r})\in \str{L}\times\str{R}$ \textit{match} for a join condition $\str{p}$ if $\mathcal{L}_{\str{p}}(\str{l}, \str{r})=1$. Finally, the \textit{true} semantic join answer is defined as~\cite{patel2024lotus}
$$
\str{Y}=\{(\str{l}, \str{r}); (\str{l}, \str{r})\in \str{L}\times \str{R}, \mathcal{L}_{\str{p}}(\str{l}, \str{r})=1\}.
$$
We define $\bar{\str{Y}}=(\str{L}\times\str{R})\setminus\str{Y}$ as the set of non-matches. We refer to the pairs in $\str{Y}$ (in $\bar{\str{Y}}$) as \textit{positive (negative) pairs}, to any $(\str{l}, \str{r})\in\str{L}\times\str{R}$ as a \textit{pair}, and to the value of $\mathcal{L}_\str{p}(\str{l}, \str{r})$ as the \textit{label} for a pair $(\str{l}, \str{r})$. 
\if 0
We formalize this notion using relational operators, which later come in handy when we describe our solution. 
Let $A$ and $B$ be two string columns in two tables $L$ and $R$, and let $P$ be a natural language prompt to join $L$ and $R$ based on $A$ and $B$. 
$P$ is a parameterized string (or a langex~\cite{patel2024lotus}) that takes in values from $A$ and $B$ as input.
We denote by $P(a, b)$ the non-parameterized string where the values of $a$ and $b$ are substituted into $P$.
$P(a, b)$ is often a natural language question that can be evaluated by and LLM. For example, \sep{discuss example}. 
Given an LLM, $L$, we denote by $L_P(a, b)$ a function that takes values $a$ and $b$ from $A$ and $B$, evaluates the predicate on an LLM and outputs true or false. \sep{somehow ``gold llm'' needs to come through somewhere}
Then, given an LLM $L$, the natural language join of $A$ and $B$ based on natural language predicate $P$ evaluates to
\begin{verbatim}
SELECT L.A, R.B
FROM L, R
WHERE L_P(A, B)=1
\end{verbatim}

We denote by $J_P(A, B)$ the set of rows in the output of the above SQL. Note that considering $L_P$ as a UDF and applying it to every pair of values from $A$ and $B$ requires $N_A\times N_B$ number of LLM calls, which is expensive. 
We aim at reducing this cost, by allowing a bounded error, as formalized below.
\fi

\textbf{Approximate Semantic Joins}. Evaluating the true semantic join exactly is expensive, as it requires invoking an LLM on every pair of rows from $\str{L}$ and $\str{R}$. To support low-cost processing while still producing answers similar to that of the true answer, we allow the users to specify quality requirements on the recall and precision of the result. For any result set $\hat{\str{Y}}\subseteq \str{L}\times \str{R}$, we define its precision, $\mathfrak{P}(\hat{\str{Y}})$, and recall, $\mathfrak{R}(\hat{\str{Y}})$, respectively, as $\mathfrak{P}(\hat{\str{Y}})=\frac{|\str{Y}\cap \hat{\str{Y}}|}{|\hat{\str{Y}}|}$ and $\mathfrak{R}(\hat{\str{Y}})=\frac{|\str{Y}\cap \hat{\str{Y}}|}{|\str{Y}|}$.
\if 0
\begin{align*}
    \mathfrak{P}(\hat{\str{Y}})&=\frac{|\str{Y}\cap \hat{\str{Y}}|}{|\hat{\str{Y}}|},\,&
    \mathfrak{R}(\hat{\str{Y}})=\frac{|\str{Y}\cap \hat{\str{Y}}|}{|\str{Y}|}.
\end{align*}
\fi
We consider the setting where user specifies a recall requirement $T_R$ and a precision requirement $T_P$, and requires a result whose precision and recall are at least $T_P$ and $T_R$, respectively, with high probability. To formalize this, let $\delta$ be a user-provided probability of failure and let $\textsc{A}(\str{L}, \str{R}, \str{p})$ be any (randomized) algorithm that outputs a join result $\hat{\str{Y}}$. We say the algorithm performs an approximate semantic join with statistical guarantees if it satisfies 
$$
\mathds{P}_{\hat{\str{Y}}\sim\textsc{A}(\str{L}, \str{R}, \str{p})}(\mathfrak{P}(\hat{\str{Y}})<T_P\;\text{or}\;\mathfrak{R}(\hat{\str{Y}})<T_R)\leq \delta,
$$
where the probability is over random runs of the algorithm. Our goal is to design an algorithm with statistical guarantees while minimizing total cost, as we discuss next. %We next discuss our cost model and available tools to minimize cost. 

\textbf{Embedding Models}. To reduce costs, we use embedding models that provide low-cost high-dimensional semantic representations of the data. Formally, an embedding model $\mathcal{E}$, which produces embedding vectors, that is, for a string $\str{s}$, $\mathcal{E}(s)$ is a high-dimensional vector. For two strings $\str{s}_1$ and $\str{s}_2$, their semantic similarity is defined as the cosine similarity between the embeddings, $\mathcal{E}(\str{s}_1)$ and  $\mathcal{E}(\str{s}_2)$. For consistency with other distance metrics, we use \textit{semantic distance} for two strings $\str{s}_1$ and $\str{s}_2$ to refer to (1$-$semantic similarity). %\sep{change this and also modify in when we talk about semantic similarity}

\textbf{Cost Model}. We focus on the cost of using LLMs and embedding models; the cost of non-LLM operations are negligible in comparison.  We define the \textit{cost of an algorithm} $\textsc{A}$, denoted by $C(\textsc{A}(\str{L}, \str{R}, \str{p}))$, as the monetary cost it incurs using LLM $\mathcal{L}$ and embedding model $\mathcal{E}$---determined using the total number of input and output tokens. %when using the LLM and the embedding model.

\textbf{Problem Definition}. Putting everything together, our goal is to solve the following problem:
\vspace{-0.2cm}
\begin{definition}[Approximate Semantic Joins with Guarantees]
    Given two datasets $\str{L}$ and $\str{R}$, a recall target $T_R$, a precision target $T_P$, a probability of failure $\delta$, an LLM $\mathcal{L}$, and an embedding model $\mathcal{E}$, design an algorithm $\textsc{A}$ to perform approximate join with statistical guarantees on precision and recall while minimizing the cost $\textsc{A}$: 
\begin{equation*}
\begin{array}{rl}
\displaystyle \min_{\textsc{A}} & C(\textsc{A}(\str{L}, \str{R}, \str{p})) \\%[2pt]
\text{s.t.} & 
\mathds{P}_{\hat{Y}\sim\textsc{A}(\str{L}, \str{R}, \str{p})}
\big(\mathfrak{P}(\hat{Y})<T_P\;\text{or}\;\mathfrak{R}(\hat{Y})<T_R\big)
\leq \delta
\end{array}
\end{equation*}
\end{definition}

\if 0
We assume the user specifies a recall requirement $T$, and our goal is to produce an output that contains at least $T\%$ of the records in $J_P(A, B)$. We allow this guarantee to hold probabilistically, that is, for any output set $S$, $\mathds{P}(\frac{|S\cap J_P(A, B)|}{|J_P(A, B)}<T)<\delta$, where $\delta$ is a user-provided probability of failure. In this setting, we want to ensure the total size is minimized, so that a second pass on the rows in $S$ with an LLM produces a high recall set. \sep{Should likely just present cost at the end, one setting cost is assuming perfect precision, another is allowing lower precision too}

The user can additionally specify a precision requirement $T_P$, where for an output set the recall requirement above has to hold, but additionally, there is a bound on how many additional elements can be returned. \sep{formalize this}

\textbf{Approximate comparisons and distance-based operations}. We assume we have access to distance-based operations: $d_E(a, b)$ is a semantic distance $d_L(a, b)$ is lexical distance. These are cheap, and assume are free. Let $d(a, b)$ be a generic distance function

\sep{not sure what to include here, and what the section should be called}

\subsection{taxonomy of semantic joins}
Can we put all the types here?
\fi

\begin{figure}
\vspace{-0.5cm}
    \centering
    \includegraphics[width=1.05\linewidth]{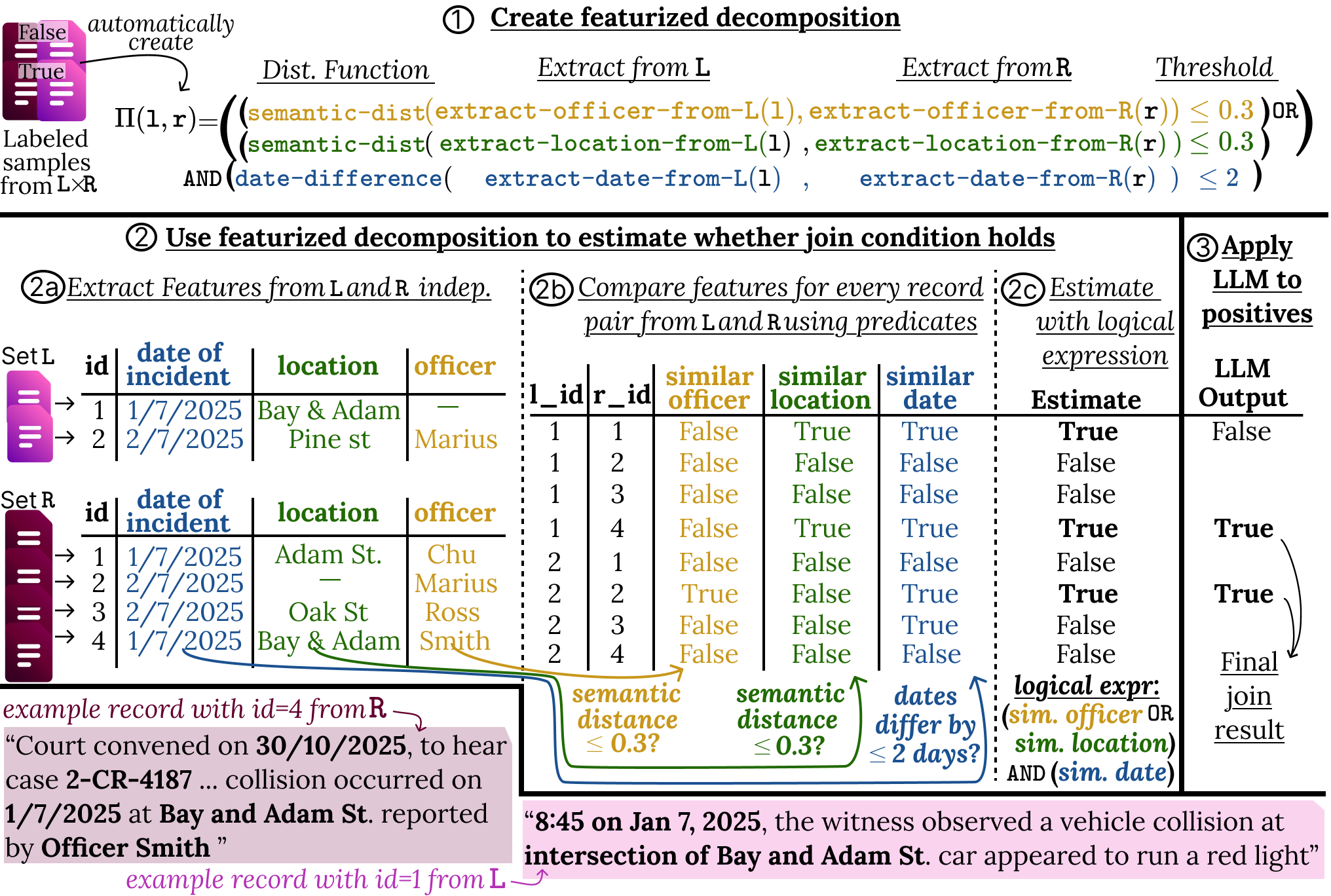}
    \caption{\name{} Example Workflow}
    \label{fig:fd_inf}
\end{figure}

\vspace{-0.2cm}
\section{\name{} Overview \& Featurized Decompositions}\label{sec:factorize_joins}
We present Featurized-Decomposition Join (\name{}), our method for performing approximate semantic joins at low cost. \name{} uses a technique we call \textit{featurized decomposition} to reduce join costs by estimating the join outcome based on features extracted from text records. In this section, we first provide an overview of how \name{} works and present the necessary terminology in Sec.~\ref{sec:overview_term}. Then, we discuss the challenges in performing joins using featurized decomposition and provide an outline for the rest of this paper in Sec.~\ref{sec:challenges_roadmap}. 
%We first provide a high-level workflow of how \name{} operates before discussing details. %to perform joins with featurized decomposition, then, we present \name{} our method that creates freatur
For now, we consider the case that $T_P=1$ and $T_R<1$, that is, we must provide 100\% precision but recall can be imperfect, and we use $T=T_R$ to simplify notation. We extend this approach to a more relaxed precision requirement in Sec.~\ref{sec:pipeline}. We note that $T_P=1$ is important for many real-world applications; for example, incorrectly showing non-matching police records together (i.e., those that refer to different incidents) to users can corrupt analysis, leading to wrong conclusions with severe consequences.   

%\subsection{\name{} Workflow and Terminology}
%\textbf{}. %\name{} first uses samples a subset of the cross produc
%We introduces a high-l technique called \textit{featurized decomposition} to perform semantic join with low cost. Here, 

\subsection{Overview and Terminology}\label{sec:overview_term}
We next provide a high-level workflow of \name{} %how to perform a join using featurized decomposition 
using Fig.~\ref{fig:fd_inf}. The figure shows the workflow for our running example where our goal is to join different police records based on whether they refer to the same incident (Fig.~\ref{fig:fd_inf} shows two toy examples of such text records). 

\textit{{\small \CircledText[]{1}} Creating Featurized Decomposition}. \name{} first samples a subset of $\str{L}\times\str{R}$ and labels them using the LLM, $\mathcal{L}$. It then uses these labeled samples to automatically create a \textit{featurized decomposition}: a Boolean function designed to estimate whether the join condition holds for a pair $(\str{l}, \str{r})$ using features extracted.  A featurized decomposition consists of (1) functions to extract features from records in $\str{L}$ and $\str{R}$, (2) logical predicates that check whether extracted features from $\str{l}\in\str{L}$ and $\str{r}\in\str{R}$ are \textit{similar} based on a feature-specific distance function; and (3) a logical expression that combines the logical predicates to estimate whether the join condition holds in conjunctive normal form (CNF). In Fig.~\ref{fig:fd_inf}, Step {\small \CircledText[]{1}}, the function $\Pi(\str{l}, \str{r})$ is a featurized decomposition; it takes $(\str{l}, \str{r})\in\str{L}\times\str{R}$ as input and produces a binary output to estimate if the join condition holds for $(\str{l}, \str{r})$. The function $\Pi$ is a logical combination of three logical predicates (each in a different color). The predicate in yellow checks whether police officer names are similar. It leverages functions that extract names of police officers from $\str{l}$ and $\str{r}$ and checks whether the semantic distance between extracted names are within  threshold 0.3. Predicates in green and blue are similarly defined for location and date features.

\textit{{\small \CircledText[]{2}} Using Featurized Decomposition to Produce Estimates}.  After constructing the featurized decomposition, we use it to estimate whether each record pair satisfies the join condition. This estimation process first applies the feature extraction functions to obtain the relevant features, and then evaluates the logical expression over these features to produce the final estimate. Step {\small \CircledText[]{2}} in Fig.~\ref{fig:fd_inf} shows an example of how the estimates are obtained. First, the feature extraction functions are applied independently to all rows in $\str{L}$ and $\str{R}$. Step {\small \CircledText[]{2a}} shows the output of this step for 2 records in $\str{L}$ and 4 records in $\str{R}$. The values under the location column for $\str{L}$ are obtained by applying \texttt{extract-location-from-L} function to every record in $\str{L}$, while the values for $\str{R}$ are obtained by applying \texttt{extract-location-from-R} to every record in $\str{R}$. After obtaining feature values, Step {\small \CircledText[]{2b}}, shows the result of each of the three predicates for each pair of records, and finally Step {\small \CircledText[]{2c}} shows the final result of the expression for every pair of records from $\str{L}\times\str{R}$, where three record pairs are estimated to satisfy the join condition.

\textit{{\small \CircledText[]{3}} Refinement}. Finally, we remove any false positives introduced, by applying the LLM $\mathcal{L}$ to any pair from {\small \CircledText[]{2c}}  estimated to satisfy the join condition. In our example, we find one false positive. The final join result output by \name{} is the set of pairs estimated to be positive by the featurized decomposition and confirmed as such by the LLM. Our approach guarantees perfect precision (since every estimated pair is evaluated by an LLM), but depending on the featurized decomposition may have imperfect recall. 

\begin{figure}
\vspace{-0.5cm}
    \centering
    \includegraphics[width=\linewidth]{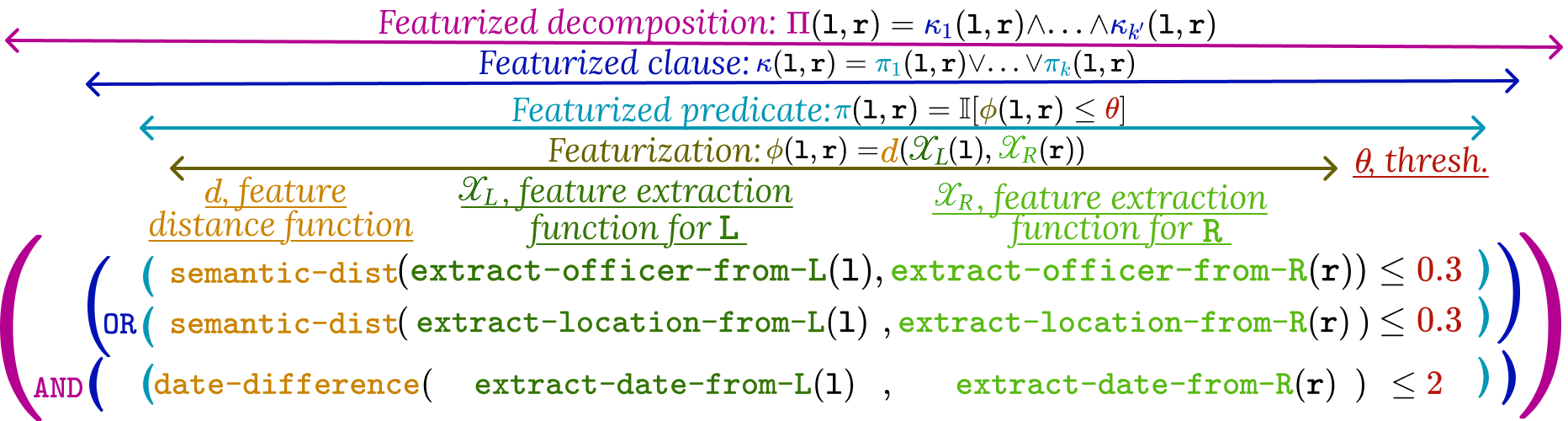}
    \caption{Terminology Summary}
    \label{fig:terminology}
\end{figure}

\textbf{Terminology.} We next introduce necessary terminology to formalize the above workflow. Fig.~\ref{fig:terminology}. shows a summary of the terminology along with the example from Fig.~\ref{fig:fd_inf}. We use the term \textit{feature extraction function} to refer to functions designed to extract features (e.g., \texttt{extract-location-from-L} in Fig.~\ref{fig:terminology}). We typically denote a feature extraction function for extracting features from $\str{L}$ ($\str{R}$) by $\mathcal{X}_L$ ($\mathcal{X}_R$, resp.)---these function can include LLM invocations and have any output signatures. We allow $\mathcal{X}_L$ to be different from $\mathcal{X}_R$ to enable dataset-specific functions. %For example, in Fig.~\ref{fig:fd_inf}, to extract the date of incident, $\mathcal{X}_R$ may contain an explicit instruction to ignore court dates to avoid mistaking it for date of incident---an instruction that is useful only for $\str{R}$. 
We use \textit{feature distance function} (or distance functions) to refer to real-valued functions designed to compute distances between features, denoted by $d$ (e.g., \texttt{semantic-dist} in Fig.~\ref{fig:terminology}). A distance function does not perform any LLM invocations and only performs lightweight comparisons on features (e.g., semantic distance, lexical difference, arithmetic operations). We use extraction and distance functions to define logical predicates which we refer to as \textit{featurized predicate}s: Boolean functions of the form $\pi(\str{l}, \str{r})=\mathds{I}[d(\mathcal{X}_L(\str{l}), \mathcal{X}_L(\str{r}))\leq \theta]$, consisting of feature extraction functions $\mathcal{X}_L$ and $\mathcal{X}_R$, a distance function $d$, and a threshold, $\theta\in \mathbb{R}$. When presenting our solution, we often discuss the functions $d, \mathcal{X}_L, \mathcal{X}_R$ separately from $\theta$, so for convenience we refer to the triplet $(d, \mathcal{X}_L, \mathcal{X}_R)$ as a \textit{featurization} and $\theta$ as a \textit{predicate threshold}. For a featurization $\phi=(d, \mathcal{X}_L, \mathcal{X}_R)$ we denote its \textit{inference function} by $\phi(\str{l}, \str{r})=d(\mathcal{X}_L(\str{l}), \mathcal{X}_R(\str{r}))$. We abuse notation and refer to the featurization and its inference function interchangeably. The three logical predicates in Fig.~\ref{fig:terminology} are featurized predicates.  

To estimate whether the join condition holds, we  combine featurized predicates into logical expressions in CNF form. Formally, we define a \textit{featurized clause} $\kappa:\str{L}\times\str{R}\rightarrow\{0, 1\}$ as a Boolean function that is a disjunction of featurized predicates, that is, of the form $\kappa(\str{L}, \str{R})=\pi_1(\str{l}, \str{r})\lor...\lor\pi_k(\str{l}, \str{r})$ for some integer $k$ and featurized predicates $\pi_1, ..., \pi_k$. Then, a \textit{featurized decomposition}, $\Pi$, is a Boolean function $\Pi:\str{L}\times\str{R}\rightarrow\{0, 1\}$ defined as a conjunction of featurized clauses, that is, 
$\Pi(\str{l}, \str{r})=\kappa_1(\str{l}, \str{r})\land...\land\kappa_{k'}(\str{l}, \str{r})$, for some integer $k'$ and featurized clauses $\kappa_1, ..., \kappa_{k'}$. We abuse notation and for any $\str{S}\subseteq\str{L}\times\str{R}$, define $\Pi(\str{S})=\{(\str{l}, \str{r})\in \str{S}; \Pi(\str{l}, \str{r})=1\}$. %Finally, we say a featurized decomposition has recall at least $T$ if $\mathfrak{R}(\Pi(\str{L}\times\str{R}))\geq T$.

\subsection{Challenges and Roadmap}\label{sec:challenges_roadmap}
Using the above terminology, to perform joins using featurized decomposition, we first create a featurized decomposition $\Pi$ (Step {\small \CircledText[]{1}} in Fig.~\ref{fig:fd_inf}), then create a candidate join result $\hat{\str{Y}}=\Pi(\str{L}\times\str{R})$ using the decomposition (Step {\small \CircledText[]{2}} in Fig.~\ref{fig:fd_inf}) and finally use the LLM $\mathcal{L}$ to obtain our final join result as $\{(\str{l}, \str{r})\in\hat{\str{Y}}; \mathcal{L}_{\str{p}}(\str{l}, \str{r})=1\}$. Recall that we want 100\% precision and recall at least $T$. Steps {\small \CircledText[]{1}} and {\small \CircledText[]{2}} guarantees the latter, while Step {\small \CircledText[]{3}} (where each $(\str{l}, \str{r})\in\hat{\str{Y}}$ is checked by $\mathcal{L}$) guarantees the former. 

\textbf{Challeges}. The main challenge in the above workflow is Step {\small \CircledText[]{1}}, wherein we automatically construct a featurized decomposition. % to reduce join cost while providing recall guarantees. 
The recall of \name{} is equivalent to the recall of its featurized decomposition, so Step {\small \CircledText[]{1}} must theoretically guarantee that the featurized decomposition found meets the recall target. Moreover, the cost of \name{} depends on the precision of the featurized decomposition---false positive pairs introduced incur additional cost of LLM calls during refinement in Step {\small \CircledText[]{3}}---in addition to the cost of finding the featurized decomposition (Step {\small \CircledText[]{1}}) and using it to produce estimates (Step {\small \CircledText[]{2}}). The rest of this paper focuses on finding featurized decompositions that minimize cost and guarantee high recall. %We next formalize this problem. 

\textbf{Roadmap}. We formalize the problem of creating minimum cost featurized decompositions in Sec.~\ref{sec:min_fd} and provide an overview of how we construct featurized decompositions; we present the details of the construction procedure in Secs.~\ref{sec:candidate_generation}-\ref{sec:threshold:all}. Finally, we present the final \name{} algorithm in Sec.~\ref{sec:pipeline}, discuss its guarantees and extensions when relaxing the precision requirement. Proofs of our theoretical results are presented in \iftoggle{techreport}{Appx.~\ref{sec:all_proof}%
}{our technical report~\cite{techrep}}.

\section{Minimum Cost Featurized Decompositions} \label{sec:min_fd}
In this section, we formalize the problem of finding minimum cost featurized decomposition, show its NP-Hardness and present an overview of our solution for creating the featurized decomposition.
%When presenting our solution, we often discuss the functions $d, \mathcal{X}_L, \mathcal{X}_R$ separately from $\theta$, so for convenience we refer to the triplet $(d, \mathcal{X}_L, \mathcal{X}_R)$ as a \textit{featurization} and $\theta$ as a \textit{predicate threshold}. For a featurization $\phi=(d, \mathcal{X}_L, \mathcal{X}_R)$ we denote the \textit{inference function} $\phi(\str{l}, \str{r})=d(\mathcal{X}_L(l), \mathcal{X}_R(r))$. We abuse notation and refer to the featurization and the inference function interchangeably. 

\subsection{Problem Statement and NP-Hardness} 
%Having recall guarantees hinges on the ability of the featurized decomposition to correctly estimate the pairs   how to constructing such a featurized decomposition is the main focus of the rest of the paper.  \sep{add}
%Finding a minimum cost featurized decomposition is the problem of searching over a space of possible featurized decompositions and choosing one with the lowest cost that meets the recall target. The space of possible featurized decomposition depends on featurizations that can be used to create the decompositions.
%First, recall that a featurized predicate consists of a distance function, $d$, two feature extraction functions, $\mathcal{X}_L$ and $\mathcal{X}_R$, and a threshold parameter $\theta$. Among these, $d$, $\mathcal{X}_L$ and $\mathcal{X}_R$ are functions that need to be automatically constructed while $\theta$ is a real number. We let $\mathbb{X}$ to be a set of all possible function
Finding a minimum-cost featurized decomposition involves searching over all possible decompositions and selecting the one that achieves the lowest cost while meeting the recall target. The search space depends on possible featurizations that can be used to create the decompositions.
Let $\mathbb{F}$ denote the set of possible featurizations where each element is a triplet $(d, \mathcal{X}_L, \mathcal{X}_R)$ such that for any $(\str{l}, \str{r})\in\str{L}\times\str{R}$, $d(\mathcal{X}_L(\str{l}), \mathcal{X}_R(\str{r}))\in\mathbb{R}$. 
In principle, $\mathbb{F}$ could include any valid combination of feature extraction functions (e.g., functions over string inputs) and distance functions (e.g., real-valued functions over two inputs). In practice, as discussed later, we restrict $\mathbb{F}$ to featurizations automatically generated by an LLM.
%In general, $\mathbb{F}$ can contain any well-defined combination of all possible feature extraction function (e.g., all possible functions that take string inputs) and distance functions (e.g., all possible functions that take two inputs and output a real number). In practice---as we discuss later---we limit $\mathbb{F}$ to the set of featurizations automatically generated by an LLM. Given $\mathbb{F}$, the space of possible featurized decompositions is the set of all possible logical expressions in CNF that contain predicates consisting of featurizations in $\mathbb{F}$ paired with threshold parameters $\theta\in \mathbb{R}$.
Given $\mathbb{F}$, the space of possible featurized decompositions consists of all logical expressions in CNF whose predicates are derived from featurizations in 
$\mathbb{F}$ paired with threshold parameters $\theta\in \mathbb{R}$. Our goal is to find a decomposition with minimum cost among such decompositions. 

\begin{definition}
\vspace{-0.1cm}
    [Minimum Cost Featurized Decomposition (MCFD)] Given a set of possible featurizations $\mathbb{F}$, two sets $\str{L}$ and $\str{R}$ and a predicate function $\mathcal{L}_\str{p}:\str{L}\times\str{R}\rightarrow\{0, 1\}$, find a featurized decomposition with recall at least $T$ while minimizing total cost. 
\end{definition}
\vspace{-0.1cm}
\noindent Unfortunately, solving MCFD optimally is computationally difficult:
\begin{theorem}\label{thm:nphard}
\vspace{-0.1cm}
    MCFD is NP-hard. 
\vspace{-0.1cm}
\end{theorem}
%\textit{Proof}. By reduction from set cover. \qed
Thm.~\ref{thm:nphard} is shown by a reduction from the Set Cover problem. The reduction maps each set in Set Cover to a featurization in the set $\mathbb{F}$ and shows that a featurized decomposition with $k$ predicates exists if and only if a set cover of size $k$ exists. Intuitively, sets in Set Cover map to featurized predicates in MCFD since a featurized predicate admitting a positive pair is akin to a set covering an element in Set Cover. Nonetheless, the reduction is non-trivial because a featurized decomposition with minimum cost may not necessarily have the lowest number of possible predicates due to its CNF form. %To address this, the reduction creates a mapping that ensures the optimal solution to MCFD only has a single clause (i.e., is a disjunction of predicates), so that there is a one-to-one mapping between predicates in the optimal solution to MCFD and sets in Set Cover.  %Intuitively, this mapping works because similar to each set in the set cover problem, each featurization \textit{covers} a set of positives, that is, adding a featurizaiton to a clause with disjunctions can . 
We defer details to \iftoggle{techreport}{Appx.~\ref{sec:all_proof}%
}{our technical report~\cite{techrep}}.

\begin{figure}
\vspace{-0.5cm}
    \centering
    \includegraphics[width=0.9\linewidth]{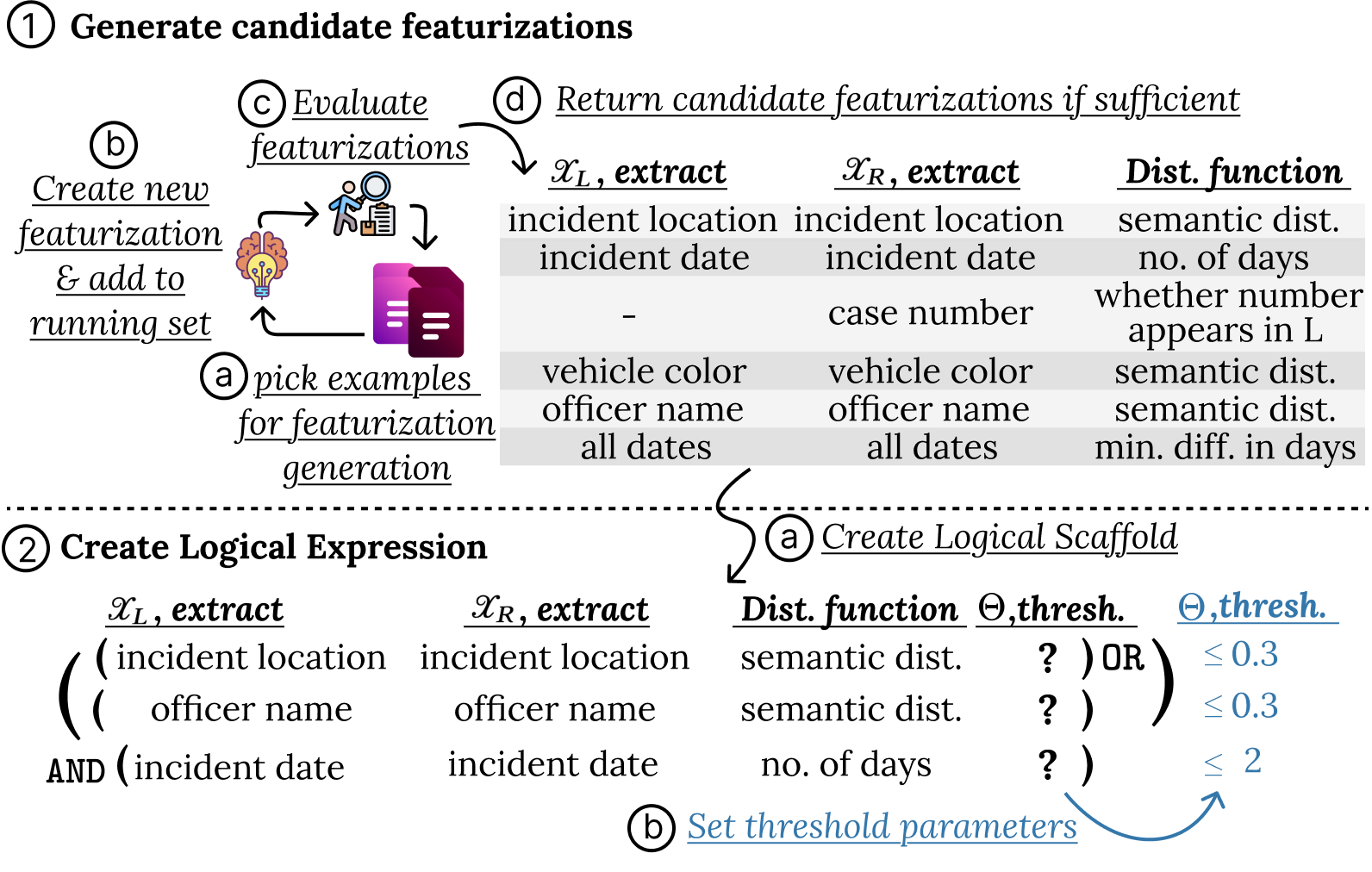}
    \caption{Finding Featurized Decompositions}
    \label{fig:decomposition_solution}
\end{figure}

\subsection{Constructing Featurized Decompositions}\label{sec:overview:jain_framework}
We next describe how we solve MCFD.  %As discussed above, this process involves consider multiple candidate decompositions and selecting the one with the lowest cost. take advantage of LLMs' reasoning abilities to construct the decomposition; we expect LLMs---if used appropriately---to be able 
At a high level, we use LLMs to obtain useful features from which we create our decomposition. LLMs help us prune the space of possible featurizations to a few relevant ones that we then algorithmically combine to create decompositions. Nonetheless, obtaining reliable featurizations from LLMs is challenging. A single-shot prompting strategy often overlooks important features or creates erroneous %featurizations definitions A naive approach is to ask an LLM to produce featurizations directly in a single-shot prompt. However, LLMs may overlook important features or make errors when defining 
extraction functions. Moreover, our goal is to guarantee that the recall of the resulting decomposition meets a user-specified target---a non-trivial statistical estimation task. 
%They furthermore cannot guarantee high recall. 
To address these challenges, we design a multi-step pipeline that leverages LLMs to generate featurizations and uses labeled samples to iteratively refine them, ensuring both reliability and coverage (i.e., to ensure all relevant featurizations are found). We then use this set of \textit{candidate featurizations} to construct a featurized decomposition with tight theoretical guarantees on recall. As such, we follow a two-step procedure for constructing featurized decompositions, where we (1) first generate a set of candidate featurizations, from which (2) we choose a subset to create the featurized decomposition. We provide an overview of our approach below with the aid of Fig.~\ref{fig:decomposition_solution}. Details of the two steps (1) and (2) are presented in Secs.~\ref{sec:candidate_generation} and \ref{sec:threshold:all}, respectively.
%To furthermore provide theoretical guarantees on output recall, we design an algorithm 
%To design a robust algorithm for constructing featurized decomosptions with statistical guarantees, we divide the process of performing creating the featurized decomposition into two steps. We first generate a comprehensive set of candidate featurization, from which we subsequently chooses a subset to create a featurized decomposition. We provide an overview of this two-step process below 
%Ensure robust decompositions
%; it does so to create a featurized decomposition that minimizes cost while guaranteeing the required recall. %The final output of the threshold selection is a featurized decomposition, so it can be directly used to perform the join.
%An example of this two step procedure is shown in Fig.~\ref{fig:decomposition_solution}. We discuss each step next. %As the figure shows, the candidate featurization set can contain many different (but possibly related) extraction and distance functions. A subset of the featurizations to create the final featurized decomposition.  

\textbf{Candidate Generation}. We begin by constructing a comprehensive set of featurizations. As shown in Fig.~\ref{fig:decomposition_solution} (top half), this process proceeds iteratively: we (1a) select a set of examples as demonstrations for the LLM; (1b) use an LLM to generate new featurizations or refine existing ones; and (1c) evaluate the resulting set of featurizations. If the current set is deemed sufficient to perform the join, (1d) we return it as the final output, or we otherwise repeat the process.  Fig.~\ref{fig:decomposition_solution} illustrates several candidate featurizations. %Otherwise, the procedure repeats to generate additional featurizations. 
At each iteration, we select positive examples on which the current featurizations performs worst---measured by the number of false positives that must be introduced to correctly classify the positive example using the current featurizations. These examples are then fed back to the LLM to generate featurizations suitable for those examples or to correct errors in feature extraction, forming a feedback loop to iteratively improve the quality of the featurizations. %, either because no existing featurization is a good indicator for such examples or because there is an error in an existing featurization (e.g., an error in the code generated for the feature extraction functions). By identifyin such examples and using them as demonstractions,  
%We describe this step in details in Sec.~\ref{sec:candidate_generation}. 

\textbf{Logical Expression Formulation.} Next, we use the set of candidate featurizations obtained from the previous step to construct a featurized decomposition that minimizes cost and meets the recall requirement. At a high-level, we evaluate possible decompositions, estimating for each whether it satisfies the recall target, and selecting the lowest-cost one among those that do. However, guaranteeing that the recall target is met is challenging, as testing multiple decompositions introduces a multiple hypothesis testing problem. A naive analysis using union bound across each test leads to loose bounds, and consequently limited cost savings. To obtain tight bounds, we divide the process of constructing the featurized decomposition into two steps. First, we build a \textit{logical scaffold} that specifies which featurizations to include and how to combine them, without specifying threshold parameters. Fig.~\ref{fig:decomposition_solution} illustrates this scaffold in our running example. Second, we set the thresholds to ensure the final decomposition meets the target recall. Theoretical analysis of recall is only needed in the latter step, where it is more tractable. %We discuss this in details in Sec.~\ref{sec:threshold:all}. 

\section{Candidate Featurization Generation}\label{sec:candidate_generation}
%Candidate generation uses LLMs to find suitable featurizations. 
As discussed in the previous section (see Fig.~\ref{fig:decomposition_solution}), we iteratively generate featurizations, evaluate them, and pick examples as demonstrations to generate new featurizations. This iterative process ensures the featurizations are reliable---the evaluation phase surfaces errors in extraction functions that the LLM can fix---and comprehensive---different examples demonstrate different useful featurizations. 

\begin{figure}[t]
    \vspace{-0.7cm}
    \centering
    \includegraphics[width=1\linewidth]{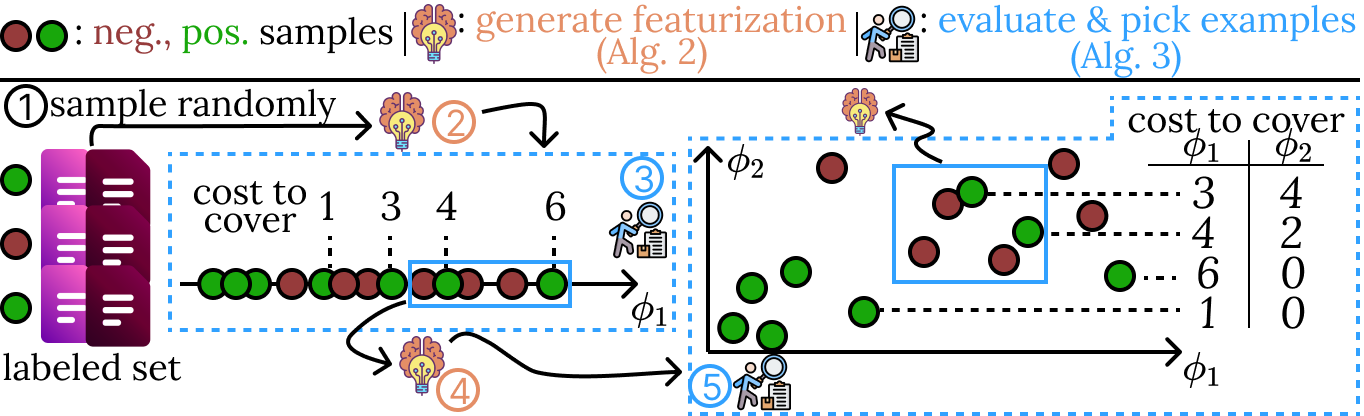}
    \caption{Candidate Featurization Generation Example}
    \label{fig:featurization_details_ex}
\end{figure}

We present this procedure in Alg.~\ref{alg:candidate_generation_all}. The algorithm takes  as input a set, $\str{S}$, containing random samples from $\str{L}\times\str{R}$ together with their labels, $Y_S=\{\mathcal{L}_{\str{p}}(\str{l}, \str{r});(\str{l}, \str{r})\in\str{S}\}$, generated by an LLM. $\str{S}$ is used to guide the generation process. Alg.~\ref{alg:candidate_generation_all} iteratively calls two subroutines (detailed later): (1) \texttt{get-featurization-from-examples} to generate new featurizations using a set of examples as in-context demonstration and (2) \texttt{evaluate-and}\texttt{-pick-examples} to evaluate the featurizations using $\str{S}$ and find examples where the featurizations cannot accurately estimate the join outcome. If \texttt{evaluate-and-pick-} \texttt{examples} decides the featurizations do accurately estimate the join outcome for the pairs in $\str{S}$, it returns an empty set and the algorithm terminates, returning the current set of featurizations. Alg.~\ref{alg:candidate_generation_all} otherwise repeats the process up to a maximum number of iterations. Note that at each iteration, a subset of size $\beta$ from $\str{S}$ is chosen to be passed to the LLM as demonstrations, where $\beta$ is a parameter determined based on the LLM's context limit. We illustrate this process in Fig.~\ref{fig:featurization_details_ex}, where starting with a random sample (Step (1)), we alternate between generating candidate featurizations (Steps (2) and (4)) and evaluating them to pick new examples as demonstrations for generation (Steps (3) and (5)). We next describe the two subroutines, generating featurizations in Sec.~\ref{sec:candidate:llm_pipeline} and evaluating and picking examples in Sec.~\ref{sec:candidate:algorithm}.

%As the figure shows, we introduce a notion called \textit{cost to cover} to evaluate featurizations and decide which examples to pick. We next describe the two  

%We first describe how we use LLMs as a component in our algorithm to generate candidate features (i.e., Step (1b) in Fig.~\ref{fig:decomposition_solution}) in Sec.~\ref{sec:candidate:llm_pipeline}, before discussing the overall algorithm (i.e., the algorithm for the entire Step 1 in Fig.~\ref{fig:decomposition_solution}) in Sec.~\ref{sec:candidate:algorithm}.
%We then show how we use this pipeline to generate candidate featurizations (Step (1) in Fig.~\ref{fig:decomposition_solution}), using labeled samples to catch errors and ensure comprehensiveness. Next, we describe the LLM pipeline in Sec.~\ref{sec:candidate:llm_pipeline} and our algorithm to generate candidate featurizations in Sec.~\ref{sec:candidate:algorithm}.  
%This step defines a function $\texttt{generate-featurizations}(\str{p}, \str{S})$ that takes in the join condition $\str{p}$ together with a labeled subset, $\str{S}$ of $\str{L}\times\str{R}$ and is asked to create a set of candidate featurizations, $\Phi$. 
{
\begin{algorithm}[t]
\footnotesize
\begin{algorithmic}[1]
%\Require Join condition $\str{p}$, the lists $\str{L}$ and $\str{R}$
%\Ensure A set of candidate factorizations
%\Procedure{\texttt{get-candidate-featuriztions}}{$\str{p}$, $\str{L}$, $\str{R}$}
\State $\str{C}$, $Y_C\leftarrow$ A sample of $\beta$ elements from  $\str{S}$ and their labels
\State $\mathbf{\Phi}\leftarrow\emptyset$
\For {$i\leftarrow 0$ \textbf{to} $max\_iter$} 
    \State $\mathbf{\Phi}\leftarrow \mathbf{\Phi}\cup \texttt{get-featurization-from-examples}(\str{C}, Y_C, \str{p})$
    \State $\str{C}, Y_C\leftarrow\texttt{evaluate-and-pick-examples}(\str{S}, Y_S, \mathbf{\Phi})$ 
    \If{$\str{C}=\emptyset$}
        \State return $\mathbf{\Phi}$ 
    \EndIf
\EndFor
\State return $\mathbf{\Phi}$ 
%\EndProcedure
\caption{\texttt{get-candidate-featurizations(\str{p}, \str{S}, $Y_S$)}}\label{alg:candidate_generation_all}
\end{algorithmic}
\end{algorithm}
}

\begin{algorithm}[t]
\footnotesize
\begin{algorithmic}[1]
%\Require $\str{p}$: join cond., $S_p$ and $S_n$: pos./neg. pairs, $\mathbf{\Phi}$: prev featurizations 
%\Ensure A set of candidate factorizations
\Procedure{\texttt{get-featurization-from-examples}}{$\str{S}, Y_S, \str{p}$}
\State $\str{U} \leftarrow \undef{get-featurization-descriptions}(\str{S}, Y_S, \str{p})$\label{alg:generator:logical}
\For{$\str{u} \in \str{U}$}\label{alg:line:for_loop}
    \State $\mathcal{X}_L \leftarrow \texttt{get-feature-extractors}(\str{S}, Y_S, \str{p}, \str{u}, \texttt{left})$\label{alg:generator:ext}
    \State $\mathcal{X}_R \leftarrow \texttt{get-feature-extractors}(\str{S}, Y_S, \str{p}, \str{u}, \texttt{right})$\label{alg:generator:ext2}
    \State $d \leftarrow \undef{get-distance-func}(\str{S}, Y_S, \str{p}, \str{u}, \mathcal{X}_L, \mathcal{X}_R)$\label{alg:generator:dist}
    \State $\mathbf{\Phi} \leftarrow \mathbf{\Phi} \cup \{(d, \mathcal{X}_L, \mathcal{X}_R)\}$\label{alg:line:end_for_loop}
\EndFor
\State \Return $\mathbf{\Phi}$
\EndProcedure
\Procedure{\texttt{get-feature-extractors}}{$\str{S}, Y_S, \str{p}, \str{u}, \texttt{col}$}
        \State $\texttt{description} \leftarrow \undef{get-feature-description-for-col}(\str{S}, Y_S, \str{p}, \str{u}, \texttt{col})$\label{alg:generator:desc}
        \If{$\undef{should-use-llm}(\str{S}, Y_S, \str{p},\texttt{description})$}\label{alg:generator:usellm}
            \State $\texttt{prompt} \leftarrow \undef{get-extraction-prompt}(\str{S}, Y_S, \str{p}, \texttt{description})$\label{alg:generator:nocode}
            \State $\mathcal{X}\leftarrow$ create LLM-powered extraction function using \texttt{prompt}
        \Else
            \State $\mathcal{X} \leftarrow \undef{get-extraction-code}(\str{S}, Y_S, \str{p}, \texttt{description})$\label{alg:generator:code}
        \EndIf
        \State \textbf{return} $\mathcal{X}$\label{alg:generator:return}
\EndProcedure
\caption{Featurization Generation}
\label{alg:generator_function}
\end{algorithmic}
\end{algorithm}
\subsection{LLM Pipeline for Featurization Generation}\label{sec:candidate:llm_pipeline}
%A single-shot prompting of LLMs to obtain featurizations may produce unreliable results. To improve reliability, 
We next describe the subroutine \texttt{get-featurization-from-examples} to generate factorizations. We use a multi-step pipeline that decomposes featurization generation into several LLM invocations, following best practices~\cite{chen2023unleashing, khot2022decomposed} to improve reliability.
%As discussed earlier, LLMs are error prone and a single-shot prompting of the LLM to design a set of featurizations leads to unreliable featurizations. To ensure reliability, we design a multi-step pipeline that breaks down the process of candidate generation into multiple LLM invocations following best practices %---this pipeline performs Step (1b) in Fig.~\ref{fig:decomposition_solution}. 
%We here describe our LLM pipeline used to generate featurizations. It consists of a few LLM calls, each designed to obtain part of the featurization. Our method 
Alg.~\ref{alg:generator_function} shows this procedure (functions in \undef{blue} are LLM-powered; we present prompts in \iftoggle{techreport}{Appx.~\ref{sec:prompts}%
}{our technical report~\cite{techrep}}). 
%We provide the LLM with a few labeled samples as in-context demonstration of cases where the join condition holds and doesn't hold. 
It takes a set of example pairs and their labels as input that are used as in-context demonstrations of when the join condition holds and doesn't hold for generation.
%We use a few labeled samples as demonstration (i.e., in-context examples) to show the LLM when the join condition holds. 

Specifically, to generate featurizations, we first ask the LLM to provide a set of high-level natural language descriptions for possible featurizations (Line~\ref{alg:generator:logical}) and then instantiate each (Line~\ref{alg:line:for_loop}-\ref{alg:line:end_for_loop}). We use the description to both generate feature extraction functions (Lines~\ref{alg:generator:ext}-\ref{alg:generator:ext2}) and a distance function (Line~\ref{alg:generator:dist}). The latter is done by asking the LLM to choose among a set of predefined distance functions (including semantic or lexical distance, and arithmetic difference)---we provide the list of predefined distance functions in \iftoggle{techreport}{Appx.~\ref{sec:prompts}%
}{our technical report~\cite{techrep}}. Designing each feature extraction function $\mathcal{X}_L$ and $\mathcal{X}_R$ (Line~\ref{alg:generator:desc}-\ref{alg:generator:return}) also involves multiple LLM invocations: first to get a detailed description of the feature extraction function for the specific column (Line~\ref{alg:generator:desc}), then to decide whether to use LLM or code for extraction (Line~\ref{alg:generator:usellm}), and finally, in each case, another LLM invocation to either obtain the prompt for LLM-powered extraction (Line~\ref{alg:generator:nocode}) or Python code (Line~\ref{alg:generator:code}). 

\textbf{Other details}. In practice we modify Alg.~\ref{alg:generator_function} to additionally take as input (1) $\Phi$, the running set of featurizations in Alg.~\ref{alg:candidate_generation_all} and (2) the output of the feature extraction functions in $\Phi$ for the demonstrations passed to the LLM. Based on these inputs, the LLM is instructed to not repeat any existing featurizations, but fix any errors in the featurizations that may now surface by observing the output of the feature extraction functions.

\subsection{Evaluation and Picking Examples}\label{sec:candidate:algorithm}
%To obtain a comprehensive set of features, we run Alg.~\ref{alg:generator_function} multiple times with different sample sets $\str{S}$, and take the union of all the featurizations created as our candidate featurization set $\Phi$. In doing so, we modify Alg.~\ref{alg:generator_function} to take into account previously generated featurizations to avoid creating duplicate featurizations across calls. We discuss the modifications to Alg.~\ref{alg:generator_function} together with how we pick the sample set across LLM calls in Sec.~\ref{sec:candidate_featurization_generation}. 

%  new featurizations where the previous ones perform poorly, which both improves the reliability of the generation as well as making sure the set of featurizations are comprehensive. 

\begin{algorithm}[t]
\footnotesize
\begin{algorithmic}[1]
%\Require Join condition $\str{p}$, the lists $\str{L}$ and $\str{R}$
%\Ensure A set of candidate factorizations
%\Procedure{\texttt{terminate-or-pick-example}}{$\str{p}$, $\str{L}$, $\str{R}$}
\State $\str{S}_p, \str{S}_n\leftarrow$ negative and positive subsets of $\str{S}$
\For{$(\str{l}, \str{r})\in \str{S}_p$}\Comment{Calculate cost to cover for all positives}\label{alg:pick_ex:calc_loop}
    \State $c_{\Phi}{(\str{l}, \str{r})}\leftarrow \min\limits_{\phi\in\mathbf{\Phi}}\sum\limits_{(\str{l}', \str{r}')\in \str{S}_n}\mathds{I}[\phi(\str{l}', \str{r}')\leq \phi(\str{l}, \str{r})]$\label{alg:pick_ex:end_calc_loop}
\EndFor
\If{$\max_{(\str{l}, \str{r})\in \str{S}_p}c_{\Phi}{(\str{l}, \str{r})}$<$\alpha$}\label{alg:generation:relaxed_t}
    \State \textbf{return} $\emptyset$ \label{alg:generation:relaxed_t_return}
\EndIf
\State $\str{C}_p\leftarrow \frac{\beta}{2}$ pairs $(\str{l}, \str{r})\in\str{S}_p$ with largest $c_{\Phi}{(\str{l}, \str{r})}$\label{alg:pick_pos_sample}
\State $\str{C}_n\leftarrow$  $\{(\str{l}', \str{r}')\in\str{S}_n;\exists (\str{l}, \str{r})\in\str{C}_p, \phi\in\Phi\;\text{s.t.}\;\phi(\str{l}', \str{r}')\leq\phi(\str{l}, \str{r})\}$\label{alg:pick_neg_sample}
\State $\str{C}_n\leftarrow$ random subset of size $\frac{\beta}{2}$ of $\str{C}_n$ if $|\str{C}_n|>\frac{\beta}{2}$\label{alg:random_pick_neg_sample}
\State \textbf{return} $\str{C}_p\cup\str{C}_n$, labels for $\str{C}_p\cup\str{C}_n$ from $Y_S$\label{alg:line:example_return}
%\EndProcedure
\caption{\texttt{evaluate-and-pick-examples}$(\str{S}, Y_S, \mathbf{\Phi})$}\label{alg:pick_example}
\end{algorithmic}
\end{algorithm}

%\textbf{}. %\label{sec:generation:alg} 
We next describe \texttt{evaluate-and-pick-examples} for evaluating featurizations using a set of labeled samples, $\str{S}$. To do so, we define the notion of \textit{cost to cover}. Let $\str{S}_p$ and $\str{S}_n$ be, respectively, the positive and negative subsets of $\str{S}$. %We use this notion to choose examples for which we expect the cost to cover will improve the most. 
For any positive pair $(\str{l}, \str{r})\in\str{S}_p$, define
$$c_{\Phi}{(\str{l}, \str{r})}=\min\limits_{\phi\in\mathbf{\Phi}}\sum\limits_{(\str{l}', \str{r}')\in \str{S}_n}\mathds{I}[\phi(\str{l}', \str{r}')\leq \phi(\str{l}, \str{r})],$$
as the \textit{minimum cost to cover}  a positive pair $(\str{l}, \str{r})$ with featurizations in $\Phi$. The summation $\sum_{(\str{l}', \str{r}')\in \str{S}_n}\mathds{I}[\phi(\str{l}', \str{r}')\leq \phi(\str{l}, \str{r})]$ represents, for any $\phi$, the number of false positives that must be admitted to correctly classify the positive pair $(\str{l}, \str{r})$ if we only used $\phi$. The minimum cost to cover a positive pair $(\str{l},\str{r})$ with the featurization set $\Phi$ is therefore, the minimum number of false positives that must be admitted to correctly classify the positive pair using any one of the featurizations in $\Phi$. If $c_{\Phi}(\str{l}, \str{r})$ is high, none of the featurizations in $\Phi$ are suitable to distinguish $(\str{l}, \str{r})$ from the negative pairs. Thus, the featurizations should be improved by taking the high cost to cover pair $(\str{l}, \str{r})$ into account. We pick such a pair as an example to be passed to the LLM during candidate generation to ensure the LLM generates new featurizations that are suitable for the pair. %, guiding the LLM to generate new featurizations suitable for $(\str{l}, \str{r})$.
Examples of cost to cover computation are shown in Fig.~\ref{fig:featurization_details_ex}. In Step 3 and after generating a single featurization $\phi_1$, we see that 4 positive pairs have non-zero cost to cover. The cost to cover value is calculated based on number of negative pairs to the left (i.e., with smaller feature distance values) of the positive pairs. A subset of these pairs are used to generate a new featurization, $\phi_2$, in Step 4. Then, in Step 5, we again compute cost to cover but now for two featurizations. We see that the cost to cover improves for some of the positive pairs when using the featurization, $\phi_2$, added in Step 4. 

Alg.~\ref{alg:pick_example} specifies this process more formally. We first calculate the minimum cost to cover for all positive pairs (Lines~\ref{alg:pick_ex:calc_loop}-\ref{alg:pick_ex:end_calc_loop}). If this cost is sufficiently small for all the pairs---according to a system parameter $\alpha$---we assume the candidate featurizatations are a suitable indicator of the join result and thus return an empty set (Lines~\ref{alg:generation:relaxed_t}-\ref{alg:generation:relaxed_t_return}). Otherwise, we return $\beta$ pairs in total ($\frac{\beta}{2}$ positives and negatives) to ensure they fit in the LLM context. To generate a candidate set that enables a high recall featurized decomposition, %so we focus on picking examples with high cost to cover for the positive pairs %. Thus, we select positive pairs with high cost to cover,  and some negative pairs to return as examples 
%(Lines~\ref{alg:pick_pos_sample}-\ref{alg:line:example_return}). 
we select positive pairs with the highest cost to cover (Line~\ref{alg:pick_pos_sample}). We select negative pairs that have smaller feature distances than at least one selected positive pair (Line~\ref{alg:pick_neg_sample}), randomly selecting $\frac{\beta}{2}$ of such pairs if there are more (Line~\ref{alg:random_pick_neg_sample}); this choice of negative pairs helps the LLM understand why the cost to cover for positive pair was high.

%any of the featurization in $\Phi$ should already be a good indicator for the examples provided but fail due to an implementation error, correct the implementation of the featurization. Further details are presented in Appx.~\ref{sec:prompts}.  

\section{Logical Expression Formulation}\label{sec:threshold:all}
We obtain a set of candidate featurizations, $\mathbf{\Phi}$, from the previous step. Here, we use a subset of these featurizations to create a featurized decomposition, $\Pi$. %To create our featurized decomposition, we first sample a set of pairs to guide how our featurized decomposition is created. We use this sample set to both estimate the cost of join and the recall using any featurized decompostion. Using these estimates, we create a featurized decomposition estimated to minimize cost while guaranteeing high recall.
%We first formalize this two step process below by introducing the necessary notation, and then discuss how we solve them in Secs.~\ref{sec:param_decomp} and \ref{sec:threshold_selection}.%We note that even in this step obtaining tight theoretical bounds on recall is challenging as we see later. 
%By dividing the problem into two steps, because the second step, where we choose thresholds is much more amenable to theoretical analysis. Nonetheless, even the problem of setting thresholds for a parameterized decomposition is non-trivial and still requires in depth mathematical analysis. 
%In the rest of this section, we first present our method creating a suitable parameterized decomposition and finally present our algorithm for setting the threshold parameters to create the final decomposition. 
As discussed in Sec.~\ref{sec:overview:jain_framework}, we construct the featurized decomposition by first generating a \textit{logical scaffold} and then generating the final featurized decomposition by setting its parameters. We state this process more formally by defining the notion of a logical scaffold in Sec.~\ref{sec:def:scaffold}, before presenting our method for creating the scaffold in Sec.~\ref{sec:param_decomp} and setting the thresholds in Sec.~\ref{sec:threshold_selection}, and finally detail our theoretical analysis in Sec.~\ref{sec:adj_target}.

\subsection{Logical Scaffolds}\label{sec:def:scaffold}
Recall that a featurized decomposition consists of a set of featurized predicates, with every featurized predicate of the form, $\phi(\str{l}, \str{r})\leq \theta$, for a featurization, $\phi$ and a threshold parameter, $\theta$. We define a \textit{predicate scaffold} similarly, except that a predicate scaffold takes the threshold parameter as an input. Thus, a predicate scaffold is a function of the form $\mathring{\pi}(\str{l}, \str{r};\theta)=\mathds{I}[\phi(\str{l}, \str{r})\leq \theta]$ for some $\phi\in \Phi$. We similarly extend the notion of featurized clauses and featurized decomposition to take the threshold parameter as input. Formally, a \textit{clause scaffold} $\mathring{\kappa}$ with predicate scaffolds $\mathring{\pi_1}, ..., \mathring{\pi_k}$ is the function
$
\mathring{\kappa}(\str{l}, \str{r};\Theta)= \mathring{\pi}_1(\str{l}, \str{r};\Theta_1)\lor ... \lor \mathring{\pi}_k(\str{l}, \str{r};\Theta_k),
$
for $\Theta\in \mathbb{R}^k$ and $\Theta_i$ its $i$-th element. A \textit{logical scaffold} with clauses $\mathring{\kappa}_1,..., \mathring{\kappa}_{k'}$ is a function
%$\mathring{\Pi}$, which is a boolean expression of the form 
$$\mathring{\Pi}(\str{l}, \str{r};\Theta)=\mathring{\kappa}_1(\str{l}, \str{r};\Theta^1)\land...\land \mathring{\kappa}_{k'}(\str{l}, \str{r};\Theta^{k'}),$$
where $\Theta=[\Theta^1, ..., \Theta^{k'}]$ is a list of parameters, each a high dimensional vector of threshold parameters for each clause. We abuse notation and define, for any $\str{S}\subseteq\str{L}\times\str{R},$ $$\mathring{\Pi}(\str{S}; \Theta)=\{(\str{l}, \str{r}); (\str{l}, \str{r})\in \str{S}, \mathring{\Pi}(\str{l}, \str{r};\Theta)=1\}.$$

A logical scaffold takes threshold parameters as input, so it does not on its own specify any predicates. Given a logical scaffold, we need to find suitable threshold parameters to create a featurized decomposition. Thus, using the above notation, to find a featurized decomposition, we first find a logical scaffold $\mathring{\Pi}(\str{l}, \str{r}; \Theta)$; we discuss how in Sec.~\ref{sec:param_decomp}. Then, we create the final featurized decomposition by finding a suitable set of threshold parameters $\Theta^*$, so we obtain $\Pi(\str{l}, \str{r})=\mathring{\Pi}(\str{l}, \str{r}; \Theta^*)$; we discuss how in Sec.~\ref{sec:threshold_selection}.

\begin{algorithm}[t]
\footnotesize
\begin{algorithmic}[1]
\State $\mathring{\Pi}(\str{l}, \str{r};\Theta)\leftarrow\texttt{True}$\Comment{Initialize $\mathring{\Pi}$ as a function always returning True}
    \State $c_{min}, \phi_{min}\leftarrow \hat{C}_S(\mathring{\Pi}), \emptyset$
\While{$|{\Phi}|\geq 0$} \label{alg:param_decomp:while}
    \For{$\phi\in\Phi$} \label{eq:parameterized_dcomp_conj_phi}
        \State $\mathring{\Pi}'(\str{l}, \str{r};\Theta\cup\{\theta\})\leftarrow\mathring{\Pi}(\str{l}, \str{r};\Theta)\land \mathds{I}[\phi(\str{l}, \str{r})\leq \theta]$
        \If{$\hat{C}_S(\mathring{\Pi}')<c_{min}$}
            \State $c_{min}, \phi_{min}\leftarrow \hat{C}_S(\Pi'), \phi$
        \EndIf
    \EndFor\label{eq:parameterized_dcomp_conj_endphi}
    \State $\Phi\leftarrow\Phi\setminus\{\phi_{min}\}$
    \If{$\phi_{min}\neq\emptyset$ \textbf{and} $c_{min}<\hat{C}_S(\mathring{\Pi})-\gamma$}
        \State $\mathring{\Pi}(\str{l}, \str{r};\Theta\cup\{\theta\})\leftarrow\mathring{\Pi}(\str{l}, \str{r};\Theta)\land \mathds{I}[\phi_{min}(\str{l}, \str{r})\leq \theta]$    \Else
        \State \textbf{break}\label{alg:param_decomp:endwhile}
    \EndIf
\EndWhile
\For{$\phi\in{\Phi}$} \label{alg:param_decomp:for}
    \For{each clause $\mathring{\kappa}$ in $\mathring{\Pi}$}
        \State $\mathring{\kappa}'(\str{l}, \str{r};\Theta\cup\{\theta\})\leftarrow\mathring{\kappa}(\str{l}, \str{r};\Theta)\lor \mathds{I}[\phi(\str{l}, \str{r})\leq \theta]$\label{alg:param_decomp:add_to_clause}
        \State $\mathring{\Pi}'\leftarrow\mathring{\Pi}$ but replace  $\mathring{\kappa}$ with $\mathring{\kappa}'$    
        \If{$\hat{C}_S(\mathring{\Pi}')<\hat{C}_S(\mathring{\Pi})-\gamma$}\label{alg:param_decomp_clause_check}
            \State $\mathring{\Pi}\leftarrow\mathring{\Pi}'$\label{alg:param_decomp:endfor}
        \EndIf
    \EndFor
\EndFor
\State \textbf{return} $\mathring{\Pi}$
\caption{\texttt{get-logical-scaffold}(${\Phi}, T, \str{S}$, $Y_S$)}\label{alg:select_form}
\end{algorithmic}
\end{algorithm}

\subsection{Creating the Logical Scaffold}\label{sec:param_decomp}
%We create decompositions that are boolean predicates in conjunctive normal form (cnf). That is, in this step, we create a parameterized decomposition $\Pi(l, r;\Theta)=(\pi_{1,1}\lor ..\lor \pi_{1, k_1})\land (\pi_{2, 1}\lor ..\lor \pi_{k_2})\land...\land(\pi_{r, 1}\lor ..\lor \pi_{r, k_r})$. Each $\pi_{i, j}(l, r;\theta)=d(\mathcal{X}_L, \mathcal{X}_R)\leq \theta$ is a parameterized predicate where $(d, \mathcal{X}_L, \mathcal{X}_R)\in\Phi$ and $\theta\in\mathbb{R}$. 
%Next, we present our method for finding a parameterized decomposition using the candidate featurization set $\Phi$, presenting the main algorithm in Sec.~\ref{sec:general_param_decomposition} with further details in Sec.~\ref{sec:param_decomposition_details}.
%\subsubsection{Creating Parameterized Decomposition\\}\label{sec:general_param_decomposition}
We use a greedy approach to construct a logical scaffold using the candidate featurizations $\Phi$ (recall that finding an optimal featurized decomposition is NP-hard). We use a set of pairs $\str{S}$ sampled from $\str{L} \times \str{R}$ together with their labels $Y_S$ obtained from an LLM. We build a logical scaffold to minimize the join cost while satisfying the recall target on these samples.
Starting with an empty scaffold, we iteratively extend the current scaffold by adding a predicate that yields the largest reduction in join cost. However, the join cost depends on the choice of threshold parameters for the scaffold. So, we optimistically estimate the join cost when using the scaffold as the minimum achievable join cost across all possible threshold settings for the scaffold.
%At a high level, at each iteration, we add to the current logical scaffold a new predicate whose addition can lower the join cost the most. Note that a logical scaffold is an intermediary representation; we must set the threshold parameters to perform joins so the final cost of using a scaffold depends on the thresholds chosen. Thus, we consider the lowest possible join cost achievable when using the scaffold (across all possible threshold settings) when comparing different scaffolds.   %we consider optimistic choices of threshold parameters to estimate the final join. % cost to create our scaffold. % by  
%that the notion of a cost of a scaffold is undefined. To
%Doing so requires a method for estimating the cost of a logical scaffold. 
We next present our solution in detail, first discussing how we estimate the join  cost when using a scaffold before presenting our algorithm for creating scaffolds.

\textbf{Cost Estimation}. %We sample and label set of pairs, $\str{S}$ from $\str{L}\times\str{R}$ to estimate the join cost when using the logical scaffold, $\mathring{\Pi}$. 
We estimate the join cost for $\mathring{\Pi}$ as the lowest cost achievable when using $\mathring{\Pi}$ while meeting the recall target on the samples, $\str{S}$. % on the sample set is at least $T$ and (2) estimate the join cost as the smallest cost  Then, we estimate the cost of performing the join when using $\mathring{\Pi}$ with threshold $\Theta$.
We use the false positive rate of $\mathring{\Pi}$ as a proxy for cost. % can achieve while meeting the recall target on the sampled set, $\str{S}$. 
False positive rate determines the cost of refining the join result with an LLM (i.e., Step 3 in Fig.~\ref{fig:fd_inf}) which in practice dominates the join cost---extensions to more fine-grained cost models are straightforward but in practice we see limited benefits. %---we defer the discussion to Appx.\ref{}.  
More formally, denote by $\str{S}_p$ and $\str{S}_n$ respectively the positive and negative subsets of $\str{S}$. Define
$$
\mathfrak{R}_S(\mathring{\Pi}, \Theta)=\frac{|\mathring{\Pi}(\str{S}_p;\Theta)|}{|\str{S}_p|},\;\mathfrak{F}_S(\mathring{\Pi}, \Theta)=\frac{|\mathring{\Pi}(\str{S}_n;\Theta)|}{|\mathring{\Pi}(\str{S};\Theta)|},
$$
respectively, the observed recall and false positive rate of a logical scaffold $\mathring{\Pi}$, when using thresholds $\Theta$. We define the cost function
\begin{align}\label{eq:minimum_cost_threshold}
\hat{C}_S(\mathring{\Pi})=\min_{\Theta\in\mathbf{\Theta}}\mathfrak{F}_S(\mathring{\Pi}, \Theta)\;\text{s.t.}\;, \mathfrak{R}_S(\mathring{\Pi}, \Theta)\geq T,
\end{align}
where $\mathbf{\Theta}$ is the set of all possible thresholds and depends on the logical scaffold, $\mathring{\Pi}$. % the minimum is over all possible thresholds for the parameterized decomposition. %This cost function denotes the minimum false positive rate possible when using $\mathring{\Pi}$ on $\str{S}$ while keeping the observed recall at least $T$. We use this cost function because in practice, the majority of the cost of the join comes from refining the join result, which is directly proportional to the false positive rate.
In practice, we find this minimum in Eq.~\ref{eq:minimum_cost_threshold} through exhaustive search since the set of all possible thresholds is typically small. Note that all possible thresholds must have recall $\geq T$ and among such thresholds we only need to consider thresholds at which cost changes. For small sample sizes, such a search space is typically small and can be exhaustively searched efficiently. We provide further details in \iftoggle{techreport}{Appx.~\ref{sec:param_decomposition_details}%
}{our technical report~\cite{techrep}}. %We also discuss potential refinements to cost estimation in Sec.~\ref{sec:param_decomposition_details}. They in practice yielded marginal gain and thus we omit them for simplicity. 

%check whether it is beneifical to add new predicates within each clause using disjunctions. The output of this set is a parameterized boolean expressions $\Pi(l, r;\theta)$, where the thresholds $\theta$ needs to be separately set. 

\textbf{Scaffold Construction Algorithm}. Using this cost function, creating the logical scaffold follows two steps. First, we create a logical scaffold only consisting of conjunction of predicates. Then, we iteratively add disjunctions to each clause if the addition reduces cost. We present this algorithm formally in Alg.~\ref{alg:select_form}. We first create the conjunctive clauses (Lines~\ref{alg:param_decomp:while}-\ref{alg:param_decomp:endwhile}). To do so, we find a new featurization whose addition to the current logical scaffold decreases the cost the most among all possible featurizations, estimated by $\hat{C}_S$ (Lines~\ref{eq:parameterized_dcomp_conj_phi}-\ref{eq:parameterized_dcomp_conj_endphi}). If such a featurization reduces cost sufficiently over the current logical scaffold, then it is added with a conjunction to the current scaffold; we check if the reduction in cost is at least equal to a threshold $\gamma$ to avoid adding featurizations that have a marginal difference. After obtaining a logical scaffold containing conjunctions, we iteratively add featurizations to each clause with disjunctions (Lines~\ref{alg:param_decomp:for}-\ref{alg:param_decomp:endfor}). %Similar to before, we iteratively add a featurization to a clause if it reduces cost by more than $\gamma$. 
We iteratively consider each featurization and evaluate for the featurization if its addition to a clause reduces cost by more than $\gamma$ (Lines~\ref{alg:param_decomp:add_to_clause}-\ref{alg:param_decomp_clause_check}), and add it the clause if so (Line~\ref{alg:param_decomp:endfor})---we consider each featurization and clause pair once. 

\begin{figure}
\vspace{-0.3cm}
    \centering
    \includegraphics[width=\linewidth]{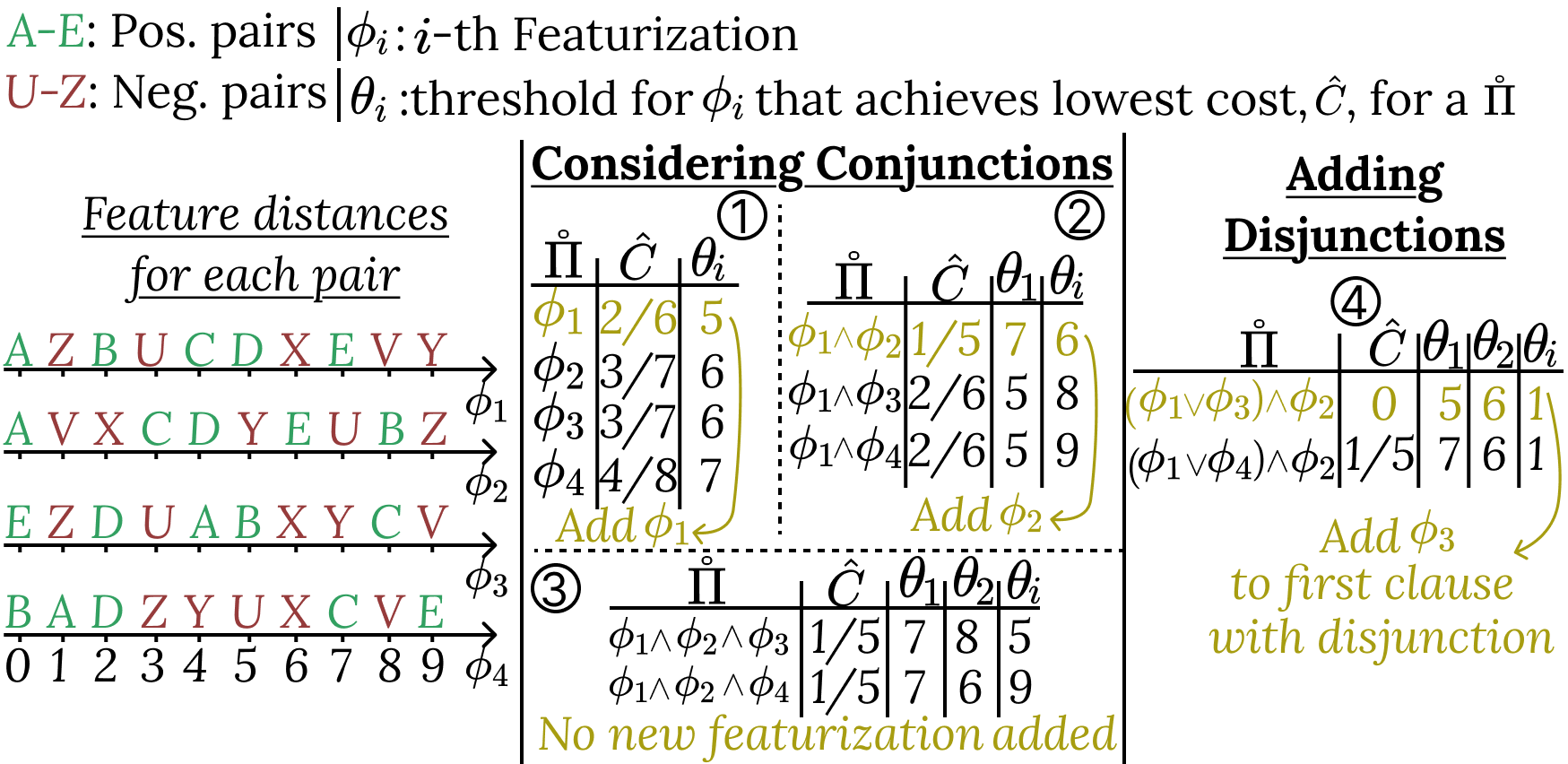}
    \caption{Example of Scaffold Construction with $T=0.8$}
    \label{fig:scaffold_ex}
\end{figure}

\textbf{Example}. We illustrate this process using Fig.~\ref{fig:scaffold_ex}. The figure constructs a scaffold from 4 featurizations $\phi_1$, ..., $\phi_4$ with target $T=0.8$ and using a sample set of 10 pairs, 5 positive (denoted by A-E) and 5 negative (denoted by U-Z). The figure (left) shows the distance for each pair for each featurization, e.g., $\phi_1(D)=5$. All distances for all featurizations are integers between 0 to 9. To construct the scaffold, at Step (1), we find a featurization, $\phi_i$, such that setting $\mathring{\Pi}(\str{l}, \str{r};\theta)=\mathds{I}[\phi_i(\str{l}, \str{r})\leq \theta]$ achieves the lowest cost, $\hat{C}(\mathring{\Pi})$. The table in Step (1) shows this cost computation for all four $\phi_i$, where the first column shows different possible scaffolds, the second column shows the cost, $\hat{C}$, of these scaffolds calculated based on Eq.~\ref{eq:minimum_cost_threshold} and the third column shows the \textit{best threshold setting} for each scaffold that achieves the minimum in Eq.~\ref{eq:minimum_cost_threshold}. For example, for the scaffold $\mathring{\Pi}(\str{l}, \str{r};\theta)=\mathds{I}[\phi_1(\str{l}, \str{r})\leq \theta]$ (the first row in the table in Step (1)), $\hat{C}(\mathring{\Pi})=\frac{2}{6}$ and that the threshold,  $\theta_1$, at which this cost is achieved is 5. %For example, when using $\phi_1$, $\hat{C}(\mathring{\Pi})=\frac{2}{6}$ is achieved when $\theta_1=5$. 
This is because negative pairs $Z$ and $U$ as well as positive pairs A, B, C, D have $\phi_1$ distance at most 5, so that false positive rate at $\theta_1=5$ is $\frac{2}{6}$, while recall is 0.8. Based on the table at Step (1), $\phi_1$ yields the lowest cost and is thus added to the scaffold. 

Next, at Step (2), we again consider the cost of adding remaining featurizations to the scaffold. Adding $\phi_2$ now leads to the lowest $\hat{C}$---the table at Step (2) shows $\hat{C}$ for different scaffolds, with corresponding best threshold settings that achieve the cost. Observe that the best value of $\theta_1$ (that achieves the cost $\hat{C}$) is different for different logical scaffolds; e.g., compare the best value for $\theta_1$ for the scaffold consisting of  conjunction of $\phi_1$ and $\phi_2$ (in Step (2)) and the scaffold only consisting of $\phi_1$ (in Step (1)). Next, at Step (3), adding new featurizations with conjunctions does not reduce the cost anymore, and thus the algorithm moves on to adding disjunctions. Step (4) shows the cost of scaffolds when adding disjunctions to the first clause (Alg.~\ref{alg:select_form} considers adding featurizations to all clauses, which we omit to simplify the figure), where now the disjunction of $\phi_1$ and $\phi_3$ reduces the cost. 

%The table shows setting $\mathring{\Phi}(\str{l}, \str{r};\theta)=\mathds{I}[\phi_1\leq \theta]$ achieves the lowest cost, $\hat{C}$. The table also shows the threshold at which this lowest occurs.  

%Specifically, let $\hat{C}_S(\mathring{\Pi}, \Theta)$, be an estimated cost of using a featurized decomposition with a specific $\Theta$, estimated based on the set $S$. 
%We first define the cost function we use, before discussing presenting the algorithm.

%\textbf{Cost function}. To formalize this procedure, we first define an estimate of cost for a parameterized decomposition. We use the minimum number of negatives admitted by the decomposition as a measure of its cost. Specifically, for any parameterized decomposition $\Pi(l, r; \Theta)$, for a value $\Theta$, let the number of negatives admitted be $N(\Theta, \Pi)=\sum_{(l, r)\in S_n}\Pi(l, r;\Theta)$. For a recall target $T'$, let the lowest number of negatives admitted be $\hat{C}(\Pi)=\min_{\Theta, R(\Theta, \Pi)\geq T}N(\Theta, \Pi)$. 

%. Since the output of this step gets refined further when we actually set the predicate parameters, we do not nee 

%We do so following a greedy algorithm. We first create a set of conjunctive clauses, and then iteratively add new disjunctions to each clause. Each decision is made by choosing a featurization that minimizes our cost estimate. To estimate the cost, we simply consider the number of negatives returned in the result set. Since this step does not require very accurate estimate, the number of negative results is fine.

\vspace{-0.1cm}
\subsection{Setting the Thresholds}\label{sec:threshold_selection}
\vspace{-0.2cm}
%. Let $\Theta=(\theta^1, ..., \theta^r)\in \mathbb{R}.$ Then, for any set of thresholds $\Theta$, we obtain a featurized decomposition. Let $\P$
%to create a featurized decomposition 
%$\bigvee\limits_{(d,\mathcal{X}_L, \mathcal{X}_R)\in\mathbf{\Phi}}d()$
%\subsection{Threshold Selection Algorithm}
The previous step returns a logical scaffold $\mathring{\Pi}(\str{l}, \str{r};\Theta)$. Next, we find the threshold parameters $\Theta^*$, to construct the final featurized decomposition $\Pi(\str{l}, \str{r})=\mathring{\Pi}(\str{l}, \str{r};\Theta^*)$. We want to find thresholds that both minimize cost and meet the recall target with high probability. %Given that the parameterized predicate

%Our threshold selection algorithm samples a set of $k$ pairs $S$ from $\str{L}\times\str{R}$ and uses it to select a threshold that guarantees recall while minimizing cost. 
%We first define some necessary notation before discussing our method.
%as the final recall of choosing thresholds $\theta$ on the whole dataset and similarly define  $\mathfrak{R}_S(\theta)$ for the dataset $S$. 
%\textbf{Setting Thresholds}. We combine all thresholds within each clause to be a single threshold. 

%\textbf{Notation}. Define $D=\str{L}\times\str{R}$ and note that $S\subseteq D$ is a random subset of $D$ taken uniformly at random. Define $\mathfrak{R}_D(\Theta)=\mathfrak{R}(\hat{Y}_\Theta)$, for $\hat{Y}_\Theta$ defined above. Let $Y^S=Y\cap S$, i.e., the set sampled positive pairs, and let $\bar{Y}^S=\bar{Y}\cap S$ be the set of sampled negatives. Define $\hat{\str{Y}}_\Theta^S=\{(\str{l}, \str{r})\in S; \bigvee_{\pi\in\Pi_\Theta}\pi(\str{l}, \str{r})\}$, and finally let $\mathfrak{R}_S(\Theta)=\frac{|\hat{\str{Y}}_\Theta^S\cap Y^S}{|Y^S|}$, that is the recall with respect to the observed samples of using the featurized decomposition $\Pi_\Theta$. We refer to $\mathfrak{R}_S(\Theta)$ as the observed recall and $\mathfrak{R}_D(\Theta)$ as the true recall. 

\subsubsection{Threshold Selection Algorithm\\}
To set our threshold, we first sample a set, $\str{S}$, uniformly at random from $\str{L}\times \str{R}$ and label them  ($\str{S}$ is a new sample set not the same set as previous sections). Let $k^+$ be the number of positive pairs observed in $\str{S}$. We refer to $\mathfrak{R}_S(\mathring{\Pi}, \Theta)$ as \textit{observed recall} and $\mathfrak{R}_{\str{L}\times\str{R}}(\mathring{\Pi}, \Theta)$ as \textit{true recall} for a featurized decomposition. For now, we assume $\mathring{\Pi}$ has $r$ clauses with no disjunctions so that $\mathring{\Pi}$ contains only conjunctions of the $r$ predicates. We extend the method to consider disjunctions in \iftoggle{techreport}{Appx.~\ref{sec:thresh:disjunctions}%
}{our technical report~\cite{techrep}}. Here, the set of all possible threshold parameters is $\mathbb{R}^r$, and we want to use $\str{S}$ to select a threshold vector $\Theta^*\in \mathbb{R}^r$, such that 
\begin{align}\label{eq:recall_requirement}
\mathds{P}_{\str{S}\sim\str{L}\times\str{R}}(\mathfrak{R}_{\str{L}\times\str{R}}(\mathring{\Pi}, \Theta^*)< T)\leq \delta.    
\end{align}

To do so, we first find an \textit{adjusted target} $T'>T$ such that any possible threshold that has observed recall more than $T'$ is guaranteed to have true recall that meet the original target, $T$ with high probability. More formally, let
$\mathbf{\bar{\Theta}}=\{\Theta;\Theta\in \mathbb{R}^r,\mathfrak{R}_{\str{L}\times\str{R}}(\mathring{\Pi},\Theta)<T\}$, be the set of thresholds whose true recall is below $T$. We find an adjusted target $T'$ such that
\begin{align}\label{eq:adj_target}
    \mathds{P}_{S\sim \str{L}\times\str{R}}(\exists\Theta\in \mathbf{\bar{\Theta}}\;\text{s.t.}\;\mathfrak{R}_{\str{S}}(\mathring{\Pi},\Theta)\geq T')\leq \delta.
\end{align}

Then, using an adjusted target $T'$ that satisfies Eq.~\ref{eq:adj_target}, we find a threshold $\Theta^*$  that has observed recall at least $T'$ and is estimated to have minimal cost. Similar to before, we use the false positive rate as a proxy for cost %Thm~\ref{thm:alg_guarantee}, we can select any threshold as long as it's observed recall is more than $T'$ and provide guarantees on the true recall. 
and choose the threshold parameters $\Theta^*$ as
\begin{align}\label{eq:opt_thresh}
\Theta^*\in\arg\min_{\Theta\in\mathbb{R}^r} \mathfrak{F}_S(\mathring{\Pi}, \Theta)\;\text{s.t.}\;\mathfrak{R}_S(\mathring{\Pi}, \Theta)\geq T'.    
\end{align}
Eq.~\ref{eq:opt_thresh} is similar to Eq.~\ref{eq:minimum_cost_threshold}---except the use of adjusted target $T'$ instead of $T$ to provide statistical guarantees---and we use the same method to find $\Theta^*$ (i.e., exhaustive search, as discussed in Sec.~\ref{sec:param_decomp}). Observe that if an adjusted target $T'$ satisfies Eq.~\ref{eq:adj_target}, and we define our featurized decomposition $\Pi(\str{l}, \str{r})=\mathring{\Pi}(\str{l}, \str{r};\Theta^*)$ for $\Theta^*$ selected from Eq.~\ref{eq:opt_thresh}, then $\Pi(\str{l}, \str{r})$ meets the recall target with high probability. 

Thus, it remains to find an adjusted target $T'$ that satisfies Eq.~\ref{eq:adj_target}. We define a target adjustment function  $\texttt{adj-target}(k^+, r, T, \delta)$ that takes the number of positive samples $k^+$, number of clauses $r$, target $T$, and probability of failure $\delta$ as input and outputs an adjusted target. We specify the details of this function---which are not straightforward---in Sec.~\ref{sec:adj_target}. Here, we first discuss its theoretical properties and use in our algorithm. The following theorem shows the theoretical guarantees the function provides. %for the case that . 

\vspace{-0.1cm}
\begin{theorem}\label{thm:alg_guarantee}
    Let $T'= \texttt{adj-target}(k^+, r, T, \delta)$, where $k^+$ is the number of positive samples in a uniform sample $\str{S}$ from $\str{L}\times\str{R}$, $r$ is the number of clauses in a logical scaffold, $\mathring{\Pi}$, with only conjunctions and $T$ and $\delta$ are user-provided recall target and probability of failure, respectively. For any threshold $\Theta^*$ where $\mathfrak{R}_S(\mathring{\Pi}, \Theta^*)\geq T'$, we have $\mathds{P}_{\str{S}\sim \str{L}\times\str{R}}(\mathfrak{R}_{\str{L}\times\str{R}}(\mathring{\Pi},\Theta^*)<T)\leq \delta$, whenever $r\leq \frac{1}{1-T}$ and $k^+> \frac{1}{1-T}$. %Moreover, for any threshold less than $T'$, there exists a dataset $D'$ such that the bound does not hold. %, where the randomness is over selection of $\theta^*$.
\end{theorem}
\vspace{-0.1cm}
Using the theorem, we first apply the function $\texttt{adj-target}$ to compute the adjusted target, and then use Eq.~\ref{eq:opt_thresh} to select the threshold parameters. Thm.~\ref{thm:alg_guarantee} guarantees that this procedure satisfies the recall requirement with high probability. The conditions on the theorem (i.e., $r\leq \frac{1}{1-T}$ and $k^+> \frac{1}{1-T}$) are an artifact of our proof technique. Although we expect extensions for relaxing them to be possible (we discuss directions for extension in \iftoggle{techreport}{Appx.~\ref{sec:all_proof}%
}{our technical report~\cite{techrep}}), we saw no practical need for them. For example, for the common case of  $T=0.9$, $k^+>10$ is needed in practice to obtain statistically significant results irrespective of Thm.~\ref{thm:alg_guarantee} (in our experiments we use 200 positive samples). Moreover, in all our experiments Alg.~\ref{alg:select_form} returned logical scaffolds with fewer than 10 clauses so that $r\leq 10$ holds without enforcing it in Alg.~\ref{alg:select_form}. To ensure theoretical validity, we enforce the constraint on $r$ in Alg.~\ref{alg:select_form} by terminating the loop in Line~\ref{alg:param_decomp:while} if number of clauses exceeds $\frac{1}{1-T}$.  %For now, we first specify our function $\hat{C}$ and then our algorithm for finding the thresholds. 

\subsection{Target Adjustment Details}\label{sec:adj_target}
%\sep{need to add the condition at the end, also dependence on number of positives}
%\sep{overall, need to clarify use of positives and how to sample}
Finally, we discuss our target adjustment function and provide an overview of the proof of Theorem~\ref{thm:alg_guarantee}. %. We discuss the setting where $\mathring{\Pi}$ only contains conjunctions, and defer the details of the case with disjunction (i.e., Lemma~\ref{lemma:adj_target_disj}) to Appx.~\ref{sec:all_proof}.

\textbf{Notation}. $\mathring{\Pi}$ is a conjunction of $r$ predicates, so let $\phi_1, ..., \phi_r$ be the featurization used in each of the $r$ predicates. Define $D\in\mathbb{R}^{n\times r}$ for $n=|\str{L}\times\str{R}|$ to be the dataset of feature distances between every pair of records and for all features. Specifically, the $j$-th column in the $i$-th row of $D$ is $D_{i, j}=\phi_j(\str{l}_i, \str{r}_i)$ for $i\in[n]$, $j\in[r]$, where $(\str{l}_i, \str{r}_i)$  is the $i$-th row of $\str{L}\times\str{R}$. Let $D^+$ contain only rows of $D$ corresponding to positive pairs and let $n^+=|D^+|$. %Note that any sampled subset, $\str{S}$, of $\str{L}\times\str{R}$ directly maps to a subset of $D$ so we focus on sampling over $D$. 
We denote by $S\in\mathbb{R}^{k\times r}$ a uniform sample without replacement from $D$ containing $k$ rows, use $S^+$ to denote its subset with positive labels, and let $k^+=|S^+|$. Note that for any $\Theta$, $\mathfrak{R}_{\str{L}\times\str{R}}(\mathring{\Pi}, \Theta)=\frac{\sum_{i\in[n^+]}\mathds{I}[\bigwedge_{j\in[r]}D^+_{i, j}\leq \Theta_j]}{|D^+|}$, which depends on $D^+$ instead of $\str{L}\times\str{R}$. We simplify our notation and define $\mathfrak{R}_{D^+}(\Theta)=\mathfrak{R}_{\str{L}\times\str{R}}(\mathring{\Pi}, \Theta)$ and similarly $\mathfrak{R}_{S^+}(\Theta)=\mathfrak{R}_{\str{S}}(\mathring{\Pi}, \Theta)$. % to denote the observed recall for any dataset $S$. 

\textbf{Minimum Adjusted Target Problem}. %. %Thus, we only focus on $S^+$ and $D^+$ . %Datasets $D$ and $S$ are sufficient to study recall. %, and we focus on $D$ in the rest of this section. 
Using the above notation to rewrite Eq.~\ref{eq:adj_target}, %when sampling $S^+$ from $D^+$ uniformly at random, 
we want a $T'$ such that  
\begin{align}\label{eq:adj_target_R}
    \mathds{P}_{S^+\sim D^+}(\exists\Theta\in \mathbf{\bar{\Theta}}_{D^+},\;\mathfrak{R}_{S^+}(\Theta)\geq T')\leq \delta,
\end{align}
where, for any $D$, $\mathbf{\bar{\Theta}}_{D}=\{\Theta\in\mathbb{R}^{r};\mathfrak{R}_{D}(\Theta)<T\}$ is the set of thresholds that don't meet the recall target. Note that $S^+\sim D^+$ denotes sampling $S^+$ uniformly from $D^+$, and the probability in Eq.~\ref{eq:adj_target_R} is equivalent to Eq.~\ref{eq:adj_target} that samples $S$ from $D$ since recall only depends on the positive pairs. This probability depends on the unknown dataset $D^+$. %Note that recall only depends on $S^+$ and $D^+$, and that $S^+$ is a uniform sample of $D^+$ so that it is equivalent to study the probability over sampling of $S^+$ from $D^+$ instead of $S$ from $D$. 
To bound it, we instead find $T'$ as
\begin{align}\label{eq:adj_target_max}
    \hspace{-0.3cm}T'=\min\limits_{\rho\in(T, 1]} \rho \;\text{s.t.} \max_{\hat{D}\in\mathbb{R}^{n^+\times d}}\mathds{P}_{\hat{S}\sim \hat{D}}(\exists\Theta\in \mathbf{\bar{\Theta}}_{\hat{D}}\;\text{s.t.}\;\mathfrak{R}_{\hat{S}}(\Theta)\geq \rho)\leq \delta.
\end{align}
Any $\rho$ in Eq.~\ref{eq:adj_target_max} that satisfies the condition in the equation is a valid adjusted target (that is, it satisfies Eq.~\ref{eq:adj_target_R}). Our goal is to find the smallest $\rho$ possible to produce a tight bound. We call the optimization problem in Eq.~\ref{eq:adj_target_max} the minimum adjusted target problem.
%which implies Eq.~\ref{eq:adj_target_R}. % that if we find a $T'$ for which Eq.~\ref{eq:adj_target_max} holds 
%Finally, to avoid trivial values for the adjusted target, $T'$, we want to find the smallest possible $T'$ such that Eq.~\ref{eq:adj_target_max} holds. Formally, our goal is to find

%consisting of $n=\str{L}\times\str{R}$ rows, and each    We refer to the quantity $\mathds{P}_{S\sim D}(\mathfrak{R}_S(\Theta)\geq T',\;\exists\Theta\in \mathbf{\bar{\Theta}})$ as \textit{probaility of failure} and denote it by $\phi_D(k, T_R, T')$.
\textbf{Finding Adjusted Target}. To solve Eq.~\ref{eq:adj_target_max}, we first identify the dataset $D^*$ that attains the maximum in the equation for a given $\rho$. Once $D^*$ is known, we can evaluate Eq.~\ref{eq:adj_target_max} by exhaustively searching over possible values of $\rho$ and computing the corresponding probabilities for $D^*$. The key challenge, addressed in the next lemma, is determining the dataset $D^*$ that achieves this maximum. To simplify the statement we present it for the setting where $n^+$ is a multiple of $r$ but the result holds in general, as stated in \iftoggle{techreport}{Appx.~\ref{sec:all_proof}%
}{our technical report~\cite{techrep}}. %. The following lemma shows what dataset achieves the maximum. 
%\vspace{-0.5cm}
\begin{lemma}\label{lemma:worst_case_ds}
    For any $r\leq \frac{n^+}{n^+(1-T)-1}$ and any $\rho>T$, define $D^i=\{x\times e_i;x\in[\frac{n^+}{r}]\}$ and $D_r^*=\cup_{i\in[r]} D^i$. We have 
    \begin{align}
        D^*_r\in \arg\max_{\hat{D}\in\mathbb{R}^{n^+\times d}}\mathds{P}_{\hat{S}\sim \hat{D}}(\exists\Theta\in \mathbf{\bar{\Theta}}_{\hat{D}}\;\text{s.t.}\;\mathfrak{R}_{\hat{S}}(\Theta)\geq \rho),
    \end{align}
    where $e_i\in\mathbb{R}^r$ is a vector with $i$-th element 1 and other elements 0.
\end{lemma}
\vspace{-0.2cm}
The proof is non-trivial and is discussed in \iftoggle{techreport}{Appx.~\ref{sec:all_proof}%
}{our technical report~\cite{techrep}}. Here we note that the important characteristic of $D^*_r$ is that data points have only one non-zero dimension which minimizes correlations across dimensions. Alg.~\ref{alg:threshold_adjust} uses the lemma to define the \texttt{adj-target} function. It considers different values of $T'$, checks the failure probability on $D^*_r$ and returns the smallest $T'$ whose probability of failure is below $\delta$. It is sufficient to consider $T'$ in increments of $\frac{1}{k}$ because $\mathfrak{R}_S(\Theta)$ is a multiple of $\frac{1}{k}$ for any $\Theta$. 

%To calculate $\phi$, we show how to find the dataset $D^*$ that achieves the maximum in the above equation. For the purpose of the lemma, $e_i$ is the $i$-th standard basis vector whose $i$-th element is 1 and other elements 0.
%Given Lemma~\ref{lemma:worst_case_ds}, we know that $\phi_{D^*}(k, T, T')$ provides an upper bound on the probability of failure for any dataset. Thus, to find a suitable adjusted target, we can compute $\phi_{D^*}(k, T, T')$ and use it in Eq.~\ref{eq:adj_target_prob} as an upper bound on the dataset specific probability of failure. 
We note that in Alg.~\ref{alg:threshold_adjust}, we need to calculate the probability of failure for dateset $D^*$ in Line~\ref{eq:calc_prob_alg}. This probability is independent of the dataset at hand and can be computed offline. Nonetheless, computing this probability is challenging for two reasons. First, it requires computing a probability over the union of a large set of possible events which creates a computationally intractable problem. Instead, we estimate its value numerically using Monte Carlo simulation. Given that this is a one-off data independent computation, the Monte Carlo simulation can be run offline and to high accuracy, so that the numerical error in computing the probability is negligible. In \iftoggle{techreport}{Appx.~\ref{appx:adj_target}%
}{our technical report~\cite{techrep}}, we discuss how to take into account the errors introduced from the Monte Carlo simulation. Second, $D^*$ still depends on $n^+$, the total number of true positives. We estimate $n^+$ through sampling and select $D^*$ based on this estimate. \iftoggle{techreport}{Appx.~\ref{appx:adj_target}%
}{Our technical report~\cite{techrep}} also discusses how to account for the error in estimating $n^+$ to ensure the final bound remains valid. %Here, we note that estimating $\phi_{D^*}(k, T, T')$ needs to be done offline and only once. Thus, we can use many trails in Monte Carlo simulation to do so to very high accuracy so that the errors from this estimation are very small and negligible. 

\begin{algorithm}[t]
\footnotesize
\begin{algorithmic}[1]
%\Procedure
%\State $S \leftarrow $ sample $k_n$ points uniformly at random
%\State $\hat{n}_p\leftarrow (\frac{|S^+|}{|S|}+\log \delta_1)\times n$
\For{$T'$ \textbf{in} $\{T+\frac{1}{k}, T+\frac{2}{k}, ..., 1\}$}\label{line:iter_over_t}
    \State $p\leftarrow\mathds{P}_{S\sim D_r^*}(\exists\Theta\in \mathbf{\bar{\Theta}}\;\text{s.t.}\;\mathfrak{R}_{S}(\Theta)\geq T')$\label{eq:calc_prob_alg}
    \If{$p\leq \delta$}
        \State\textbf{return} $T'$
    \EndIf
\EndFor
\State\textbf{return} $\infty$
%\EndProcedure
\caption{{\texttt{adj-target}}($k$, $r$, $T$, $\delta$)}\label{alg:threshold_adjust}
\end{algorithmic}
\end{algorithm}

\begin{algorithm}[t]
\footnotesize
\begin{algorithmic}[1]
%\Require Desired recall threshold $T$, probability of failure $\delta$, datasets $L$ and $R$, join prompt $\mathcal{P}$
%\Ensure join result
\State $\str{S}\leftarrow$ sample subset of size $k$ from $\str{L}\times\str{R}$ uniformly at random\label{eq:line_sampling}
\State $Y_S\leftarrow\{\mathcal{L}_{\str{p}}(\str{l}, \str{r});(\str{l}, \str{r})\in\str{S}\}$
\State $\Phi\leftarrow \texttt{get-candidate-featurizations}(\str{p}, \str{S}, Y_S)$ 
\State $\mathring{\Pi}\leftarrow \texttt{get-logical-scaffold}(\Phi, T, \str{S}, Y_S)$ 
\State $\str{S}'\leftarrow$ sample a subset of size $k'$ from $\str{L}\times\str{R}$ uniformly at random\label{eq:sample_2}
\State $\str{Y}_{S'}\leftarrow\{\mathcal{L}_{\str{p}}(\str{l}, \str{r});(\str{l}, \str{r})\in\str{S}'\}$
\State $T'\leftarrow$ \texttt{get-adjusted-target}($k', d, T, \delta$)\Comment{$d$ is number of clauses in $\mathring{\Pi}$}
\State $\Theta^*\leftarrow\arg\min_{\Theta} \mathfrak{F}_{S'}(\mathring{\Pi}, \Theta)\;\text{s.t.}\;\mathfrak{R}_{S'}(\mathring{\Pi}, \Theta)\geq T'$ \label{alg:find_thresh}\Comment{$\str{Y}_{S'}$ used here}
\State $\Pi(\str{l}, \str{r})\leftarrow \mathring{\Pi}(\str{l}, \str{r};\Theta^*)$\Comment{define featurized decomposition function}
\State $\hat{\str{Y}}\leftarrow \{(\str{l}, \str{r}); (\str{l}, \str{r})\in \str{L}\times \str{R}, \Pi(\str{l}, \str{r})=1\}$\label{line:est_join}
\State\Return $\{(\str{l}, \str{r}); (\str{l}, \str{r})\in \hat{\str{Y}}, \mathcal{L}_{\str{p}}(\str{l}, \str{r})=1\}$\label{line:alg:refine}
\caption{\texttt{\name{}}($\str{L}, \str{R}$, $\str{p}$, $T$, $\delta$)}\label{alg:join_overall}
\end{algorithmic}
\end{algorithm}

\vspace{-0.1cm}
\section{Final Algorithm and Extensions}\label{sec:pipeline}
\vspace{-0.05cm}
We next present the final \name{} algorithm and discuss extensions.

%\subsection{\name{} Algorithm and Guarantees}
\textbf{Final Algorithm}. Alg.~\ref{alg:join_overall} presents the final \name{} algorithm. We first sample and label a subset of $k$ pairs from $\str{L}\times\str{R}$. We pass this set to \texttt{get-candidate-featurizations} defined in Alg.~\ref{alg:candidate_generation_all} to generate candidate featurizations, $\Phi$. Then, we use $\Phi$ to create a logical scaffold---we use the same sample set $\str{S}$ in this step. To set the threshold parameters, we sample and label another subset, $\str{S}'$ of size $k'$, $\str{L}\times\str{R}$, obtain the adjusted target from Alg.~\ref{alg:threshold_adjust}
and find the threshold with the lowest false positive rate that has observed recall more than the adjusted target (Line~\ref{alg:find_thresh}). Using the threshold parameters, we now have a featurized decomposition that we use to create a candidate join result, $\hat{\str{Y}}$. The join result is then refined using $\mathcal{L}_{\str{p}}$ to remove false positives and returned to the user. The output of Alg.~\ref{alg:join_overall} provides the required recall guarantees:
\vspace{-0.1cm}
\begin{theorem}
    Let $\bar{\str{Y}}$ be output of Alg.~\ref{alg:join_overall}. Then, $\mathds{P}(\mathfrak{R}(\bar{\str{Y}}))<T)\leq \delta.$
\end{theorem}
\vspace{-0.1cm}
%uses the function \texttt{select-threshold} to select the corresponding thresholds for the candidate featurizations. These two steps together are used to create a featurized decomposition which is then applies to the dataset to produce a first candidate result set $\hat{Y}$. The LLM is then used on the results in $\hat{Y}$ to remove any false positives, after which the output is returned as the join result. %Note that, \name{} meeting the recall target follows directly from Thm.~\ref{thm:alg_guarantee}. Next, we analyze the cost complexity of this \name{}, and then discuss various potential extensions. 
We provide an analysis of the cost complexity of \name{} as well as impact of system parameters on cost in \iftoggle{techreport}{Appx.~\ref{sec:system_params}%
}{our technical report~\cite{techrep}}. In practice, we set $k$ and $k'$ to ensure a sufficient number of positive pairs to enable statistically significant estimates. % without oversampling too many records to avoid unnecessary labeling costs. In practice, 
In our experiments, we see fixed values perform well across datasets and \name{} is not sensitive to the parameter values.

\textbf{Considering Precision Target.} To take advantage of a relaxed precision target ($T_P < 1$), we introduce a post-processing step after generating the estimated result $\hat{\str{Y}}$. This step identifies positive pairs in $\hat{\str{Y}}$ using the existing featurizations $\Phi$, without invoking $\mathcal{L}_{\str{p}}$, thereby reducing the join cost.
Specifically, for each featurization $\phi \in \Phi$, we create a high-precision subset of $\hat{\str{Y}}$ as $\hat{\str{Y}}_\phi=\{(\str{l}, \str{r})\in\hat{\str{Y}};\phi(\str{l}, \str{r})\leq \theta\}$ where $\theta$ is a threshold  on the feature distances chosen in order to guarantees $\mathfrak{P}(\hat{\str{Y}}_\phi) \geq T_P$ with high probability. %; doing so we obtain
%$\hat{\str{Y}}_\phi=\{(\str{l}, \str{r})\in\hat{\str{Y}};\phi(\str{l}, \str{r})\leq \theta\}$.
Determining $\theta$ is a a one-dimensional threshold selection problem given a precision target, for which we directly apply the solution proposed by~\citet{zeighami2025cut}.
We then combine these subsets to obtain
$\hat{\str{Y}}'=\cup_{\phi\in\Phi}\hat{\str{Y}}_\phi
$
and return the final result as
$\{(\str{l}, \str{r}); (\str{l}, \str{r})\in \hat{\str{Y}}\setminus \hat{\str{Y}}', \mathcal{L}_{\str{p}}(\str{l}, \str{r})=1\}\cup \hat{\str{Y}}'$. 
To preserve theoretical guarantees, we account for the cumulative failure probability across multiple applications of~\cite{zeighami2025cut} and ensure that the subsets $\hat{\str{Y}}_\phi$ do not overlap. Details are provided in \iftoggle{techreport}{Appx.~\ref{sec:precision}%
}{our technical report~\cite{techrep}}.

\section{Experiments}\label{sec:exps}
We empirically evaluate \name{} in this section. Sec.~\ref{sec:exp:setup} presents our experimental setup, Sec.~\ref{sec:exp:across_ds} shows results comparing \name{} with baselines and Sec.~\ref{sec:exp:across_targets} studies sensitivity of \name{} to various parameters. We analyze the impact of data characteristics in Sec.~\ref{sec:exp:data_char}. % and finally present discussion of featurizations found in our experiments in Sec.~\ref{sec:exp:featureizations}.

\if 0

Datasets:
- Citations
- Police records
- Product Categories
- Biodex
- Product EM
Maybe these can just be small datasets where sampling matters more?
- Movies
- News articles

Perhaps add:
- more entity resolution type data
- more traditional semantic join type
\fi

\subsection{Setup}\label{sec:exp:setup}
\textbf{Datasets and Tasks}. We perform experiments on 6 different real-world datasets and multiple synthetic datasets. We discuss the real datasets here, and synthetic ones in Sec.~\ref{sec:exp:data_char}. We used datasets covering various domains and applications. \textit{Citations}~\cite{mahari2024lepard} is a  dataset of legal arguments and the task is to perform a self-join to find legal arguments that cite each other. \textit{Categorize} \cite{mcauley2013hidden, varma2019extreme} and \textit{BioDEX} \cite{d2023biodex, patel2024lotus} are multi-label classification tasks; the former classifying a product given its description into a list of categories and the latter classifying patient notes into a list of medical reaction terms. \textit{Products} is a classic entity resolution dataset from \cite{kopcke2010evaluation, mudgal2018deep} that contains product descriptions from different online listings and the task is to match listings that refer to the same product. \textit{Movies} is a dataset of Wikipedia pages of different actors and movies---we created this dataset by crawling Wikipedia pages of popular movies and actors in IMDB \cite{imdbdataset}. The task is to join movie pages with pages of actors based on whether an actor acts in a given movie. Finally, \textit{Police Records} is as discussed in Sec.~\ref{sec:intro}, where the task is to join records that belong to the same incident. For all datasets except BioDEX, Categorize and Citations, we use them as is; we describe our join prompts in \iftoggle{techreport}{Appx.~\ref{sec:prompts}%
}{our technical report~\cite{techrep}}. For those three datasets, the original data have million of records. To stay within our budgetary constraints, we sampled 20,000 ground-truth pairs, and used only the column values from the 20,000 pairs as the columns to be joined. Table~\ref{tab:datasize} shows data sizes and number of ground-truth pairs, $n^+$.

\begin{table}[t]
%\vspace{-0.1cm}
\footnotesize
\setlength{\tabcolsep}{2pt}
\begin{minipage}{0.6\columnwidth}
    \centering
    \begin{tabular}{cccc}
    \toprule
     & $|\str{L}|$ & $|\str{R}|$ & $n^+$ \\
    \midrule
    \textbf{Products} & 973 & 956 & 616 \\
    \textbf{BioDEX} & 8,103 & 3,718 & 20,000 \\
    \textbf{Citations} & 16,161 & 16,161 & 20,000 \\
    \textbf{Movies} & 4,043 & 4,983 & 10,475 \\
    \textbf{Police Records} & 2,093 & 2,093 & 56,531 \\
    \textbf{Categorize} & 19,101 & 3,144 & 20,000 \\
    \bottomrule
\end{tabular}
    \caption{Dataset Size}
    \label{tab:datasize}
\end{minipage}
\begin{minipage}{0.39\columnwidth}
\begin{tabular}{ccc}
\toprule
 & \textbf{Avg.} & \textbf{\% Failed} \\
\midrule
\textbf{LOTUS} &  75.4 & 100 \\
\textbf{BARGAIN} & 91.1 & 9 \\
\textbf{\name{}} & 91.2 & 7  \\   
%\midrule
\bottomrule
\end{tabular}
\caption{Observed recall and failure rate, $T_R=90\%,\delta=10\%$}
\label{tab:obs_recall}
\end{minipage}
\end{table}

\begin{table*}[t]
\vspace{-0.2cm}
%\hspace*{-1cm}
%\hspace*{-0.5cm}
%\vspace{-0.3cm}
\centering
\begin{tabular}{cccccccc}
\toprule
 & \textbf{Citations} & \textbf{Police Records} & \textbf{Categorize} & \textbf{BioDEX} & \textbf{Movies} & \textbf{Products}  \\
\midrule
\textbf{BARGAIN} & 28.0 & 81.3 & 27.6 & \textbf{73.9}& 69.9& 36.2  \\
\textbf{\name{}} & 
\textbf{6.80}~(\textcolor{cadmiumgreen}{\textbf{0.24$\times$}}) & 
\textbf{44.7}~(\textcolor{cadmiumgreen}{\textbf{0.55$\times$}}) & 
\textbf{23.0}~(\textcolor{cadmiumgreen}{\textbf{0.83$\times$}}) & 
\textbf{73.8}~(\textcolor{cadmiumgreen}{\textbf{0.99$\times$}}) & 
\textbf{6.70}~(\textcolor{cadmiumgreen}{\textbf{0.09$\times$}}) &
\textbf{25.0}~(\textcolor{cadmiumgreen}{\textbf{0.69$\times$}}) 
\\\midrule\midrule
\textit{optimal cascade} & \textit{19.1} & \textit{66.6} & \textit{12.5} & \textit{72.6} & \textit{52.5}& \textit{30.1}  \\
%\midrule
\bottomrule
\end{tabular}
\caption{Cost ratio at $T=90\%$ (\textcolor{cadmiumgreen}{number} in parenthesis shows cost reduction factor compared with BARGAIN).}
\label{tab:across_ds}
%\end{table*}
\end{table*}

\textbf{Baselines}. We compare \name{} with BARGAIN~\cite{zeighami2025cut} the state-of-the-art model cascade method with statistical guarantees. BARGAIN is designed for filter and classification tasks, but can be applied to joins by treating a join as a filter on the cross product of the columns. We use semantic similarity between the vector embedding of records as the proxy score for BARGAIN. %Furthermore, we report results for choosing \textit{the optimal} cascade threshold, that is, choosing the best possible threshold on semantic similarity between vectors.
Additionally, to understand the limit of cascade-based approaches that rely on semantic similarity alone, we consider \textit{optimal cascade} as a baseline. This approach chooses the \textit{best possible} cascade threshold by looking at \textit{all} the ground-truth pairs; i.e., it selects the threshold that prunes the most pairs while satisfying the quality target based on the ground-truth pairs. For this method we report the cost of performing the join given the cascade threshold, that is, we ignore the cost of finding the thresholds. Optimal cascade provides a lower bound on the cost of any model-cascade approach as it is allowed to find the best possible cascade threshold for free---this is not possible in practice. 
Finally, we note that we considered using LOTUS~\cite{patel2024lotus} as a baseline that performs model cascades with some statistical guarantees. However, as was observed by \cite{zeighami2025cut}, LOTUS~\cite{patel2024lotus} that uses SUPG~\cite{kang2020approximate} only provides guarantees in the limit as data size goes to infinity, and can fail to meet the target on any finite dataset. In practice, as reported by \cite{zeighami2025cut},  SUPG~\cite{kang2020approximate} may fail to meet the target---\cite{zeighami2025cut} reports that SUPG misses the target up to 75\% of the time with a failure probability of only 10\%. We observed a similar trend in our experiments. We report supporting results for BioDEX dataset in Table~\ref{tab:obs_recall}, where LOTUS fails to meet the target recall of $90\%$ \textit{on every run}, and returns a result with average recall of less than $80\%$, while both \name{} and BARGAIN meet the recall target. We exclude LOTUS from experiments since it does not provide statistical guarantees. %it is not possible to match its performance in practice since threshold selection requires knowing the correct join result in advance. 

\textbf{Metrics}. We are primarily concerned with the LLM cost of semantic joins. %For all the approaches we . The cost of using LLMs is proportional to the number of tokens used by the LLMs. 
By default and for all the datasets except Products (see below), we report the \textit{cost ratio} of the methods that normalizes the cost of a method by the cost of a naive all pairs comparison---normalization enables comparison across datasets and data sizes.  
%\sep{just make this to be the actual cost ratio? why bother with all these token stuff?}. 
Formally, for an algorithm $A$ that costs $C_A$ to join dataset $\str{L}$ and $\str{R}$, the \textit{cost ratio} is $\frac{C_A}{C}$, where $C$ is the cost of performing the join by calling an LLM on all the pairs in $\str{L}\times \str{R}$. We use OpenAI models---GPT-4.1 for feature extraction and performing join, O3 for generating candidate featurizations; and text-embedding-large as the embedding model---so the cost ratio is calculated using the corresponding monetary costs. 
%We calculate the cost based on the OpenAI cost model~\cite{openaipricing}. 
We note that all the datasets used contain ground-truth labels. To reduce costs, for every invocation of $\mathcal{L}_p$, instead of actually performing the LLM call, we simulate it by returning the known ground-truth and calculate the cost by creating the prompt---containing text records and join instruction---that would have been sent to the LLM, counting the number of tokens in the prompt and computing monetary cost based on OpenAIs cost model.
%However, performing the algorithms exactly may require millions of LLM calls---especially in the refinement stage where the LLM is called on pairs where join result was not determined with featurized decompositon (for \name{}) or semantic seimilarity (for BARGAIN). Thus, for both \name{} and baselines, we estimate the cost of this last step by randomly sampling 1,000 pairs---from the ones that would need to be performed by the LLM---and estimating the cost of the LLM invocations at the refinement step based on the cost on this sample set. 

Products dataset is a traditional entity matching dataset \cite{mudgal2018deep} that uses training and test pairs for evaluation. To maintain the original evaluation procedure \cite{mudgal2018deep}, we evaluate the join condition on the test set and use the training set to create the featurized decomposition (for \name{}) and set model cascade thresholds (for BARGAIN). For this dataset, we calculate the cost of the methods as the cost of finding the featurized decomposition or model cascade threshold on the training set plus the cost of using them to perform the join on the test. Cost ratio is defined as this cost divided by the cost of only using the LLM to perform the join on the test set. 

% each a labeled subset of $\str{L}\times \str{R}$. For this dataset, we use the training set for offline processing and report fraction of records in the test that that would be correctly pruned using each method.  

\textbf{Parameters}. For both BARGAIN and \name{}, the total number of samples is set so that we obtain 250 positive pairs for their offline procedures. For BARGAIN, the samples are used only to determine the cascade thresholds, for \name{}, they are used both for generating featurizations, creating the logical scaffold, and setting the thresholds. Featurizations and logical scaffolds are determined on a subset using 50 of the positives, and the rest of the samples are used to set the predicate thresholds. By default we set $T_R=90\%$, $T_P=100\%$ and $\delta=10\%$ in our experiments---we saw limited cost saving from using a more relaxed precision target, $T_P$, for both BARGAIN and \name{} and thus we keep $T_P=100\%$ to ensure better output quality. We use BARGAIN with $\beta=0$ which provides the same theoretical guarantees as \name{}.

\begin{figure*}
%\vspace{-0.5cm}
    \centering
    \begin{minipage}{0.49\linewidth}
    \includegraphics[width=\linewidth]{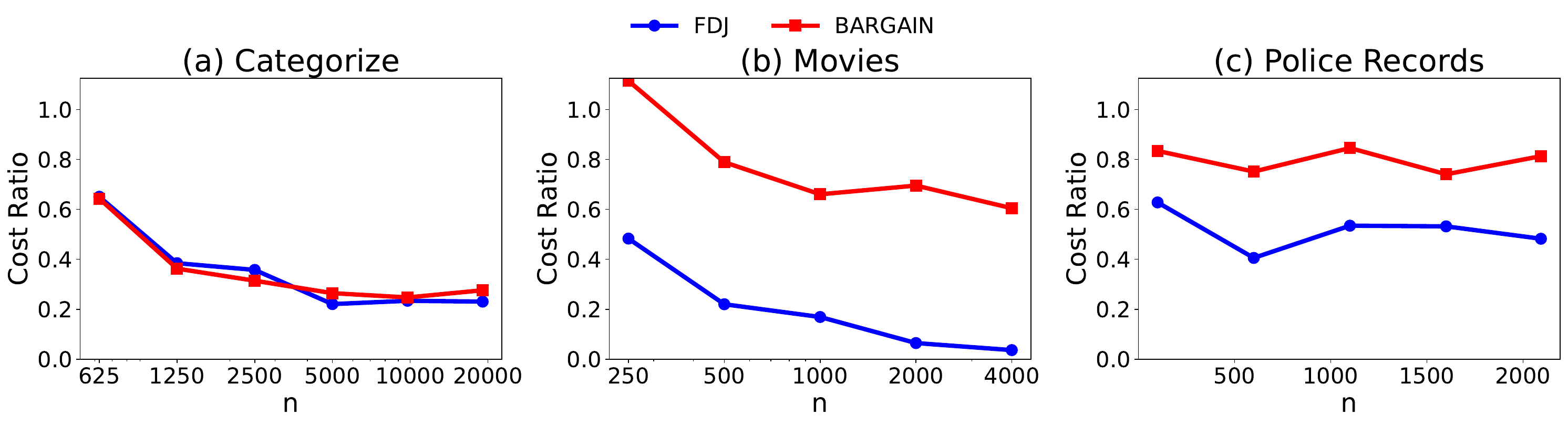}
    \caption{Impact of Data size}
    \label{fig:data_size}
    \end{minipage}
    \centering
    \begin{minipage}{0.49\linewidth}
    \includegraphics[width=\linewidth]{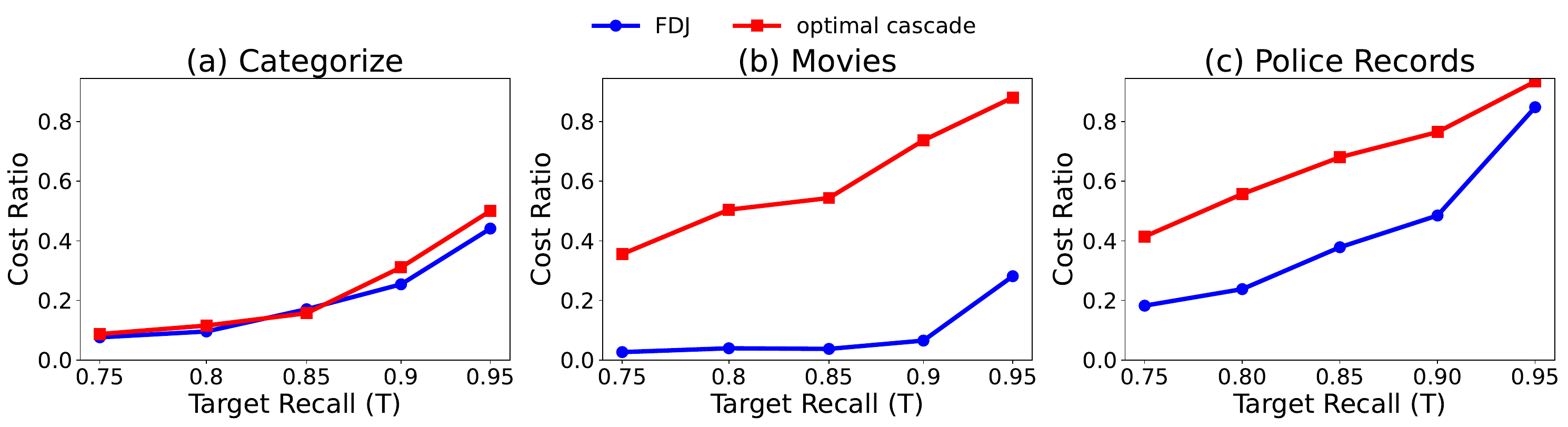}
    \caption{Impact of Target Recall}
    \label{fig:ts}
    \end{minipage}
\end{figure*}

\subsection{Baseline Comparisons}\label{sec:exp:across_ds}
Table~\ref{tab:across_ds} shows the comparison between \name{} and baselines on the real-world datasets. \textbf{\name{} improves over BARGAIN across all datasets, with gains up to 10 times over BARGAIN, and up to 8 times over the optimal cascade}.

The relative performance of \name{} compared with BARGAIN depends on dataset characteristics. In our experiments, the datasets can be divided into three categories: (1) Movies and Citations have a feature that is an accurate indicator of the join outcome; \name{} provides the highest benefits here. In Movies, a featurization  that extracts movie names and actor names from the corresponding Wikipedia pages represents the join condition accurately. However, BARGAIN and the optimal cascade perform poorly because the pages contain a lot of information beyond a single actor or movie name---we systematically analyze the failure modes of embedding-based solutions in Sec.~\ref{sec:exp:data_char}.  (2) In Police Records and Products, relevant features exist but are a weaker indicator of the join outcome. \name{} also provides significant gains in both, but less so compared to Movies and Citations. In Products, model numbers in product listings is a featurization that can be used as an indicator of the join results. However, model numbers may not exist for every record or may be noisy (e.g., one description may have complete model numbers but another only a few digits) requiring reliance on other features (e.g., color or brand name) which may also not conclusively imply a match. (3) Categorize and BioDEX are classification tasks. In such datasets the join relationship may not be easily decomposable into a set of featurizations---classification may require a complex mapping across many features. Nonetheless, \name{} still provides improvements over BARGAIN by extracting relevant information.

\subsection{Sensitivity Analysis}\label{sec:exp:across_targets}
We next study the methods' sensitivity to data size and quality target, using one dataset from each category described above.

\textbf{Impact of Data Size}. We next investigate the impact of data size. In this experiment, we consider different values, $n$, for $|\str{L}|$, the size of the table $\str{L}$. To construct our datasets, we then sample $n$ records, $\str{S}_L$, from $\str{L}$ and also select a subset, $\str{S}_R$, from $\str{R}$ that includes any record in $\str{R}$ that has a match in  $\str{S}_L$. We use $\str{S}_L$ and $\str{S}_R$ as datasets to be joined. The result of this experiment for different values of $n$ is shown in Fig.~\ref{fig:data_size}. Overall, observe similar trends as before across dataset sizes, with \name{} outperforming BARGAIN.

\textbf{Impact of Quality Target}. Fig.~\ref{fig:ts} shows the impact of recall target for different datasets, varying the target from $0.75$ to $0.95$. \name{} maintains its benefit over BARGAIN across datasets and targets. Moreover, we see that in Categorize, \name{} improves over BARGAIN at higher targets, showing that the embedding based approaches struggles more when asked to provide a comprehensive result set.

\begin{figure}
    \centering
    \includegraphics[width=\linewidth]{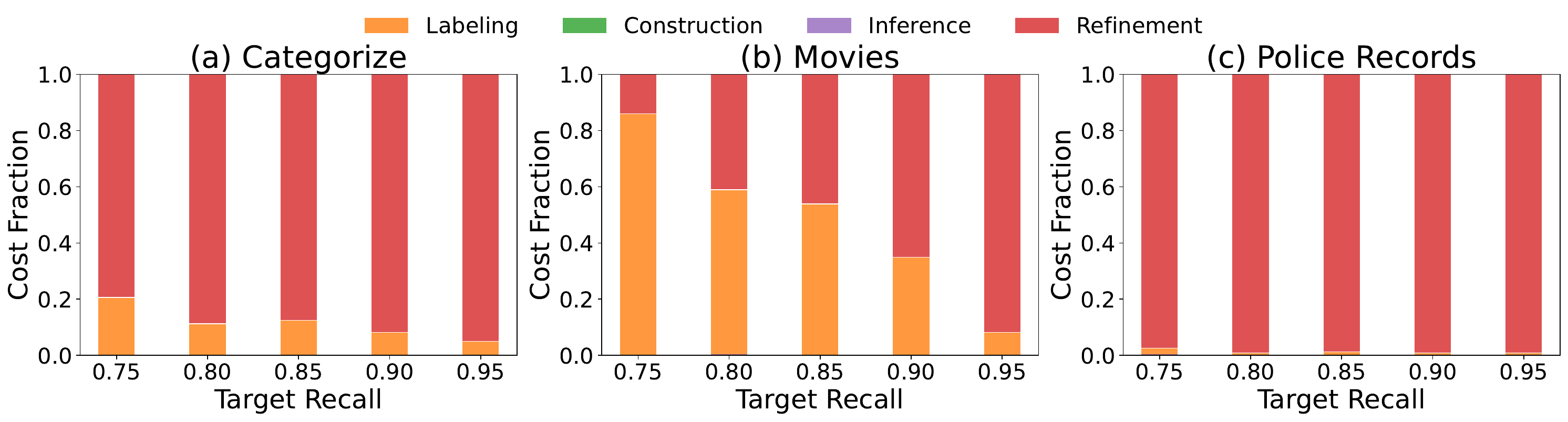}
    \caption{Cost breakdown at different targets for \name{}}
    \label{fig:breakdown}
\end{figure}

\textbf{Cost Breakdown}. We next provide a breakdown of the costs of \name{} across datasets and targets in Fig.~\ref{fig:breakdown}. In the figure, Labeling refers to the cost of obtaining labeled pairs for all of featurization generation, creating the scaffold, and setting predicate thresholds. Construction refers to the cost of constructing the featurized decomposition using the samples, excluding the cost of any feature extraction performed in the process. Inference is the total cost of performing feature extraction across all operations as well as creating embeddings (i.e.,  estimating join outcome with a featurized decomposition) and Refinement is the cost of performing LLM calls on the estimated join result $\hat{\str{Y}}$. As Fig.~\ref{fig:breakdown} shows, in general, the dominant cost is refinement. This is because the output of the featurization is often a high recall but low-precision set. For Movies specifically, the featurization has much higher precision, so less refinement is needed. In such cases, the dominant cost becomes the cost of labeling. Fig.~\ref{fig:breakdown} shows the cost of Construction and Inference are negligible---the cost of Construction is a small constant, and the cost of Inference is at most a few passes over the data.

\subsection{Impact of Data Characteristics}\label{sec:exp:data_char}
We next empirically study the impact of data characteristics on the methods. We generate synthetic data to systematically evaluate the impact of data characteristics. We first discuss the data generation process before discussing our results.

\textbf{Setup}. To generate synthetic data, we start with two tables, one containing the column \texttt{person names} and another the column \texttt{movie names} from the IMDB dataset \cite{imdbdataset}. We combine these two datasets by creating a mapping such that any person is mapped to exactly two movies and every movie is mapped to exactly two people. Thus, starting with a set of $n$ \texttt{movie names} and $n$ \texttt{person names}, we obtain a dataset,  $D$, containing $2n$ rows, each containing a value from \texttt{movie names} and \texttt{person names}. We create different datasets from $D$ to perform joins by applying templatized strings to each row to obtain new text datasets. As the base setting, we apply \texttt{``\{person-name\} likes the movie \{movie-name\}''} to every row to generate a string list $\str{L}$ containing a list of records such as \texttt{``Alex Lopez likes the movie The Shawshank Redemption''}. We perform a self-join on this list based on whether two records mention a movie liked by the same person. That is, the output of the join is pairs of sentences showing the movies liked by the same person. Across experiments, we keep the join condition the same, but modify the templatized string applied to create the sentences to inject additional information. All sentences still contain the original information that a person likes a specific movie.

\begin{figure}[t]
    \vspace{-0.5cm}
    \centering
    \includegraphics[width=0.64\linewidth]{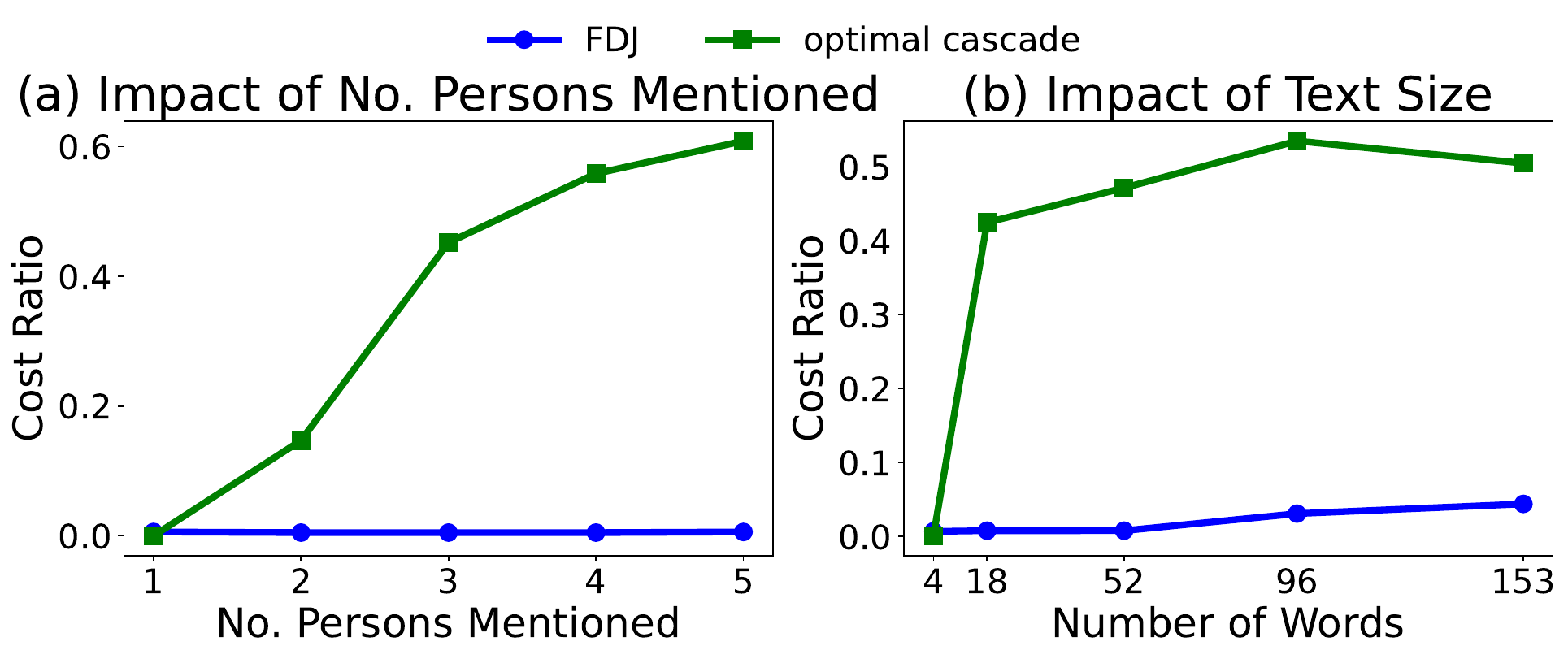}
    \caption{Impact of Data Characteristics}
    \label{fig:synthetic}
\end{figure}

\textbf{Varying number of persons mentioned}. In this experiment, we vary the number of people mentioned in each sentence. For example, if the number of people is 3, our templatized string is \texttt{``\{person-name1\}, \{person-name2\} and \{person-name3\} like the movie \{movie-name\}''}, which leads to records such as \texttt{``Alex Lopez, Adam Smith, and Ali Joe like the movie The Shawshank Redemption''}. The results of \name{} and optimal cascade to perform this task is presented in Fig.~\ref{fig:synthetic} (a). 

First, observe that this is an easy task for \name{}, as a featurized decomposition can accurate estimate the join condition by extracting names and comparing if a same name is mentioned across a pair of rows.  \name{} finds this featurization which leads to cost ratio of close to zero (almost no post-filtering is needed to be done). Furthermore, observe that when the number of attribute values is 1, the task is also simple for the optimal cascade and it achieves cost close to zero. However, as the number of attribute values increases, the cascade approach becomes significantly worse. This behavior shows that the embeddings struggle to meaningfully represent multiple attribute values and the embedding similarity becomes a poor indicator of the join outcome. As such, a featurized decomposition, using \name{} is needed to extract the relevant information.

\textbf{Varying text length}. We also experiment with how existence of other text in the data that is not directly related to the join condition impacts the performance. In this experiments, our templetized strings have the form \texttt{``\{text-1\}. For example, \{person-name\} likes the movie \{movie-name\}. \{text-2\}''} where we have additionally added two parameters \texttt{\{text-1\}} and \texttt{\{text-2\}}. These parameters are populated with text about how people choose movies but do not mention any movie or persons name. Using this templatized string, we generate our dataset as follows. (1) we create \texttt{\{text-1\}} and \texttt{\{text-2\}} strings of various lengths, ranging from one sentence to a paragraph. (2) for each length, we create two candidate values for each \texttt{\{text-1\}} and \texttt{\{text-2\}} and apply one at random. That is, to generate our final text dataset for the join, we first pick a size for the texts  (e.g., one sentence) and for every row, we pick one of \texttt{\{text-1$^1$\}} and \texttt{\{text-2$^1$\}} or \texttt{\{text-1$^2$\}} and \texttt{\{text-2$^2$\}} to apply to the row. This creates sentences that are longer but are not homogeneous. The result of this experiment is shown in Fig.~\ref{fig:synthetic}. We see that, similar to before, \name{} performs well across different lengths of the text. However, the performance of the optimal cascade drops, even when only two additional sentences (of less than 20 words) are added to the original text. The result shows that embeddings are unable to represent the part of the text relevant to the join appropriately, and thus feature extraction is needed to extract the relevant portion of the text. %This is because the embeddings start considering sentence that have the same  \texttt{\{text-1\}} and \texttt{\{text-2\}} values as similar, as opposed to considering the content of the specific feature needed for the join condition. 

\section{Related Work}\label{sec:rel_work}
We next survey the related work, discussing methods that perform AI-powered data management, semantic joins and entity resolution.  

\textbf{LLM-Powered Data Management.} LLMs have now been applied to a large array of data management problems: data discovery~\cite{freirelarge, wang2023solo}, data extraction and cleaning~\cite{arora2023language, narayan2022can, vos2022towards, naeem2024retclean, lee2025semantic}, query planning~\cite{urban2024demonstrating}, and text to SQL~\cite{pourreza2024chase}; while other work have built systems to flexibly process data with LLMs in open-ended ways~\cite{jo2024thalamusdb, shankar2024docetl, patel2024lotus, liu2024declarative, anderson2024design, urban2024eleet, lin2024towards, wang2025aop, zeighami2025llm, sun2025quest}. \name{} reduces costs while guaranteeing quality, making it applicable to any LLM-based system that supports semantic joins, including \cite{shankar2024docetl, patel2024lotus, snowflake25, alloydb, duckdb-llm}.

\textbf{Cost-Efficient AI-Powered Data Processing}. Many techniques have been used to reduce the cost of ML-powered data processing, some examples include: optimizing aggregations through approximation~\cite{jo2024thalamusdb, kang13blazeit}, building indexes to reduce online query processing time~\cite{kang2022tasti, bastani2022otif}, performing profiling to estimate model accuracies to decide which model to use \cite{jo2024smart}, and ensembling various models to generate the final output \cite{huang2025thriftllm}. {\em Model cascade}, which routes queries through cheaper proxy models before using expensive oracle models, has been widely adopted in traditional ML and deep learning, particularly for video analytics~\cite{kang2017noscope, lu2018accelerating, kang13blazeit, anderson2019physical, cao2022figo, ding2022efficient}. In particular, recent work \cite{chen2023frugalgpt,patel2024lotus, zeighami2025cut} has used this framework for LLM-powered data processing, specifically for filters. Featurized decomposition in \name{} parallels model cascades by using a cheap logical expression to approximate outputs before invoking an expensive model. We automatically construct such logical expressions and show they empirically outperforms existing cascade methods for joins.

\textbf{Semantic Joins}. Joining tables with semantically related columns \cite{he2015sema, wang2017synthesizing, dargahi2024dtt, abedjan2016dataxformer} pre-dates LLMs. Recently, LLMs have become the standard solution for joining tables based on a user-specified semantic relationship \cite{snowflake25, alloydb, databricks-llm, duckdb-llm, pg-ai,flock, shankar2024docetl, patel2024lotus}. Most work does not optimize such joins, except \cite{patel2024lotus} which uses model cascades. As discussed above and empirically shown, \name{} outperforms cascade-based solutions. A recent preprint \cite{trummer2025implementing} uses batching to reduce join costs. Batching is orthogonal to our work, and can be combined with \name{} to further reduce costs.

\textbf{Entity Resolution}. Entity resolution (ER) is a well-established problem in data management \cite{michelson2006learning, kejriwal2013unsupervised, bilenko2006adaptive, marcus2011human, wu2023blocker, thirumuruganathan2021deep, ebraheem2017deeper, zhang2020autoblock, papadakis2019survey}, which can be seen as a special case of semantic joins. Most related work on ER focuses on a relational setting and does not support
%A common technique for entity resolution is defining blocking rules \cite{michelson2006learning, kejriwal2013unsupervised, bilenko2006adaptive, marcus2011human, wu2023blocker, thirumuruganathan2021deep, ebraheem2017deeper, zhang2020autoblock}, see \cite{papadakis2019survey} for a survey. None of such work support 
arbitrary natural language joins on text data. More specifically, related work can be categorized into non-deep learning, e.g.,  \cite{kejriwal2013unsupervised, bilenko2006adaptive, kim2017scaling} and deep learning work  e.g., \cite{ wu2023blocker, thirumuruganathan2021deep, ebraheem2017deeper, zhang2020autoblock}. The former category assumes existence of a set of attributes (or features) that can be used for blocking, different from \name{} that operates on raw text with any natural language join instruction and thus must automatically discover and extract attributes. Furthermore, \cite{kejriwal2013unsupervised, bilenko2006adaptive, kim2017scaling} leverage blocking rules on pre-specified attributes to remove as many false positive matches without introducing any false negatives, unlike our setting that allows a more relaxed recall target (and therefore much lower costs) but requires statistical guarantees on meeting this user-provided target. %We operate on raw text and thus must automatically discover and extract attributes, and additionally require statistical guarantees on meeting a user-provided recall target. 
Thus, \name{} is not comparable to \cite{kejriwal2013unsupervised, bilenko2006adaptive, kim2017scaling}. The latter category trains task-specific models to learn blocking rules but once again operates on tabular data \cite{wu2023blocker, thirumuruganathan2021deep, ebraheem2017deeper, zhang2020autoblock} and designs method to encode such tabular rows, unlike \name{} that operates on raw text columns with any natural language join instruction. Furthermore, such methods require heavy infrastructure for training and hosting ML models. Instead, \name{} can be used as a wrapper with any LLM.

\section{Conclusion}\label{sec:conclusion}
We presented featurized decomposition, a new technique for optimization semantic joins by defining logical expressions using feature extraction functions. Using this technique, we developed \name{}, a new method for performing semantic joins efficiently and reliably. \name{} automatically constructs featurized decompositions that accurately estimate whether the join condition holds, leading to substantial cost savings. We furthermore performed an in-depth theoretical analysis of \name{} to provide tight theoretical bounds that guarantee meeting user-specified quality requirements. Our results across real-world datasets demonstrate up to 10 times lower cost than state-of-the-art methods. %, with benefits that scale asymptotically from as data grows.

\bibliographystyle{ACM-Reference-Format}
\bibliography{references}

\if 0
\begin{acks}
 We acknowledge support from grants DGE-2243822, IIS-2129008, IIS-1940759, and IIS-1940757 awarded by the National Science Foundation, funds from the State of California, an NDSEG Fellowship, funds from the Alfred P. Sloan Foundation, as well as EPIC lab sponsors: Adobe, Google, G-Research, Microsoft, PromptQL, Sigma Computing, and Snowflake. Compute credits were provided by Azure, Modal, NSF (via NAIRR), and OpenAI.
\end{acks}
\fi

\iftoggle{techreport}{
\appendix
\section{Overview}
This appendix is organized as follows.
\begin{itemize}
    \item Appx.~\ref{appx:adj_target} presents details of our target adjustment procedure.
    \item Appx.~\ref{sec:precision} presents details for \name{} extension to precisions
    \item Appx.~\ref{sec:thresh:disjunctions} discusses how to extend CNF construction procedure to consider disjunctions
    \item Appx.~\ref{sec:system_params} discusses cost complexity of our approach and how to set system parameters.
    \item Appx.~\ref{sec:cost_model} discusses extensions to our cost model.
    \item Appx.~\ref{sec:prompts} presents details to our LLM-Powered functions
    \item Appx.~\ref{sec:all_proof} presents proofs of our theoretical results
\end{itemize}

\section{Calculating Adjusted Target}\label{appx:adj_target}
%\sep{need a new algorithm that has estimated adjusted target}
To calculate adjusted target, we need to estimate the probability 
\begin{align}\label{eq:prob_ws_nr}
P_{T'}=\mathds{P}_{S\sim D_{r, n^+}^*}(\exists\Theta\in \mathbf{\bar{\Theta}}\;\text{s.t.}\;\mathfrak{R}_{S}(\Theta)\geq T').    
\end{align}
where $D_{r, n^+}^*$ is the worst-case dataset in Lemma~\ref{lemma:worst_case_ds} with $n^+$ ground-truth pairs and in $r$ dimensions (we add subscript $n^+$ compared with statement in Lemma~\ref{lemma:worst_case_ds} to make dependence on $n^+$ clear). We first discuss how we estimate $P_{T'}$ and then present our algorithm for selecting adjusted target using this estimate.

\subsection{Estimating failure probability}
Calculating $P_{T'}$ is difficult for two reasons. First, it is computationally expensive. It requires finding all feasible solutions to a set of inequalities and this number increases with $n^+$. Thus, instead of exact computation, we use Monte Carlo simulation to estimate this probability. Second, $n^+$ is unknown in practice. To estimate Eq.~\ref{eq:prob_ws_nr}, we estimate a range where $n^+$ belongs with high probability and bound $P_{T'}$ as the largest probability across all $n^+$ values within this range. 

\if 0
For an $r$-dimensional dataset, $D$, we need to calculate the quantity 
$$
\mathds{P}_{S^+\sim D^+}(\forall \Theta\in\mathbf{\Theta}, N_{S^+}(\Theta)<k^+\rho).
$$
where $N_S(\Theta)$ is the total number of data points in $S$ estimated positive with threshold $\Theta$. 
\fi 
\textbf{Monte Carlo Simulation.} We sample $S$ points from $D^*_{r, n^+}$, and at every iteration calculate if any threshold that has $\mathfrak{R}_{D^*_{r, n^+}}(\Theta)<T$ has $\mathfrak{R}_{S}(\Theta)\geq T'$. Let $Z_i$ be an indicator random variable denoting whether a random sample $S$  of $k^+$ points has a threshold $\Theta$ with $\mathfrak{R}_{D^*_{r, n^+}}(\Theta)<T$ has $\mathfrak{R}_{S}(\Theta)\geq T'$. Note that 
$
\mathds{E}[Z_i]=P_{T'}.
$
Thus, to estimate $P$, we randomly sample $N$ times a set $S$ and empirically observe $Z_i$ and use its empirical mean. We estimate the probability in Eq.~\ref{eq:prob_ws_nr} as $\hat{P}_{n^+, T'}=\sum_i\frac{Z_i}{N}$.
By Hoeffding's inequality~\cite{hoeffding1994probability} we have
\begin{align}\label{eq:monte_carlo_im}
\mathds{P}(P_{T'}-\hat{P}_{n^+, T'}\geq \sqrt{\frac{1}{2N}\log\frac{1}{\delta_1}})\leq \delta_1,     
\end{align}
for a parameter $\delta_1$ we set later. Note that $N$ is a parameter that allows us to get more accurate estimates by using more compute to refine our bound.

\textbf{Estimating $n^+$}. We estimate $n^+$ using the set of samples already obtained by \name{} for finding adjusted target. Specifically, recall that  $k'$ samples were taken in Line~\ref{eq:sample_2} in Alg.~\ref{sec:pipeline}, and let $k^+$ be the number of positive records among this sample. To bound the total number of positives, we again use the Hoeffding's inequality. Define $W_i$ to be an indicator random variable denoting whether the $i$-th sample is positive. Then, $n\times E[W_i]$ is the total number of positives in the dataset. Using Hoeffding's inequality we have, 
$$
\mathds{P}(|n\mathds{E}[W_i]-n\sum_i\frac{W_i}{k'}|\geq n\sqrt{\frac{1}{2k'}\log\frac{1}{\delta_2}})\leq \delta_2,
$$
for a parameter $\delta_2$ set later. The above gives a high-probability lower bound, $\hat{n}_l^+=n\sum_i\frac{W_i}{k'}-n\sqrt{\frac{1}{2k'}\log\frac{1}{\delta_2}}$ and upper bound, $\hat{n}_u^+=n\sum_i\frac{W_i}{k'}+n\sqrt{\frac{1}{2k'}\log\frac{1}{\delta_2}}$ bound on total number of positives.

\textbf{Final Estimate}. Putting the above two together, we estimate $P_{T'}$ in Eq.~\ref{eq:prob_ws_nr} as $\hat{P}_{T'}=\max_{\hat{n}\in[\hat{n}_l^+, \hat{n}_u^+]}\hat{P}_{\hat{n}, T'}+\sqrt{\frac{1}{2N}\log\frac{1}{\delta_1}}$. Note that this provides a high probability upper bound on $P_T'$, that is, $\mathds{P}(P_{T'}>\hat{P}_{T'})\leq \delta_2+(2n\sqrt{\frac{1}{2k}\log\frac{1}{\delta_2}})\delta_1$ by taking union bound across failure probabilities of the Monte-Carlo simulation as well as estimating $n^+$.

\begin{algorithm}[t]
\footnotesize
\begin{algorithmic}[1]
%\Procedure
%\State $S \leftarrow $ sample $k_n$ points uniformly at random
%\State $\hat{n}_p\leftarrow (\frac{|S^+|}{|S|}+\log \delta_1)\times n$
\State $\hat{n}_l^+, \hat{n}_u^+\leftarrow$ lower and upper bounds on $n^+$
\For{$T'$ \textbf{in} $\{T+\frac{1}{k}, T+\frac{2}{k}, ..., 1\}$}
    \State $\hat{P}_{T'}\leftarrow\max_{\hat{n}\in[\hat{n}_l^+, \hat{n}_u^+]}\hat{P}_{\hat{n}, , T'}+\sqrt{\frac{1}{2N}\log\frac{1}{\delta_1}}$
    \If{$\hat{P}_{T'}\leq \delta_3$}
        \State\textbf{return} $T'$
    \EndIf
\EndFor
\State\textbf{return} $\infty$
%\EndProcedure
\caption{{\texttt{adj-target-est}}($k$, $r$, $T$, $\delta$)}\label{alg:threshold_adjust_est}
\end{algorithmic}
\end{algorithm}

\subsection{Calculating adjusted target}
Finally we return the adjusted target as $\min\{T+\frac{1}{i}; i\in[k^+], \hat{P}_{T'}\leq \delta_3\}$ for a parameter $\delta_3$ set later. The target adjustment algorithm is shown in Alg.~\ref{alg:threshold_adjust_est}. Note that we use $\delta_3<\delta$ now because we need to also account for the probability that our estimate $\hat{P}_{T'}$ is wrong. Thus, the total probability the the algorithm meets the recall target is now $\delta_3+\delta_2+(2k^+n\sqrt{\frac{1}{2k}\log\frac{1}{\delta_2}})\delta_1$ by taking union bound across all applications of Eq.~\ref{eq:monte_carlo_im}, estimating $n^+$ as well as choosing the CNF threshold parameters using the adjusted target. 
%In practice we may also not know the data size $n$. Theoretically, we need to consider all possible values---which is possible---but in practice we observed the probaiblity is increasing in $n$ so we just use the maximum.
%Furthermore, given an unknown $n^+$, we need to estimate it from a sample. 
%Putting everything together, for each of $k$ possible adjusted target $T'$, we iteratve over all $n^+$ values from $n_l^+$ to  $n_u^+$, and use the Monte Carlo estimation to obtain a bound on probability of failure and take the maximum across all $n$ values as our bound on the total probabilty of failure. 

To ensure this probability is at most $\delta$, we set $\delta_2=\frac{\delta}{10}$ and $\delta_1=\frac{\delta}{10\times (2k^+n\sqrt{\frac{1}{2k}\log\frac{1}{\delta_2}})}$ and $\delta_3=\frac{8\delta}{10}$. Note that using these values, the total probability of failing to meet the true recall, $T$ when using adjusted recall, $T'$ when selecting thresholds is at most $\delta$, ensuring the final recall requirement is met.
To ensure low estimation error during Monte Carlo simulation, we set $N$ to the smallest value so that $\sqrt{\frac{1}{2N}\log\frac{1}{\delta_1}}\leq 0.01$.

\if 0
Also note that we obtain an estimate for $k^+$ number of different values in Line~\ref{line:iter_over_t} of Alg.~\ref{line:iter_over_t}, so taking an additional union bound over use of  $\hat{P}_r$, the total probability of obtaining a bound with 
Next, we analyze the probability that the bound $\hat{P}_{r}$

, let $$c=n_u^+-n_l^+=2\sqrt{\frac{1}{2k}\log\frac{1}{\delta_2}}$$.  Note that the algorithm returns a valid bound with probability $1-(\delta_1+ck\delta_2)$. Thus, in total the probaility of failure becomes $(\delta_1+ck\delta_2+\delta_3)$. Thus, we need to set $\delta_1, \delta_2, \delta_3$ and $N$, the number of runs in Monte Carlo simulation. 

$\delta_1=(0.001)/ck^+$, $\delta_2=0.001$, $\delta_3=0.088$, $\epsilon=0.01$. We need to set $N$. Let $\sqrt{\frac{1}{2N}\log\frac{1ck^+}{(0.01)}}=\epsilon$. We get $\frac{1}{2\epsilon^2}\log\frac{ck^+}{(0.01)}=N$

%N*c*k^+=\frac{1}{2\epsilon^2}\log\frac{ck^+}{(0.001)}*c*k^+

%N, n, k^+
%N*n^u*

%sqrt(n)*(n^+/k^+)*log(1/\delta)

\sep{needs work adding discussion with $n$}\fi

\section{Setting Thresholds for Precision}\label{sec:precision}
Recall that we have access to featurizations $\phi_1$, ..., $\phi_r$ for some integer $r$. Our goal is to create a result set with precision at least $T_P$ given a high-recall set $\hat{\str{Y}}$ with recall $T_R$. To do so, we consider the featurizations in order and iteratively use each featurization to remove false positives from a subset of $\hat{\str{Y}}$. To do so, denote by $\str{Y}'_i$ the set of pairs labeled up until the $i$-th iteration with $\str{Y}'_0=\emptyset$. At the $i$-th iteration, consider the set $\bar{\str{Y}}=\hat{\str{Y}}\setminus\str{Y}'_i$. We run BARGAIN \cite{zeighami2025cut} with precision target $T_P$ and probability of failure $\delta_1$---for a parameter $\delta_1$ set later---on this set, which outputs a set $\bar{\str{Y}}'\subset \bar{\str{Y}}$ which has precision $T_P$ with probability at least $\delta_1$. Thus, we have $\str{Y}'_{i+1}=\str{Y}'_{i}\cup \bar{\str{Y}}'$. We repeat this process using all featurization, after which, if $\str{Y}'_{r+1}$ is non-empty, we use the oracle only to remove any false positives. 

Note that the set of pairs estimated with BARGAIN has precision $T_P$ with probability $1-\delta_1$. Furthermore, the set of points across multiple call of BARGAIN are mutually exclusive, so that $\str{Y}'_{r+1}$ also has precision at least $T_P$, and with probability of failure $r\times \delta_1$ using union bound. Setting $\delta_1=\frac{\delta}{2\times r}$ ensure failure to meet precision target is bounded by $\frac{\delta}{2}$. Finally, we run our algorithm for recall also with failure probability $\frac{\delta}{2}$ so that the total probability of failing to meet precision or recall target is at most $\delta$.
\section{Extensions to Disjunctions}\label{sec:thresh:disjunctions}
Here, we discuss how to extend our method to consider cases when there may be disjunctions in the parameterized decomposition, $\mathring{\Pi}$. In such cases, $\mathring{\Pi}$ contains $r$ clauses combined with conjunction, and the $i$-th clause contains $r'_i$ predicates, combined with disjunctions. Thus, we need to set the threshold parameters $\Theta^1, ..., \Theta^r$, where $\Theta^i\in \mathbb{R}^{r_i'}$ denotes the threshold parameters for the $i$-th clause. To set these parameters, rather than considering all possible combinations for all parameters, we simplify our parameter selection method. We only consider parameter settings where within any clause, all the thresholds are the same (thresholds across clauses may still be different). That is $\Theta^i_j=\Theta^i_{j'}$ for all $j,j'\in[r_i']$, where $\Theta^i_{j}$ is the threshold for the $i$-th predicate in the $j$-th clause. We normalize distance values across featurizations to ensure distances from different distance functions are comparable.
As we discuss later, this simplification allows us to extend Thm~\ref{thm:alg_guarantee} to guarantee recall in this setting as well. Following this approach, let $\mathbf{\Theta}$ be the set of all possible threshold parameters where all thresholds within a clause are equal. We select thresholds for our decomposition as
\begin{align}\label{eq:opt_thresh_disj}
\Theta^*\in\arg\min_{\Theta\in\mathbf{\Theta}} \mathfrak{F}_S(\mathring{\Pi}, \Theta)\;\text{s.t.}\;\mathfrak{R}_S(\mathring{\Pi}, \Theta)\geq T'.    
\end{align}
Eq.~\ref{eq:opt_thresh_disj} is the same as Eq.~\ref{eq:opt_thresh} except for the search space of the thresholds. The following lemma shows that the same adjusted target as in Thm.~\ref{thm:alg_guarantee} can stil be used to provide statistical guarantees.
\begin{lemma}\label{lemma:adj_target_disj}
    Let $\mathring{\Pi}$ be a logical scaffold in CNF form with $r$ clauses; $\mathbf{\Theta}$ be the set of possible threshold parameters for $\mathring{\Pi}$ where the parameters within the same clause are equal; and $T'=\texttt{adj-target}(k^+, r, T, \delta)$ for the same function in Theorem~\ref{thm:alg_guarantee}.  For any threshold $\Theta^*\in \mathbf{\Theta}$ such that $\mathfrak{R}_S(\mathring{\Pi}, \Theta^*)\geq T'$ we have $\mathds{P}_{\str{S}\sim \str{L}\times\str{R}}(\mathfrak{R}_{\str{L}\times\str{R}}(\mathring{\Pi},\Theta^*)<T)\leq \delta$, under the same conditions as Theorem~\ref{thm:alg_guarantee}. 
\end{lemma}
Thus, selecting thresholds according to Eq.~\ref{eq:opt_thresh_disj} and using the adjusted target $T'$ from Lemma~\ref{lemma:adj_target_disj} yields a featurized decomposition that meets the recall requirement. Note that applying a common threshold within each clause allows us to extend Theorem~\ref{thm:alg_guarantee} to disjunctions. %However, extensions to handle arbitrary thresholds across disjunctions is non-trivial as we further discuss in Appx~\ref{sec:proof:conjunction}.
\section{Cost Analysis and System Parameters}\label{sec:system_params}
\textbf{Cost Complexity}. We analyze the cost of \name{} in terms of the number of tokens passed by \name{} to the LLM.
\begin{proposition}[Cost Complexity]\label{prop:cost_complexity}
The number of tokens passed by \name{} to the LLM is at most $$\varkappa t(k+k' +|\hat{\str{Y}}|+|\Phi|(|\str{L}|+|\str{R}|)),$$
where $t$ is the maximum token length across records in $\str{L}$ and $\str{R}$ and $\varkappa$ is a constant dependent on number of tokens in system prompts and independent of data size.
\end{proposition}
As Prop.~\ref{prop:cost_complexity} shows, there are three main factors impacting the cost of \name{}: number of samples $k+k'$, size of candidate join result created by the featurized decomposition, and the number of featurizations. $k$ and $k'$ are systems parameters; we discuss how to set them below. We note that $|\Phi|$ is the number of candidate featurizations, and $|\hat{\str{Y}}|$ depends on their quality. Both are dependent on the data and LLM capabilities; both are typically small. For example, if for a dataset a single feature determines the join result, then $|\Phi|$ can be 1 and, furthermore if the LLM can reliably extract the feature, then $|\hat{\str{Y}}|$ can be $\mathcal{O}(n^+)$, where $n^+$ is the total number of true positives in the dataset. Thus, in settings where $n^+=\mathcal{O}(|\str{L}|+|\str{R}|)$, (i.e., the size of join result is linear in data size) and a constant number of featurizations that can be reliably extracted by LLMs are sufficient to perform the join, the join cost can be reduced to $\mathcal{O}(|\str{L}|+|\str{R}|)$.

\textbf{System Parameters}. To generate a comprehensive set of candidate featurizations, we need to provide the LLM in Alg.~\ref{alg:generator_function} with sufficient positive examples. These positive examples show the LLM different logical reasons the join condition can hold across the dataset; which are represented as featurizations in \name{}. The number of such positive examples needed depend both data characteristics and LLM capabilities. Data characteristics determine how many different useful featurizations exist and whether observing more positive examples can lead to discovering more featurizations, while the LLM needs to be able to create new featurizations when given more positive examples. In practice, across real-world datasets, we observed that LLMs will stop creating new featurizations after observing at most 50 positive samples across all datasets. %Thus, we set $k'$ so that we observe 20 positive samples. 

Moreover, we also set  $k$ based on number of positives observed because the function $\texttt{get-adjusted-target}$ only depends on the number of observed positives. In practice, with 200 samples, the adjusted target is sufficient to obtain meaningful statistical estimation independent of the dataset. We note that, for both $k$ and $k'$ we need to sample enough pairs until we observe a given total number of positives. Let $k^+$ be the total number of positive samples needed. Then, the expected sample size is $k^+\frac{|\str{Y}\times\str{R}|}{n^+}$ which depends on the true positive rate in the dataset. Note that we set $k^+$ to be a constant, and in many realistic settings $n^+=\mathcal{O}(|\str{L}|+|\str{R}|)$, i.e., every record in $\str{L}$ only matches a constant number of records in $\str{R}$. Thus, the total number of samples needed are linear in data size.%, so that the size of the join result is linear in data size

Finally, note that we use $\gamma$ as another system parameter, which we set to a fixed small value ($0.05$ in our experiments). $\gamma$ is used to avoid having too many featurizations each with marginal impact. In cost expression in Prop.~\ref{prop:cost_complexity}, it helps avoid unnecessarily increasing $|\phi|$. 

\begin{algorithm}[t]
\small
\begin{algorithmic}[1]
\Procedure{\texttt{select-thresholds}}{$\mathbf{\Phi}$}
\State $T'\leftarrow$\texttt{adj-target}($k$, $T$, $\delta$)
\If{$r < c$}
    \State $\textbf{return} \arg\min_{\Theta} \hat{C}(\Theta)\;\text{s.t.} \mathfrak{R}_S(\Theta)>T'$\Comment{exhaustive search}
\EndIf
\State $\Theta\leftarrow (\infty, ..., \infty)$
\State $T_c=0$
\While {$T_c<T'$} \Comment{greedy search}
    \State $w^* \leftarrow 0$
    \For{$i\leftarrow0$ to $r$} \Comment{Find best one-dimensional change to $\Theta$}
        \For{$\theta<\theta^i$}
        \State $\Theta'\leftarrow$ $\Theta$ with $i$-th dimension replaced with $\theta$
        \If{$\frac{\mathfrak{R}(\Theta')-R(\Theta)}{C(\Theta')-C(\Theta)}>w^*$}
            \State $\Theta^*, w^*\leftarrow \Theta', \frac{\mathfrak{R}(\Theta')-R(\Theta)}{C(\Theta')-C(\Theta)}$
        \EndIf
        \EndFor
    \EndFor
    \State $\Theta\leftarrow \Theta^*$ \Comment{Update $\Theta$ to increase recall}
\EndWhile
\State\textbf{return} $\Theta$
\EndProcedure
\caption{Selecting Thresholds}\label{alg:threshold_select}
\end{algorithmic}
\end{algorithm}

\section{Extending cost model}\label{sec:cost_model}
Here we discuss a more accurate cost model for estimating the cost of using a threshold. As discussed earlier in Sec.~\ref{sec:param_decomp}, we observed  in our experiments that the number of false positives is sufficient to obtain reliable estimates of cost. Here, for completeness, we provide a more comprehensive discussion of the cost. For any set of thresholds $\Theta$, its cost is composed of (1) total number of tokens used to extract all the features and (2) total number of tokens to process the join result after the features are created. %Note that for any threshold, the cost of the join is the cost of (1) feature extraction and (2) refinement after performing the join. 
Regarding (1) for the $i$-th featurizaion $(d^i, \mathcal{X}_L^i, \mathcal{X}_R^i)$, if $\theta^i<\infty$, then we need to apply the extraction functions to all the records in $\str{L}$ and $\str{R}$, so that the total cost of extraction is proportional to the number of tokens in $\str{L}$ and $\str{R}$. Let $C_E$ be the total number of tokens in $\str{L}$ and $\str{R}$. If a featurized decomposotion with thresholds $\Theta$ contains $u_\Theta$ number of predicates that need to be extracted with an LLM, then the cost of extraction is estimated as $u_\Theta\times C_E$. To estimate (2), note that the cost of refinement depends on the total number of tokens corresponding to the pairs considered positive after applying the featurized decomposition. We use our samples to estimate this cost. Specifically, recall that we sampled a set $\str{S}$ of $k$ pairs and let $\hat{\str{Y}}^S$ be the set of pairs estimated to be positive after applying the featurized decomposition with a threshold $\Theta$ on $\str{S}$. Let $C^S_\Theta$ be the total number of tokens needed to process the pairs in $\hat{\str{Y}}^S$, i.e., total number of input tokens to perform with $\mathcal{L}_\str{p}(\str{l}, \str{r})$ for all $(\str{l}, \str{r})\in \str{S}$. Using this, we estimate the total number of tokens for refinement after using a featurized decomposition as $\frac{C_\Theta^S}{k}\times |\str{R}\times \str{L}|$. Finally, the total cost estimate is calculated as $\hat{C}(\Theta)=\frac{C_\Theta^S}{k}\times |\str{R}\times \str{L}|+u_\Theta\times C_E$.

\section{Calculating Minimum Cost Thresholds}\label{sec:param_decomposition_details}
Here, we discuss how to find the minimum cost threshold, i.e., to solve the optimization problem in Eq.~\ref{eq:opt_thresh}. Recall that our goal is to find thresholds with lowest false positive rate while meeting recall $T$ for $r$ featurized predicates combined together with a conjunction. 
Define $S\in\mathbb{R}^{k\times r}$ for $k=|\str{S}|$ to be the dataset of feature distances between every pair of records and for all features. Specifically, the $j$-th column in the $i$-th row of $S$ is $S_{i, j}=\phi_j(\str{l}_i, \str{r}_i)$ for $i\in[k]$, $j\in[r]$, where $(\str{l}_i, \str{r}_i)$  is the $i$-th row of $\str{L}\times\str{R}$. Furthermore, let $S^+$ be the subset of $S$ with only the positive records. Note that we only need to consider thresholds in $S_{1, :}^+\times ...\times S_{r, :}^+$. We sort each of $S_{r, :}^+$ in descending order, and iteratively reduce threshold values and calculate false positive rates and recall for each threshold. Note that when for a threshold set, $\Theta$ the calculated recall is below $T$, then the recall for any threshold set, $\Theta'$ with all thresholds less than $\Theta$ is also below $T$. Thus, we prune away such threshold settings from computation to reduces computation complexity. 

%Using the above cost function, our algorithm is presented in Alg.~\ref{alg:threshold_select}. We first obtain the adjusted recall $T'$. If the number of candidate featurized predicates is below a user specified threshold $c$ (where $c$ represents the latency requirement from the user), then we perform an exhaustive search to find a $\Theta$ with minimum cost. On the other hand, if the dimensionality is more than $d$, we use a greedy heuristic to find the threshold. Our greedy heuristic is very similar to that of the greedy algorithm for weighted set cover. At every iteration, we choose we modify one of dimension of our threshold set such that the increase in observed recall divided by increase in cost is minium. Note that this algorithm covers at least one new element per iteration, and thus terminates after iterations equal to the number of observed positives (which is less than the sample size). 
\section{Proofs}\label{sec:all_proof}
\subsection{Proof of Theorem~\ref{thm:nphard}}\label{sec:np-hardness}
First, recall that the cost of a featurized decomposition is the total cost of using LLMs to perform the join using the featurized decomposition. We assume the following LLM  behavior for the following two prompts. For any integer $row\_id$, the prompt $\str{p}_1=$``\texttt{row\_id \{row\_val\}, is 1=1?}'', we assume $\mathcal{L}$ returns True, for $\str{p}_2=$``\texttt{row\_id \{row\_val\}, is 2=1?}'', we assume $\mathcal{L}$, returns False. We furthermore assume the monetary cost of calling the LLM with any of the two prompts above is $c$. We have verified the LLM output for the LLM used in our experiments. We note that the following proof can be easily generalized to any LLM for which there exists two prompts that both cost equal and the LLM output is True for one and False for the other. We use the above two prompts for concreteness. 

We show a reduction from set cover to MCFD. We first state the set cover problem formally.

\begin{definition}[Set Cover]
    Let $U$ be a set of elements and $\mathcal{S}$ a collection of sets, where each set $S\in \mathcal{S}$ is a subset of $U$. Does there exist a set of at most $k$ sets $S^*$, $S^*\subseteq \mathcal{S}$  such that $\cup_{S\in S^*}S=U$?
\end{definition}

Given an instance of the set cover problem, we show how to construct an instance of MCFD in polynomial time. Note that an MCFD instance is defined by sets $\str{L}$ and $\str{R}$, a set of possible featurizations $\mathbb{F}$ and a join prompt \str{p}. Each featurization in $(d, \mathcal{X}_L, \mathcal{X}_R)\in\mathbb{F}$ is a triplet and there is a cost associated with each application of $\mathcal{X}_L$ and $\mathcal{X}_R$. There is furthermore a cost associated with each application of $\mathcal{L}$. Our reduction specifies the sets $\str{L}$, $\str{R}$, $\mathbb{F}$ and the prompt \str{p}.%, the costs associated with each invocation of the featurization as well as the function $\mathcal{L}$ and the cost associated with it's invocation.

\textit{Reduction}. For the purpose of the reduction, let $\str{L}_1$ be the set of strings $\{\texttt{``row\_id \{row\_val\}, is 1''}; row\_val\in[|U|]\}$, %numbers in $[|U|]$, where the $i$-th element in $U$ is associated with the $i$-th element in $\str{L}$ 
and let $\str{R}=\{\texttt{``1''}\}$. 
Let $\str{L}_2$ be  $\{\texttt{``row\_id \{|U|+1\}\}, is 2''}; i\in|\mathcal{S}|+1\}$, that is, a set of $|\mathcal{S}|+1$ strings. 
Define $\str{p}=\texttt{``\{l\}=\{r\}?''}$. We pad all string representations of $row\_val$ so that all rows in $\str{l}$ contain the same number of tokens.
For each set $S\in\mathcal{S}$, construct a featurization $\phi=(d, \mathcal{X}_L,\mathcal{X}_R)$, where $\mathcal{X}_L(\str{l})$ is a function that extracts the value of $row\_id$ from $\str{l}\in\str{L}$ using a regular expression and returns $0$ if and only if the $row\_id$-th element in $U$ is in the set $S$ and 1 otherwise. Also define $\mathcal{X}_R(\str{r})$ that a function that calls $\mathcal{L}$ using any of the prompts but disregards the output and always returns 1, i.e., a function identically equal to 1 that costs $c$ tokens. Finally, define $d$ as a function that returns the output of $\mathcal{X}_L$. Intuitively, a featurization returns 0 if it is applied to a pair covered by its corresponding set cover, and 1 otherwise. Let the set of possible featurizations be the collection of all such featurizations, one for each set in $\mathcal{S}$. 
%Furthermore, define $\mathcal{X}_L(\str{l})$ to cost 0 and $\mathcal{X}_R(\str{r})$ to cost 1 for all such featurizations. Finally, assume the function $\mathcal{L}(\str{l}, \str{r})=1$ for all $\str{r}\in\str{R}$ and $\str{l}\in\str{L}_1$ while $\mathcal{L}(\str{l}, \str{r})=0$ for $l\in\str{L}_2$ and the cost of every invocation of $\mathcal{L}$ is 1 irrespective of $\str{l}$ and $\str{r}$. 
Finally, to answer the Set Cover instance, we return True if the optimal solution has cost at most $ck+c|U|$ and False otherwise. 

Next, we show that there exists a set cover of size at most $k$ if and only if there exists an FD that costs at most $c|U|+ck$. To show this, first note that if there exists a set cover of size at most $k$, then there exists an FD that costs at most $c|U|+ck$. This is because the featurized decomposition that uses the corresponding featurizations for the at most $k$ predicates (with threshold 0.5) achieves cost at most $c|U|+ck$. This is because using each featurization costs $c$ and the decomposition returns a set of $|U|$ elements with no false positives. Thus, the refinement step costs $c|U|$ to call the LLM for every output of the decomposition. Conversely, assume there exists a featurized decomposition that costs at most $c|U|+ck$. Note that if $k\geq|\mathcal{S}|$, than there trivially exists a set cover of size at most $k$. Thus, consider the case that $k<|\mathcal{S}|$. 
Then, we note that such a featurized decomposition must have at least one clause whose predicates all have threshold $< 1$, because otherwise the cost would be more than $c|U|+ck$. Then, because for the clause with thresholds $<1$, none of the featurized  predicates admit any negatives and every disjunctive clause must have 100\% recall, then taking only this single clause will not increase the cost and remains a valid solution. Note that if the cost of such a disjunctive clause is at most $c|U|+ck$, such a clause contains at most $k$ featurization, implying that there are at most $k$ featurizations which cover all the positives. Taking the sets corresponding to those featurizations yields a set cover of size at most $k$, thus completing the proof. 

Note that constructing the sets $\str{L}$ requires a pass over all the data points in $U$ plus an additional $O(|\mathcal{S}|)$. Furthermore, constructing the featurizations requires a pass over all the sets in $S$, so
the reduction takes polynomial time in the input size, thus completing the proof. %\sep{representing integers as string is constant time}

\subsection{Proof of Lemma~\ref{lemma:worst_case_ds}}
%\subsubsection{Proof of Worst-Case Dataset\\\nopunct}\label{sec:proof}
First, we state the lemma more generally. Let $u=\lceil n^+(1-T)\rceil-1$, and let $D_0$ be a dataset consisting of $n^+-ur$ number of $r$-dimensional vectors with all dimensions equal to zero.
\begin{lemma}\label{lemma:worst_case_ds_gen}
    For any $r\leq \frac{n^+}{u}$ and any $\rho>T$, define $D^i=\{x\times e_i;x\in[u]\}$ and $D_r^*=(\cup_{i\in[r]} D^i)\cup D_0$. We have 
    \begin{align}
        D^*_r\in \arg\max_{\hat{D}\in\mathbb{R}^{n^+\times d}}\mathds{P}_{\hat{S}\sim \hat{D}}(\exists\Theta\in \mathbf{\bar{\Theta}}_{\hat{D}}\;\text{s.t.}\;\mathfrak{R}_{\hat{S}}(\Theta)\geq \rho),
    \end{align}
    where $e_i\in\mathbb{R}^r$ is a vector with $i$-th element 1 and other elements 0.
\end{lemma}
%Recall that our goal is to show when sampling $k$ records, $S$, 
%$$D^*_r\in \arg\max_{{D}\in\mathbb{R}^{n^+\times d}}\mathds{P}_{{S}\sim {D}}(\exists\Theta\in \mathbf{\bar{\Theta}}_{{D}}\;\text{s.t.}\;\mathfrak{R}_{{S}}(\Theta)\geq \rho),$$
\textit{Proof}. We next prove the above lemma. We use $D^*$ instead of $D^*_r$ to simplify our notation here. 
 
First, we show that it is sufficient to consider \textit{rank normalized} datasets, which we define next to reduce the sample of possible the datasets we need to consider. A rank normalized dataset is defined as follows
\begin{definition}
A dataset, $D$ is rank normalized if it satisfies the following properties:
\begin{enumerate}
    \item $D_{i, j}\in \{0\}\cup[u]$ for any $i\in [n^+]$, $j\in[r]$.
    \item $D_{i, j}$ for any dimension $j$ and any two rows $i$ and $i'$, $i\neq i'$ and whenever $D_{i, j}\neq 0$, then $D_{i, j}\neq D_{i', j}$. That is, two different rows do not have the same value for the same dimension, unless that value is zero.  
    \item For every dimension $j$ and every value $v\in[u]$, there exists a point $D_{i, j}=v$ for some $i\in[n^+]$.
\end{enumerate}
\end{definition}
%, where a dataset, $D$, is set to be rank normalized if (1) $D_{i, j}\in[n]$ for all $i\in [n]$, $r\in[r]$ and (2) for any dimensions $j$ and any two rows $i$ and $i'$, $D_{i, j}\neq D_{i', j}$, that is, no two points share the same value in the same dimension. 
Let $\mathbf{D}$ be the set of all possible rank normalized datasets, we have:

\begin{lemma}\label{lemma:rank_norm}
    The maximum probability of failure is achieved by a rank normalized dataset. That is, \begin{align*}
        \max_{{D}\in\mathbb{R}^{n^+\times d}}\mathds{P}_{{S}\sim {D}}(\exists\Theta\in \mathbf{\bar{\Theta}}_{{D}}\;\text{s.t.}\;\mathfrak{R}_{{S}}(\Theta)\geq \rho)=\\\max_{{D}\in \mathbf{D}}\mathds{P}_{{S}\sim {D}}(\exists\Theta\in \mathbf{\bar{\Theta}}_{{D}}\;\text{s.t.}\;\mathfrak{R}_{{S}}(\Theta)\geq \rho).
    \end{align*}
\end{lemma}

%\sep{$D^*$ needs to round $u$, which means some dimensions will have fewer points than other dimensions}
Using this lemma, we now only need to show that 
$$D^*\in \arg\max_{{D}\in\mathbf{D}}\mathds{P}_{{S}\sim {D}}(\exists\Theta\in \mathbf{\bar{\Theta}}_{{D}}\;\text{s.t.}\;\mathfrak{R}_{{S}}(\Theta)\geq \rho).$$

We refer to $\mathds{P}_{{S}\sim {D}}(\exists\Theta\in \mathbf{\bar{\Theta}}_{{D}}\;\text{s.t.}\;\mathfrak{R}_{{S}}(\Theta)\geq \rho)$ as probability of success for a dataset ${D}$ and probability of failure for the dataset is (1-probability of success) for that dataset. 

\begin{algorithm}[t]
\small
\begin{algorithmic}[1]
%\Ensure A set of candidate factorizations
%\Procedure{\texttt{get-candidate-featuriztions}}{$\str{p}$, $\str{L}$, $\str{R}$}
\State $\bar{D}\leftarrow D^*$
\State Order $D$ by values in dimensions 1, ..., $r$
\State Order $\bar{D}$ by values in dimension 1
\For {$j\leftarrow 2$ \textbf{to} $r$} 
    \State $I=\{i; i\in[n], D_{i, j}\geq 1, D_{i, :j-1}\neq\mathbf{0}\}$
    \For{$i\in I$ in decreasing order of $D_{i, j}$}
        \While{$D_{i, j} > \bar{D}_{i, j}$}
            %\State $v\leftarrow \bar{D}_{i, j}$
            \State $i'\leftarrow$ row in $\bar{D}$ s.t., $\bar{D}_{i', j}=\bar{D}_{i, j}+1$\label{alg:const:swap:begin}
            \State \textbf{assert} $\bar{D}_{i', :}=(\bar{D}_{i, j}+1)\times e_j$\label{alg:construct:asser:one_hot_swap}
            \State $\bar{D}_{i, j}\leftarrow \bar{D}_{i, j}+1$
            \State $\bar{D}_{i', j}\leftarrow \bar{D}_{i', j}-1$\label{alg:const:swap:end}
        \EndWhile
    \EndFor
    \State Order $\bar{D}$ by values in dimension 1, ..., $j$
    \State \textbf{assert} $D_{:, :j}=\bar{D}_{:, :j}$\label{alg:const:assert}
\EndFor
%\EndProcedure
\caption{Procedure to obtain $D$ from $D^*$}\label{alg:construction_algorithm}
\end{algorithmic}
\end{algorithm}

We need to show $D^*$ has lower probability of success than any rank normalized dataset. To do so, let $D$ be any rank normalized dataset not equal to $D^*$. We iteratively modify $D^*$ to eventually obtain $D$, while showing probability of success is non-decreasing for every modification. 

We show how we iteratively modify $D^*$ to obtain $D$ in Alg.~\ref{alg:construction_algorithm}. We use $\bar{D}$ to denote the running dataset after iterative modifications, where initially, $\bar{D}=D^*$. We iteratively consider dimensions $j$, $j\geq 2$, for each dimension consider the set of records $D$ that have non-zero $j$-th dimension and not all their previous dimensions are zero. For each of those records considered in the decreasing order of their $j$-th dimension value, we iteratively modify $\bar{D}$ to obtain the same value for the corresponding records in $\bar{D}$. After the modifications we reorder $\bar{D}$ based on the values in its first $j$ dimension (first sort by dimension 1 and break ties by dimension 2, ..., $j$).  As the algorithm shows, at the end of every iteration, we have $\bar{D}_{:, :j}=D_{:, :j}$ as our loop invariant, so that at the end of the algorithm, we have $\bar{D}=D$. The following lemma shows that the loop invariant holds in the algorithm.

\begin{lemma}\label{lemma:correct_construction}
    For any rank normalized dataset $D$ in $r$ dimensions, the assertions in lines~\ref{alg:const:assert} and \ref{alg:construct:asser:one_hot_swap} of Alg.~\ref{alg:construction_algorithm} hold.
\end{lemma}

Note that modifications to $\bar{D}$ are only made in Lines~\ref{alg:const:swap:begin}-\ref{alg:const:swap:end}, where each modification decreases the value of some record in $\bar{D}$ by one and increases the value of another one. We call this modification an incremental swap. Thus, if we show every incremental swap creates a dataset with increased probability of success, then by the above construction, we show that our proposed dataset $D^*$ has the lowest probability of success; which is the subject of the following lemma.

\begin{lemma}\label{lemma:incremental_swap}
    Let $\bar{D}$ be an $r$-dimensional rank normalized dataset with $n$ points such that its $i$-th row and $j$-th column is $\bar{D}_{i, j}=v-1$ for some integer $v\geq 1$ and its $i'$-th row is $D_{i', :}=v\times e_j$. Now let $\bar{D}'$ be the same as $\bar{D}$, except that $\bar{D}'_{i', j}=\bar{D}_{i', j}-1=v-1$ and that $\bar{D}'_{i, j}=\bar{D}_{i, j}+1=v$. $\bar{D}'$ has a lower probability of failure than $\bar{D}$. That is, 
    $$\mathds{P}_{S\sim \bar{D}}(\exists\Theta\in \mathbf{\bar{\Theta}}_{\bar{D}}\;\text{s.t.}\;\mathfrak{R}_{S}(\Theta)\geq \rho)\geq \mathds{P}_{S\sim \bar{D}'}(\exists\Theta\in \mathbf{\bar{\Theta}}_{\bar{D}'}\;\text{s.t.}\;\mathfrak{R}_{S}(\Theta)\geq \rho)$$
\end{lemma}

This lemma completes the proof. \qed

\textbf{Relaxing condition on $r$.} We note that the condition on $r$ in the statement of the lemma is because our construction places $u$ points along every dimension, so that total data size must be more than $u\times r$ points. To extend the result to higher dimensions, we note that our swapping argument holds in general, which would imply that all the records even in when $n^+<u\times r$ must have values in $[u]\times e_i$ for $i\in[r]$ (i.e., the set $\{i\times e_i; i\in [u]\}$). However, in higher dimensions, there is more flexibility in how a dataset can be constructed using only points from $\cup_{i\in r}[u]\times e_i$. Thus, the argument needs to additional study, among datasets with records in $\cup_{i\in r}[u]\times e_i$, which one has the lowest probability of success.

\textbf{Notation for proof of lemmas.} We use the following notation in our proof. Define $P_D(\Theta)=\{p_i; p_i\in D, \mathds{I}[p_i^1\leq \Theta^1, ..., p_i^d\leq \Theta^r]\}$, where $\Theta^i$ is the $i$-th dimension of $\Theta$ and let $N_D(\Theta)=|P_D(\Theta)|$, the total number of positives in $D$ that will be returned if $\Theta$ is used. Note that
$$
\mathds{P}_{{S}\sim {D}}(\exists\Theta\in \mathbf{\bar{\Theta}}_{{D}},\;\mathfrak{R}_{{S}}(\Theta)\geq \rho)=\mathds{P}_{{S}\sim {D}}(\exists\Theta\in \mathbf{\bar{\Theta}}_{{D}},\;|P_{D}(\Theta)\cap S|\geq k\rho).
$$
We often use the right hand side in our proofs.

\subsubsection{Proof of Lemma~\ref{lemma:rank_norm}\\}
Note that trivially, 
\begin{align*}
        \max_{{D}\in\mathbb{R}^{n^+\times d}}\mathds{P}_{{S}\sim {D}}(\exists\Theta\in \mathbf{\bar{\Theta}}_{{D}}\;\text{s.t.}\;\mathfrak{R}_{{S}}(\Theta)\geq \rho)=\\\max_{{D}\in \mathbf{D}}\mathds{P}_{{S}\sim {D}}(\exists\Theta\in \mathbf{\bar{\Theta}}_{{D}}\;\text{s.t.}\;\mathfrak{R}_{{S}}(\Theta)\geq \rho).
\end{align*}
So we only need to show 
\begin{align*}
        \max_{{D}\in\mathbb{R}^{n^+\times d}}\mathds{P}_{{S}\sim {D}}(\exists\Theta\in \mathbf{\bar{\Theta}}_{{D}}\;\text{s.t.}\;\mathfrak{R}_{{S}}(\Theta)\geq \rho)\leq\\\max_{{D}\in \mathbf{D}}\mathds{P}_{{S}\sim {D}}(\exists\Theta\in \mathbf{\bar{\Theta}}_{{D}}\;\text{s.t.}\;\mathfrak{R}_{{S}}(\Theta)\geq \rho).
\end{align*}
To show this result, let $D$ be any non-rank normalized dataset. We show that there exists a rank normalized dataset $D'$ whose probability of success less than or equal to $D$. That is, we show there exists a $D'\in\mathbf{D}$ s.t.
\begin{align*}
        \mathds{P}_{{S}\sim {D}}(\exists\Theta\in \mathbf{\bar{\Theta}}_{{D}}\;|P_{D}(\Theta)\cap S|\geq k\rho)\leq\\\mathds{P}_{{S}\sim {D'}}(\exists\Theta\in \mathbf{\bar{\Theta}}_{{D'}}\;|P_{D'}(\Theta)\cap S|\geq k \rho).
\end{align*}

We first show that using rank of values leads to that same probability of success. %Then, we discuss using unique values. 
Given a dataset $D$, create a new dataset $\bar{D}$ with points $\bar{D}_{i, j}=\sum_{i'\in[n]}\mathds{I}[D_{i, j}>D_{i', j}]$, that is, the number of points in $D$ whose $j$-th dimension is less than $D_{i, j}$. We have 
\begin{align*}        \mathds{P}_{{S}\sim {D}}(\exists\Theta\in \mathbf{\bar{\Theta}}_{{D}}\;|P_{D}(\Theta)\cap S|\geq k\rho)=\\\mathds{P}_{{S}\sim {\bar{D}}}(\exists\Theta\in \mathbf{\bar{\Theta}}_{{\bar{D}}}\;|P_{\bar{D}}(\Theta)\cap S|\geq k \rho).\end{align*}
To see why, note that for every $\Theta\in \mathbf{\bar{\Theta}}_{{D}}$ there exists a $\Theta'\in\mathbf{\bar{\Theta}}_{\bar{D}}$ such that $P_{\bar{D}}(\Theta')=P_{D}(\Theta)$ and vice versa. Furthermore, there is a one-to-one mapping between every data point in $D$ and $\bar{D}$. Thus, for any random sample $S$ from $D$, we have $\exists\Theta\in \mathbf{\bar{\Theta}}_{{D}}\;|P_{D}(\Theta)\cap S|\geq k\rho$ holds if and only if $\exists\Theta\in \mathbf{\bar{\Theta}}_{{\bar{D}}}\;|P_{\bar{D}}(\Theta)\cap S|\geq k \rho$ holds for the corresponding sample set $\bar{S}$ from $\bar{D}$. Both sample set have the same probability, so the total probability of failure remains the same.
%for the above mapping keeps the ordering of the points in any dimension unchanged. \sep{do we need to talk about this in terms of sets?}

Now, assume in $\bar{D}$, there is a set of indices $I$ s.t. $\bar{D}_{i, j}=\bar{D}_{i', j}$ for any $j$ and two rows $i, i'\in I$. 
%to show uniqueness, let $D$ any database constructed as above. For any point $p_i^r$, if there exists another points $p_j^r$ with $p_j^r=p_i^r$, 
Construct a new dataset $\bar{D}'$ with all the points the same as $\bar{D}$, except that for $k$-th value of $I$ for $k\in[|I|]$, $i_k\in I$, let $\bar{D}'_{i_k, j}=\bar{D}_{i_k, j}+k-1$. Note that by construction there cannot be any point in $\bar{D}$ with value $\bar{D}_{i, j}+k-1$ for $k\in[|I|]$, $k>1$ in the $j$-th dimension.  Now for every $\Theta\in \mathbf{\bar{\Theta}}_{\bar{D}}$ there exists a $\Theta'\in\mathbf{\bar{\Theta}}_{\bar{D}'}$ such that $P_{\bar{D}}(\Theta')=P_{\bar{D}'}(\Theta)$ and furthermore, there is a one-to-one mapping for every element in $\bar{D'}$ and $\bar{D}$. Thus, for any random sample $\bar{S}$ from $\bar{D}$, if  $\exists\Theta\in \mathbf{\bar{\Theta}}_{\bar{D}}\;|P_{\bar{D}}(\Theta)\cap S|\geq k\rho$ then we also have $\exists\Theta\in \mathbf{\bar{\Theta}}_{{\bar{D}'}}\;|P_{\bar{D}'}(\Theta)\cap S|\geq k \rho$ holds for the corresponding sample set $\bar{S}'$ from $\bar{D}$. Thus, 
%Note that the probability of success does not increase, because any failure even when sampling from $\bar{D}$ happens still with the same probability when now sampling from $\bar{D}'$. Thus, %\sep{probabiliy need a better way to talk about "new events", this set of all possible events doesn't make sense, I kinda like talking about sets}
\begin{align*}        \mathds{P}_{{S}\sim {\bar{D}}}(\exists\Theta\in \mathbf{\bar{\Theta}}_{{\bar{D}}}\;|P_{\bar{D}}(\Theta)\cap S|\geq k \rho)\leq\\\mathds{P}_{{S}\sim {\bar{D}'}}(\exists\Theta\in \mathbf{\bar{\Theta}}_{{\bar{D}'}}\;|P_{\bar{D}'}(\Theta)\cap S|\geq k \rho).\end{align*}
Iteratively repeating this process for any point that has duplicate values, we obtain a dataset with no duplicate values in the same dimension.

Next, let $I^j$ be the index of records with the $u$ largest values in the $j$-th dimension in $\bar{D}'$. We construct another dataset $\mathring{D}$ where, for any $i\in [n^+]$, we set %create $p'$ as
$$
\mathring{D}_{i, j}=\mathds{I}[i\in I^j](\bar{D}'_{i, j}-(n^+-u))%+\mathds{I}[p\in D^r],
$$
%$$
%p'^r=\mathds{I}[p\in D^r]p^r+(1-\mathds{I}[p\in D^r]),
%$$
that is the value of a dimension is set to zero unless it is among the largest $u$ values in that dimension---in which case it's value is set to $\bar{D}'_{i, j}-(n^+-u)$ where the $n^+-u$ term ensures all values for $\mathring{D}$ are in $[u]$. %Let $D'$ be the dataset of all the $p'$ values.Then, 
$\mathring{D}$ has the same probability of success as $\bar{D}'$. 
This holds using a similar argument as before, since the above modification does not change $P_{\bar{D}}(\Theta)$ for any $\Theta$.
%any threshold $\Theta$ with $N_D(\Theta)=nT-1$ must have $\theta^r\geq\min_{p\in D^r} p^r$. Otherwise, it will exclude more than $nT-1$ points only based on the $r$-th dimension. This means, for any $p\in P_D(\Theta)$, we have $p'\in P_{D'}(\Theta)$. 

Finally, note that $\mathring{D}$ is now a rank normalized dataset and has a lower probability of success than $D$. Note that because there are no duplicate values in any dimension, $\mathring{D}$ contains all the values in $[u]$ for any dimension.

\subsubsection{Proof of Lemma~\ref{lemma:correct_construction}\\}
\if 0
We let $\bar{D}=D^*$ at the beginning and iteratively modify $\bar{D}$. The modification proceeds as follows. It iteratively considers For every dimension $j$ 

let $i_1, ...., i_k$ be the index of records in $D$ such that (1) $D_{i_t, j}\geq 1$ for $t\in[u]$ and that (2) $D_{i_t, :j-1}\neq \mathbf{0}$ ordered by the value $D_{i_t, j}$ in decreasing order---there are at most $u$ such records for a rank normalized dataset $D$. 

To obtain $D$ from modifications to $D^*$, first define $D_{:,j}$ as an $n$-dimensional vector of the values $D_{i, j}$ for all $i\in [n]$. Furthermore, define $D_{:,:j}$ as an $n\times j$-dimensional matrix consisting of values $D_{i, j'}$ for all $i\in [n]$ and $j'\in[j]$. Note that for any dimension $j$, in a rank normalized dataset $D$, the vector $D_{:, j}$ consists of all the values in $[u]$ in addition to $n-u$ zeros. Thus, reorder the points in $D$ such that $D_{:, j}=\bar{D}_{:, j}$ for $j=1$. We next iteratively consider each dimension $j>1$ and modify the vector $\bar{D}_{:, j}$ such that at the end of each iteration we have $\bar{D}_{:, :j}=D_{:, :j}$. Thus, at the beginning of every iteration, $j$, we have  $\bar{D}_{:, :j-1}=D_{:, :j-1}$ as our loop invariant.

\fi

To see why the first assertion holds, note that initially, for every dimension, the records with $D_{:, j}>0$ are all from $e_j\times[u]$. However, as incremental swaps happen other rows also may have  $D_{:, j}>0$. Nonetheless, because the incremental swaps are done in increasing order, no point that had been decremented before is ever incremented, and thus the assertion always holds. 

%To do so, for a dimension $j$,  let $i_1, ...., i_k$ be the index of records in $D$ such that (1) $D_{i_t, j}\geq 1$ for $t\in[u]$ and that (2) $D_{i_t, :j-1}\neq \mathbf{0}$ ordered by the value $D_{i_t, j}$ in decreasing order---there are at most $u$ such records for a rank normalized dataset $D$. Note that if no such record exists, then $\bar{D}_{:, :j}$ and $D_{:, :j}$ are equivalent up to a reordering of rows. To see how, first note that for every row in $D$ such that $D_{i, :j-1}\neq \mathbf{0}$ has  $D_{i_t, j}=0$, and by the loop invariant, $\bar{D}_{i, :j-1}=D_{i, :j-1}$ while $\bar{D}_{i, j}=0$, so for every $i$ such that $D_{i, :j-1}\neq \mathbf{0}$ we have $\bar{D}_{i, :j}=D_{i, :j}$.  Furthermore, since $D$ is rank normalized, then there exists exactly $u$ where for each $v\in[u]$, $D_{i, j}=v$. Such rows have $D_{i, :j-1}=\mathbf{0}$. By construction of $\bar{D}$, for every such $i$, there also exists exactly one row with $\bar{D}_{i, j}=v$ and $D_{i, :j-1}=\mathbf{0}$. such that such rows also can be mapped one to one. Finally, for every other row $D_{i, :j}=\bar{D}_{i, :j}=0$. 

Now Let the indexes in $I$ be $i_1$, ..., $i_k$ ordered by the value $D_{i_t, j}$ in decreasing order for $t\in [k]$. Note that for these indexes we have (1) $D_{i_t, j}\geq 1$ for $t\in[u]$ and that (2) $D_{i_t, :j-1}\neq \mathbf{0}$. We iteratively consider $i_t$ starting from $i_1$ (recall that $i_1$ has the largest value among all $D_{i_t, j}$). For every such row, note that $D_{i_t, :j-1}=\bar{D}_{i_t, :j-1}$ by the loop invariant, but we have $\bar{D}_{i_t, j}=0$ while $D_{i_t, j}\geq 0$. Next, we iteratively modify $\bar{D}_{i_t, j}=0$. Let $\bar{D}_{i_t, j}=v$ ($v=0$ initially) and let $i$ be a row in  $\bar{D}$ such that $\bar{D}_{i, j}=v+1$. At each iteration, we modify $\bar{D}$ so that $\bar{D}_{i_t, j}=v+1$ and $\bar{D}_{i, j}=v$. Note that for any such row $i$, we always have $D_{i, :j-1}=\mathbf{0}$. Observe that after all the incremental swaps, we will have $D_{i_t, :j}=\bar{D}_{i_t, :j}$. Furthermore, for any row such that  $D_{i, j}\geq 1$ and $D_{i_t, :j-1}= \mathbf{0}$, after the incremental swaps (as well as when there was no swaps) there  exists a point $\bar{D}_{i', j}=0$ for some $i'$; and that every other row in both $\bar{D}_{i, :j}=D_{i, :j}=\mathbf{0}$ so that after the incremental swaps, a permutation of the rows ensures that $\bar{D}_{:, :j}=D_{:, :j}$, maintaining the loop invariant. As such, after $r-1$ iterations, we will have $\bar{D}=D$.

\subsubsection{Proof of Lemma~\ref{lemma:incremental_swap}\\}
Note that an incremental swap modifies two points: one point $A$, whose  $j$-th dimension decreases by one, i.e., it has $A=v\times e_i$ becomes $A'=(v-1)\times e_j$  and another point $B$ whose $j$-th dimension increases, i.e., it has $B[j]=v-1$ which becomes $B'=B+e_j$. %, where the value $A$ along the $i$-th dimension decreases by one and the value of $B$ along the $i$-th dimension increases by one---they become swapped along the $i$-th axis. 
We call the original dataset $D$ and the dataset after the swap $D'$. 

To prove the lemma, let $\mathbf{\Theta}$ be the set of possible thresholds. We divide the set of all possible threshold into five set:

\begin{enumerate}
    \item $\mathbf{\Theta}_1=\{\Theta; A\in P_D(\Theta), B\in P_D(\Theta)\}$
    \item $\mathbf{\Theta}_2=\{\Theta; A\not\in P_D(\Theta), B\not\in P_D(\Theta), \Theta^i<j-1\}$
    \item $\mathbf{\Theta}_3=\{\Theta; A\not\in P_D(\Theta), B\in P_D(\Theta)\}$
    \item $\mathbf{\Theta}_4=\{\Theta; A\in P_D(\Theta), B\not\in P_D(\Theta)\}$
    \item $\mathbf{\Theta}_5=\{\Theta; A\not\in P_D(\Theta), B\not\in P_D(\Theta), \Theta^i=j-1\}$
\end{enumerate}

%$\theta_{A,B}$, $\theta_{\bar{A},B}$,$\theta_{A,\bar{B}}$ and $\theta_{\bar{A},\Bar{B}}$, depending on whether $A\in P_D(\Theta)$ and $B\in P_D(\Theta)$ or not. 
We make the following observations:

\textit{(1)}. For any $\Theta\in\mathbf{\Theta}_{1}$, $P_{D'}(\Theta)=(P_{D}(\Theta)\setminus\{A, B\})\cup\{A', B'\}$.

\textit{(2)}. For any $\Theta\in\mathbf{\Theta}_{2}$ $P_{D'}(\Theta)=P_D(\Theta)$.

\textit{(3)}. For any $\Theta\in\mathbf{\Theta}_{3}$, $P_{D'}(\Theta)=(P_D(\Theta)\cup\{A'\})\setminus\{B\}$.

\textit{(4)}. For any $\Theta\in\mathbf{\Theta}_{4}$, $P_{D'}(\Theta)=(P_D(\Theta)\cup\{A'\})\setminus\{A\}$.

\textit{(5)}. $P_{D'}(\mathbf{\Theta_{5}})=P_{D'}(\mathbf{\Theta}_{3})$.

Let $\bar{\mathbf{\Theta}}_i'=\{\Theta;\Theta\in \mathbf{\Theta}_i, |P_{D'}(\Theta)|=\lceil nT\rceil-1\}.$ We are interested in the probability
\begin{align*}
\mathds{P}_{S\sim D'}(\forall \Theta\in\cup_{i\in[5]}\bar{\mathbf{\Theta}}_i', |P_{D'}(\Theta)\cap S|<k\rho)=\\
\mathds{P}_{S\sim D'}(\forall \Theta\in\cup_{i\in[4]}\bar{\mathbf{\Theta}}_i', |P_{D'}(\Theta)\cap S|<k\rho).
\end{align*}
Moreover, let $\bar{\mathbf{\Theta}}_i=\{\Theta;\Theta\in \mathbf{\Theta}_i, |P_D(\Theta)|=\lceil nT\rceil-1\}$. We consider 
\begin{align*}
\mathds{P}_{S\sim D}(\forall \Theta\in\cup_{i\in[5]}\bar{\mathbf{\Theta}}_i, |P_{D}(\Theta)\cap S|<k\rho)\leq\\
\mathds{P}_{S\sim D}(\forall \Theta\in\cup_{i\in[4]}\bar{\mathbf{\Theta}}_i, |P_{D}(\Theta)\cap S|<k\rho).
\end{align*}
We note that for $i\in[4]$ we have $\bar{\mathbf{\Theta}}_i=\bar{\mathbf{\Theta}}'_i$, since for any $\Theta\in\mathbf{\Theta}_i$, $|P_D(\Theta)|=|P_{D'}(\Theta)|$. Thus, our goal is to show the following is at least zero
\begin{align*}
\sum_{a, b\in\{0,1\}}\mathds{P}&(A'=a, B'=b)\times\\
&\mathds{P}_{S\sim D'}(\forall \Theta\in\cup_{i\in[4]}\bar{\mathbf{\Theta}}_i', |P_{D'}(\Theta)\cap S|<k\rho|A'=a, B'=b)-\\
\sum_{a, b\in\{0,1\}}\mathds{P}&(A=a, B=b)\times\\
&\mathds{P}_{S\sim D}(\forall \Theta\in\cup_{i\in[4]}\bar{\mathbf{\Theta}}_i, |P_{D}(\Theta)\cap S|<k\rho|A=a,B=b),
\end{align*}
Where $A=a$ and $B=b$ for $a, b\in\{0, 1\}$ denote the events that $A$ or $B$ are sampled or not, and we analogously define events $A'=a$ and $B'=b$. Note that $\mathds{P}(A=a, B=b)=\mathds{P}(A'=a, B'=b)$, for any $a$ and $b$. Furthermore, denote $Z$ as the set of $D\setminus\{A, B\}$ which is equal to $D'\setminus\{A', B'\}$. Given that all probabilities are conditioned on whether $A, B, A', B'$ are sampled, all sampling procedures are over the remainder of the dataset, i.e., sampling from $Z$. 
\if 0
Thus, we show the following summation is less than zero:  
\begin{align*}
&\sum_{A', B'\in\{0,1\}}\mathds{P}(A', B')\mathds{P}_{S\sim Z}(\forall \Theta\in\cup_{i\in[4]}\bar{\mathbf{\Theta}}_i', |P_{D'}(\Theta)\cap S|<k\rho|A', B')-\\
&\sum_{A,B\in\{0,1\}}\mathds{P}(A, B)\mathds{P}_{S\sim Z}(\forall \Theta\in\cup_{i\in[4]}\bar{\mathbf{\Theta}}_i, |P_{D}(\Theta)\cap S|<k\rho|A,B).
\end{align*}
\begin{align*}
\sum_{a, b\in\{0,1\}}\mathds{P}&(A'=a, B'=b)\times\\
&\mathds{P}_{S\sim Z}(\forall \Theta\in\cup_{i\in[4]}\bar{\mathbf{\Theta}}_i', |P_{D'}(\Theta)\cap S|<k\rho|A'=a, B'=b)-\\
\sum_{a, b\in\{0,1\}}\mathds{P}&(A=a, B=b)\times\\
&\mathds{P}_{S\sim Z}(\forall \Theta\in\cup_{i\in[4]}\bar{\mathbf{\Theta}}_i, |P_{D}(\Theta)\cap S|<k\rho|A=a,B=b),
\end{align*}
\fi
Next, we study the summation in 4 cases, depending on whether $A, A', B, B'$ are sampled.

\textbf{Case 1. $A=B=1$ and $A'=B'=1$.} We show
\begin{align*}
\mathds{P}_{S\sim Z}(\forall \Theta\in\cup_{i\in[4]}\bar{\mathbf{\Theta}}_i, |P_{D'}(\Theta)\cap (S\cup\{A',B'\})|<k\rho| A'=1, B'=1)\\
-\mathds{P}_{S\sim Z}(\forall \Theta\in\cup_{i\in[4]}\bar{\mathbf{\Theta}}_i, |P_{D}(\Theta)\cap (S\cup\{A,B\})|<k\rho| A=1, B=1)=0.
\end{align*}
Note that for any $S$ and $\Theta$, $$|P_{D}(\Theta)\cap (S\cup\{A,B\})|=|P_{D'}(\Theta)\cap (S\cup\{A',B'\})|.$$ This is because $\{A, B\}\notin S$ so $$|P_{D'}(\Theta)\cap (S\cup\{A',B'\})|=|P_{D'}(\Theta)\cap S|+|P_{D'}(\Theta)\cap\{A',B'\}|$$
and that $$|P_{D'}(\Theta)\cap S|=|P_{D}(\Theta)\cap S|$$ while $$|P_{D'}(\Theta)\cap\{A',B'\}|=|P_{D}(\Theta)\cap\{A,B\}|.$$ The former follows because $S$ does not have any $A, B$ while the latter follows because from observations 1-4.

\textbf{Case 2. $A=B=0$ and $A'=B'=0$.} A similar argument shows
\begin{align*}
\mathds{P}_{S\sim Z}(\forall \Theta\in\cup_{i\in[4]}\bar{\mathbf{\Theta}}_i, |P_{D'}(\Theta)\cap S|<k\rho| A'=0, B'=0)\\
-\mathds{P}_{S\sim Z}(\forall \Theta\in\cup_{i\in[4]}\bar{\mathbf{\Theta}}_i, |P_{D}(\Theta)\cap S|<k\rho| A=0, B=0)=0.
\end{align*}
This is similarly because for any $S\sim Z$ we have $$|P_{D'}(\Theta)\cap S|=|P_{D}(\Theta)\cap S|.$$

\textbf{Case 3. $A'=1, B'=0$ and $A=0, B=1$}. Next consider
\begin{align}
\mathds{P}_{S\sim Z}(\forall \Theta\in\cup_{i\in[4]}\bar{\mathbf{\Theta}}_i, |P_{D'}(\Theta)\cap (S\cup\{A'\})|<k\rho| A'=1, B'=0)\notag\\
-\mathds{P}_{S\sim Z}(\forall \Theta\in\cup_{i\in[4]}\bar{\mathbf{\Theta}}_i, |P_{D}(\Theta)\cap (S\cup\{B\})|<k\rho| A=0, B=1).\label{eq:case3}
\end{align}
Note that for any $\Theta\in \Theta_1\cup \Theta_2$, $$|P_{D'}(\Theta)\cap (S\cup\{A'\})|=|P_{D'}(\Theta)\cap (S\cup\{B\})|.$$ For any $\Theta\in \Theta_3$, $$|P_{D'}(\Theta)\cap (S\cup\{A'\})|=|P_{D}(\Theta)\cap (S\cup\{B\})|=|P_{D}(\Theta)\cap S|+1.$$ Finally, for $\Theta\in \Theta_4$, $$|P_{D}(\Theta)\cap (S\cup\{A'\})|=|P_{D}(\Theta)\cap (S\cup\{B\})|+1.$$
Thus, rewriting, let the event $E$ be the event that  we have $$\forall \Theta\in\cup_{i\in[2]}\bar{\mathbf{\Theta}}_i, |P_{D'}(\Theta)\cap (S\cup\{A'\})|$$ which occurs if and only if $$\forall \Theta\in\cup_{i\in[2]}\bar{\mathbf{\Theta}}_i, |P_{D}(\Theta)\cap (S\cup\{B\})|.$$ For simplicity let $V$ denote the event $$\forall \Theta\in\bar{\mathbf{\Theta}}_3, |P_{D}(\Theta)\cap S|<k\rho-1,$$ 
Which occurs if and only if 
$$\forall \Theta\in\bar{\mathbf{\Theta}}_3, |P_{D'}(\Theta)\cap S|<k\rho-1.$$ 
Thus, we can rewrite Eq.~\ref{eq:case3} as
\begin{align*}
\mathds{P}_{S\sim Z}&(E, V, \forall \Theta\in\bar{\mathbf{\Theta}}_4, |P_{D}(\Theta)\cap S|<k\rho-1)\\
&-\mathds{P}_{S\sim Z}(E, V,  \forall \Theta\in\bar{\mathbf{\Theta}}_4, |P_{D}(\Theta)\cap S|<k\rho)=\\
\mathds{P}_{S\sim Z}&(E, V)\times \\
&\big(\mathds{P}_{S\sim Z}(\forall \Theta\in\bar{\mathbf{\Theta}}_4, |P_{D}(\Theta)\cap S|<k\rho-1|E, V)-\\
&\qquad \mathds{P}_{S\sim Z}(\forall \Theta\in\bar{\mathbf{\Theta}}_4, |P_{D}(\Theta)\cap S|<k\rho|E, V)\big),
\end{align*}
Where we can replace $P_D$ with $P_{D'}$ because $P_D(\Theta)\cap S=P_{D'}(\Theta)\cap S$ when $S$ is sampled from $Z$. The above is equal to
\begin{align*}
-\mathds{P}_{S\sim Z}&(E, V)\times \\
&\mathds{P}_{S\sim Z}(\forall \Theta\in\bar{\mathbf{\Theta}}_4,|P_{D}(\Theta)\cap S|<k\rho,\\ &\qquad\exists \Theta\in\bar{\mathbf{\Theta}}_4, |P_{D}(\Theta)\cap S|=k\rho-1|E, V)=\\
-\mathds{P}_{S\sim Z}&(E, V, \forall \Theta\in\bar{\mathbf{\Theta}}_4, |P_{D}(\Theta)\cap S|<k\rho,\\&\exists \Theta\in\bar{\mathbf{\Theta}}_4, |P_{D}(\Theta)\cap S|=k\rho-1)
\end{align*}

\textbf{Case 4. $A'=0, B'=1$ and $A=1, B=0$.} Next consider
\begin{align*}
\mathds{P}_{S\sim Z}(\forall \Theta\in\cup_{i\in[4]}\bar{\mathbf{\Theta}}_i, |P_{D'}(\Theta)\cap S\cup\{B'\}|<k\rho| A'=0, B'=1)\\
-\mathds{P}_{S\sim Z}(\forall \Theta\in\cup_{i\in[4]}\bar{\mathbf{\Theta}}_i, |P_{D}(\Theta)\cap S\cup\{A\}|<k\rho| A=1, B=0).
\end{align*}
Note that for any $\Theta\in \Theta_1\cup \Theta_2$, $$|P_{D'}(\Theta)\cap (S\cup\{B'\})|=|P_{D'}(\Theta)\cap (S\cup\{A\})|.$$ 
Furthermore, for any $\Theta\in \Theta_3$, 
$$|P_{D'}(\Theta)\cap (S\cup\{B'\})|=|P_{D}(\Theta)\cap (S\cup\{A\})|=|P_{D}(\Theta)\cap S|.$$
Finally, for $\Theta\in \Theta_4$, $$|P_{D}(\Theta)\cap S\cup\{A'\}|=|P_{D}(\Theta)\cap S\cup\{B\}|+1.$$ 
Thus, rewriting the above, we have
\begin{align*}
&\mathds{P}_{S\sim Z}(E, \forall \Theta\in\bar{\mathbf{\Theta}}_3 |P_{D}(\Theta)\cap S|<k\rho, ,\\&\qquad\qquad \forall \Theta\in\bar{\mathbf{\Theta}}_4, |P_{D}(\Theta)\cap S|<k\rho)\\
&-\mathds{P}_{S\sim Z}(E, \forall \Theta\in\bar{\mathbf{\Theta}}_3, |P_{D}(\Theta)\cap S|<k\rho,\\&\qquad\qquad  \forall \Theta\in\bar{\mathbf{\Theta}}_4, |P_{D}(\Theta)\cap S|<k\rho-1)=\\
&\mathds{P}_{S\sim Z}(E, V', \forall\Theta\in\bar{\mathbf{\Theta}}_4, |P_{D}(\Theta)\cap S|<k\rho,\\&\qquad\qquad\exists \Theta\in\bar{\mathbf{\Theta}}_4, |P_{D}(\Theta)\cap S|=k\rho-1)
\end{align*}
Using an argument similar to before but with $V'$ defined as  $$V'= \forall \Theta\in\bar{\mathbf{\Theta}}_3, |P_{D}(\Theta)\cap S|<k\rho.$$

\textbf{Combining the results}. Finally, putting everything together we have, 
\begin{align*}
\sum_{a, b\in\{0,1\}}\mathds{P}&(A'=a, B'=b)\times\\
&\mathds{P}_{S\sim D'}(\forall \Theta\in\cup_{i\in[4]}\bar{\mathbf{\Theta}}_i', |P_{D'}(\Theta)\cap S|<k\rho|A'=a, B'=b)-\\
\sum_{a, b\in\{0,1\}}\mathds{P}&(A=a, B=b)\times\\
&\mathds{P}_{S\sim D}(\forall \Theta\in\cup_{i\in[4]}\bar{\mathbf{\Theta}}_i, |P_{D}(\Theta)\cap S|<k\rho|A=a,B=b)=\\
&-\mathds{P}_{S\sim Z}(E, V, \forall \Theta\in\bar{\mathbf{\Theta}}_4, |P_{D}(\Theta)\cap S|<k\rho,\\&\qquad\qquad\exists \Theta\in\bar{\mathbf{\Theta}}_4, |P_{D}(\Theta)\cap S|=k\rho-1)+\\
&\mathds{P}_{S\sim Z}(E, V', \forall\Theta\in\bar{\mathbf{\Theta}}_4, |P_{D}(\Theta)\cap S|<k\rho,\\&\qquad\qquad\exists \Theta\in\bar{\mathbf{\Theta}}_4, |P_{D}(\Theta)\cap S|=k\rho-1)\geq 0.
\end{align*}%\sep{actually not equality because of $\mathds{P}(A', B')$}.

Which prove 
\begin{align*}
\mathds{P}_{S\sim D'}(\forall \Theta\in\cup_{i\in[4]}\bar{\mathbf{\Theta}}_i', |P_{D'}(\Theta)\cap S|<k\rho)\geq    \\
\mathds{P}_{S\sim D}(\forall \Theta\in\cup_{i\in[5]}\bar{\mathbf{\Theta}}_i, |P_{D'}(\Theta)\cap S|<k\rho).
\end{align*}
as desired.

\subsection{Proof of Theorem~\ref{thm:alg_guarantee}}\label{sec:proof:final_guarantee}
First, note that $k^+\leq n^+$, so that whenever $k^+>\frac{1}{1-T} $ and $r\leq \frac{k^+}{k^+(1-T)-1}$ then $r\leq \frac{k^+}{k^+(1-T)-1}$. Note that $\frac{k^+}{k^+(1-T)-1}\geq \frac{1}{1-T}$ and thus, whenever $k^+>\frac{1}{1-T}$ and $r\leq \frac{1}{1-T}$ we have $r\leq\frac{n^+}{n^+(1-T)-1}$. Thus, the algorithm \texttt{adj-target} returns a threshold $T'$ such that 
\begin{align}
    \mathds{P}_{S\sim \str{L}\times\str{R}}(\exists\Theta\in \mathbf{\bar{\Theta}}\;\text{s.t.}\;\mathfrak{R}_{\str{S}}(\mathring{\Pi},\Theta)\geq T')\leq \delta,
\end{align}
Which implies the the selected threshold meets the recall target.

\subsection{Proof of Lemma~\ref{lemma:adj_target_disj}}\label{sec:proof:conjunction}
To consider conjunctions, first consider the $i$-th cluase scaffole $\mathring{\kappa_i}$ and let $\phi_{i, j}$ be the $j$-th featurization used in the $i$-th clause scaffold of the decomposition. Recall that $\mathring{\kappa_i}(\str{l}, \str{r}; \theta)=\bigvee_j\mathds{I}[\phi_{i, j}(\str{l}, \str{r})\leq \theta]$. Now define a new featurization $\phi'_i(\str{l}, \str{r})=\min_{j}\phi_{i, j}(\str{l}, \str{r})$, and define a clause scaffold $\kappa'(\str{l}, \str{r}; \theta)=\mathds{I}[\phi'_i(\str{l}, \str{r})\leq \theta]$. Now observe that for any $\theta$, $\kappa'_i(\str{l}, \str{r}; \theta)=\kappa_i(\str{l}, \str{r}; \theta)$. Thus, we can construct a logical scaffold consisting only of conjunctions that is equivalent to the original conjunction. Applying Theorem~\ref{thm:alg_guarantee} to this scaffold proves the desired result.

\if 9
\subsection{Proof of Prop.~\ref{prop:cost_complexity}}
\sep{rewrite}
The LLM calls made by \name{} can be categorize into 4 categories. (1) \textit{Labeling}: the cost of using LLM to label a set of sampled pairs (Lines~\ref{eq:line_sampling} and \ref{eq:sample_2} in Alg.~\ref{alg:join_overall}).  (2) \textit{Discovery}: the cost of performing LLM calls in to generate candidate factorizations, $\Phi$, (i.e., the cost of LLM calls in Alg.~\ref{alg:generator_function}). (3) \textit{Extraction}: the cost of applying the LLM to obtain extracted features, (i.e., the cost of calling extraction functions $\mathcal{X}_L$ and $\mathcal{X}_R$ if they are LLM powered both for creating the parameterized decomposition and selecting the final threshold parameters). (4) \textit{Refinement}: the cost of refining the predicted join set from the faeturization to obtain the final join result. 

Next we analyze each component. The Labeling cost $C_L$, is the cost of labeling $k$ pairs selected uniformly at random from $\str{L}\times\str{R}$, and its cost is proportional to $k$ times the average token length in the set $\str{L}\times\str{R}$ and thus grows linearly with $k$. Let $t$ be the average token length, so that $C_L=O(k\times t)$. The Discovery cost, $C_D$ depends on the number of factorizations created in the candidate generation phase, $|\Phi|$. There are a constant number of featurizations, so that $C_D$ is a constant multiple of $|\Phi|$. The Extraction cost, $C_E$ also depends on $|\Phi|$, since we perform feature extraction potentially for all the candidate featurizations when creating the parameterized decomposition or when selecting threshold parameters. Thus, this step costs number of tokens proportional to $t|\Phi|(|\str{L}|+|\str{R}|)$. Finally, refinement cost, $C_R$ is proportional to the number of tokens in predicted set created by \name{} in line~\ref{line:alg:refine} of Alg.~\ref{alg:join_overall}. Note that $\hat{\str{Y}}$ contains at least $T\times n^+$ pairs with high probability, but may contain additional false positives. Let $|\hat{\str{Y}}|=f\times n^+$, where $f$ is a false positive factor quantifying how many false positives are created. \qed
\fi
\section{Details of LLM-Powered Functions}\label{sec:prompts}
Detailed prompts for all our LLM powered functions are available in our source code. Here, we provide a summary of the prompts for all the functions in \undef{blue} in Alg.~\ref{alg:generator_function}. We note that our actual prompts include additional content for passing examples and defining output format and instruction. Additionally, our pipeline includes validation LLM calls that observe input/output pairs and feed them back to the LLM to fix any error. 

\undef{\texttt{get-featurization-descriptions}}:
%\begin{minipage}{0.5\textwidth}
  \begin{lstlisting}
Design a set of features that are useful for deciding if the join condition is satisfied. Provide a high level discussion of the type of information needed and how it *logically* help decide a join condition is satisfied. The information should contain all possibly relevant information to perform the join, that is, using them we should be able to decide if the join condition holds. Use the examples to decide what information is useful, and include all features that are needed. 
\end{lstlisting}
%\end{minipage}

\undef{\texttt{get-feature-description-for-col}}:
%\centering
  %\begin{minipage}{0.5\textwidth}
  \begin{lstlisting}
Let's use the following feature: {feature}   
Provide a high-level discussion of what needs to be obtained from each of {left_column_name} and {right_column_name}. Do not specify the operation type or provide type hints. Just discuss in natural language what needs to be done and what the expected output is. Provide the discussion for each feature independently (i.e., do not refer to the discussion for {left_column_name} in your discussion of {right_column_name}) so that each can be understood without knowing the other. 
\end{lstlisting}%\end{minipage}

\texttt{\undef{should-use-llm}}:
  \begin{lstlisting}
Decide the type of the operation that needs to be performed among the following operations:
- Extract: A value or set of values **explicitbly stated in the text** should be extracted and retained **as stated in the text**.
- Infer: A value or set of values should be inferred based on the text. This includes cases where we need to enrich the data to contain additional information based on general knowledge, reasoning and the join condition
- Split: We need to make comparisons with all the content of the text, possibly using rolling windows andy by splitting up  the text into sentences or paragraphs.
\end{lstlisting}
If the LLM decides the values need to inferred, then we use LLMs and otherwise use code-based extraction. Additionally, if the extraction is only short keywords and not sentences, we also use LLM-based extraction that we observed to be more robust.

\texttt{\undef{get-extraction-prompt}}:
\begin{lstlisting}
Provide a detailed description of the operation that needs to be performed on {column_name} obtain the feature, along with the expected output type. Recall that the goal is to extract the relevant information from the text of {column_name}. Do not provide a solution (or any details on how the output should be obtained), only specify the task so that if your description is given to a human along with the {column_name} value, the human can obtain the expected output on their own. Make sure to instruct the model about the additional information that needs to be inferred besides from what is explicitly in the text. 
\end{lstlisting}

\texttt{\undef{get-extraction-code}}:
\begin{lstlisting}
To perform the task, I want to design a function that extracts the required values. Write a python function that goes over every paragraph and uses characters or keywords to extract the relevant information.
\end{lstlisting}

\texttt{\undef{get-distance-func}}:
\begin{lstlisting}
Based on the description of the feature below, which distance function should be used to measure the distance between the features? Choose between:
    - word_overlap_similarity: applicable to features with similar content written with the same words
    - semantic_similarity: applicable to features with similar semantic content but not necessarily the same words
    - arithmetic_similarity: applicable to numerical featureas
    - date_similarity: applicable to features that contain dates
\end{lstlisting}
We use pre-defined functions for each of the four categories. 

\textbf{Join prompts}. Next, we describe prompts for our joins.

\textit{Citations}: 
\begin{lstlisting}
Is the product described in {left_column_name} column the same product described in {right_column_name}?
\end{lstlisting}

\textit{Police} Records:
\begin{lstlisting}
Does the police report in {right_column_name} refer to the same incident as the police report in {left_column_name}?
\end{lstlisting}

\textit{Categorize}:
\begin{lstlisting}
Can the product described in {left_column_name} column be classified with the category in {right_column_name}?
\end{lstlisting}

\textit{BioDEX}:

\hspace*{-0.4cm}
\begin{minipage}{\columnwidth}
\begin{lstlisting}
Does the medical reaction term in {right_column_name} apply to the patient discussed in {left_column_name}?
\end{lstlisting}    
\end{minipage}

\textit{Movies}:
\begin{lstlisting}
Is the person mentioned in {right_column_name} a cast or crew member in the movie in the {left_column_name}?
\end{lstlisting}

\textit{Products}:
\begin{lstlisting}
Is the product described in {left_column_name} column the same product described in {right_column_name}?
\end{lstlisting}
}
{}

\end{document}